\documentclass[pra,showpacs,amsmath,amssymb,amsfonts,lengthcheck,longbibliography,superscriptaddress]{revtex4-2}

\usepackage{mathrsfs}  
\usepackage[usenames,dvipsnames]{color}
\usepackage{soul}
\usepackage{cancel}
\usepackage[
verbose,
colorlinks=true,
naturalnames=true,
linkcolor=blue,
]{hyperref}
\usepackage{braket}

\usepackage{graphicx}
\usepackage{subfigure}
\usepackage{amsthm}
\usepackage{verbatim}
\usepackage{dcolumn}
\usepackage{bm}
\usepackage{epsf}

\hypersetup{colorlinks,linkcolor={blue},citecolor={blue},urlcolor={blue}} 

\newcommand{\tr}[1]{\mathrm{tr}\left\{#1\right\}}
\newcommand{\ptr}[2]{\mathrm{tr_{#1}}\left\{#2\right\}}

\newcommand{\etal}{\textit{et al. }}
\newcommand{\e}[1]{\exp{\left(#1\right)}}
\newcommand{\lo}[1]{\ln{\left(#1\right)}}

\newcommand{\bla}{bla\\bla\\bla\\bla\\bla}

\newcommand{\mrm}[1]{\mathrm{#1}}

\def\dbar{{\mathchar'26\mkern-12mu {\rm d}}}

\DeclareMathOperator*{\sumint}{%
\mathchoice%
  {\ooalign{$\displaystyle\sum$\cr\hidewidth$\displaystyle\int$\hidewidth\cr}}
  {\ooalign{\raisebox{.14\height}{\scalebox{.7}{$\textstyle\sum$}}\cr\hidewidth$\textstyle\int$\hidewidth\cr}}
  {\ooalign{\raisebox{.2\height}{\scalebox{.6}{$\scriptstyle\sum$}}\cr$\scriptstyle\int$\cr}}
  {\ooalign{\raisebox{.2\height}{\scalebox{.6}{$\scriptstyle\sum$}}\cr$\scriptstyle\int$\cr}}
}

\renewcommand{\vec}[1]{\boldsymbol{#1}}

\newcommand{\draftmode}{1}    
\newcommand{\notetoself}[1]{\ifnum \draftmode=1 {\color[rgb]{0,0,0.8} [#1]} \fi}  
\newcommand{\cuttext}[1]{\ifnum \draftmode=1 {\color[rgb]{0,0.5,0} [#1]} \fi}  
\newcommand{\warntext}[1]{\ifnum \draftmode=1 {\color[rgb]{0.9,0.6,0} #1} \else {#1} \color{black} \fi}

\begin{document}

\title{Thermodynamics of quantum information in noisy polarizers}

\author{Maxwell Aifer}
\email{maifer1@umbc.edu}
\affiliation{Department of Physics, University of Maryland, Baltimore County, Baltimore, MD 21250, USA}

\author{Nathan M. Myers}
\email{myersn1@vt.edu}
\affiliation{Department of Physics, Virginia Tech, Blacksburg, VA 24061, USA}

\author{Sebastian Deffner}
\email{deffner@umbc.edu}
\affiliation{Department of Physics, University of Maryland, Baltimore County, Baltimore, MD 21250, USA}

\date{\today}

\begin{abstract}
Among the emerging technologies with prophesied quantum advantage, quantum communications has already led to fascinating demonstrations---including quantum teleportation to and from satellites. However, all optical communication necessitates the use of optical devices, and their comprehensive quantum thermodynamic description is still severely lacking. In the present analysis we prove several versions of Landauer's principle for noisy polarizers, namely absorbing linear polarizers and polarizing beamsplitters. As main results we obtain statements of the second law quantifying the minimal amount of heat that is dissipated in the creating of linearly polarized light. Our findings are illustrated with an experimentally tractable example, namely the temperature dependence of a quantum eraser.
\end{abstract}

\maketitle

\section{\label{sec:intro}Introduction}

There are generally three applications in which quantum advantage is expected to be of technological significance---computation, sensing, and communication \cite{Raymer2019QST}. Despite the distinct and unique technological challenges of each area, devices designed for each of these applications can be considered quantum devices that process information. Hence, they can be described by means of quantum thermodynamics \cite{Deffner2019book}.

The development of classical thermodynamics of information \cite{Parrondo2015NP} originated in the formulation of Landauer's principle \cite{landauer_irreversibility_1961}. This statement of the second law asserts that any computational task requiring the erasure of information must result in dissipation of heat, and the amount of heat produced is at least $k_B T \ln(2)$ times the number of logical bits erased. Recent years have seen intense research efforts in generalizing the bound to a wide variety of physical situations, such as classical systems with both discrete and continuous state spaces \cite{piechocinska_information_2000, deffner_information_2013,Boyd2016PRL,Boyd2016NJP,Boyd2018PRX}, as well as quantum systems undergoing Markovian and non-Markovian dynamics \cite{hilt_landauers_2011, lorenzo_landauers_2015, van_vu_finite-time_2022, goold_nonequilibrium_2015}. Landauer's principle has even been experimentally verified in microscale systems using an overdamped colloidal particle in a double-well potential \cite{Berut2012}. However, given the rather heuristic nature of the original formulation \cite{landauer_irreversibility_1961}, it is still being debated whether the original statement can really be shown in all generality \cite{maroney_generalizing_2009, norton_waiting_2011, maroney_absence_2005}. Notably, some authors have recently interpreted certain non-Markovian processes as violations of Landauer's principle \cite{Bylicka2016, Pezzutto2016, Zhang2021, hu_relation_2022, man_validity_2019}.

Curiously, most of the current discussion is focused on statements of Landauer's principle for computation. Yet, the communication of quantum information also obviously incurs thermodynamic costs, which can be determined with versions of the Landauer bound. Such ``dynamical'' formulations of Landauer's principle can be traced back to Bremermann \cite{Bremermann1967}, who proposed that any computational device must obey the fundamental laws of physics namely \emph{special relativity}, \emph{quantum mechanics}, and \emph{thermodynamics}. Then, identifying Shannon's noise energy with the $\Delta E$ in Heisenberg's uncertainty relation \cite{Heisenberg1927} for energy and time, $\Delta E \Delta t\geq \hbar$, he found an upper bound on the rate with which information can be communicated. A more rigorous argument was put forward by Bekenstein \cite{Bekenstein1981}  in the context of black hole thermodynamics \cite{Bekenstein1973,Bekenstein1974,Hawking1975,Bekenstein1981PRD,Bekenstein1990}. However, the Bremermann-Bekenstein bound does not seem to enjoy the same prominence as Landauer's principle, and in fact many different statements for the maximal rate with which entropy and information can be communicated have been formulated \cite{Pendry1983,Landauer1987,Bekenstein1988,Caves1994,Blencowe2000,Lloyd2004,Garbaczewski2007,Pei_Rong2010,Deffner2010PRL,Guo2011,Guo2012,Bousso2017,Lewis2019,Deffner2020PRR}.

In the present work, we will be analyzing the thermodynamics of some aspects of quantum optical communication. This is motivated by the fact that light is an attractive physical platform, given that it can carry information at the greatest possible speed and that it interacts relatively weakly with the environment \cite{Gisin2007NP,Chen2021JP}. The most significant technical challenge in creating optical quantum networks is overcoming attenuation \cite{wehner_quantum_2018, gisin_quantum_2002}. Entanglement swapping schemes can extend the range of quantum communication \cite{Lu2022RMP}, but the performance of this method is still limited by the amount of dissipation caused by the individual optical components used \cite{briegel_quantum_1998, munro_inside_2015, jiang_optimal_2007}. Hence, a comprehensive thermodynamic characterization of noisy optical elements appears urgently needed.

Most communication schemes employ linearly polarized light \cite{eldada_optical_2004}. Such light can be produced by sending unpolarized light through a linear polarizer, which absorbs one component of the electric field and transmits the component perpendicular to it. We will call such optical elements absorbing linear polarizers (ALP). It is also possible to produce linearly polarized light using a polarizing beamsplitter (PBS), which transmits one component of the electric field and reflects the component which is perpendicular to it. While it is clear that some heat must be generated by the ALP, it is less obvious that there is any minimum amount of dissipation which is caused by the PBS \cite{jones_ultimate_1962, im_microscale_2018}. In what follows we will show that both devices in fact are responsible for dissipation of heat when an assumption of locality is made, which can be understood through the concept of modularity dissipation \cite{Boyd2018PRX}.
Most light sources found in nature can be accurately described by the classical laws of electrodynamics, and their description does not benefit from quantization of the field. However, there are engineered sources available which can produce single or entangled pairs of photons on demand, and the results of experiments done with these sources cannot always be explained by classical electrodynamics \cite{lounis_single-photon_2005, edamatsu_entangled_2007,takeuchi_recent_2014}. In our analysis, we will first address the case of classical light. The ``classicality" of light can be characterized in a number of ways, such as by the negativity of the Wigner function \cite{konrad_quantum_2019, konrad_quantum_2019, kenfack_negativity_2004}. For our purposes, we will consider classical light to be a statistical mixture of approximate coherent states at large photon number. A derivation is given of a version of Landauer's principle for ALPs acting on classical sources followed by a derivation of a similar result applicable to to PBSs, assuming the reflected light is inaccessible.
Then, we will provide insight into more exotic sources with nonclassical behavior. We show that the findings for the PBS acting on classical light can be seen as stemming from local nonconservation of globally conserved quantities, and that an analogous dissipation cost applies to quantum information processing tasks. Having generalized the PBS result to the quantum case, we investigate the ALP acting on quantum sources, and argue that an ALP can be modeled as a collection of PBSs strung together with intervening thermal reservoirs. Using this model we are able to probe the dynamics of the polarization process, rather than just its end result. Interestingly, our model becomes formally equivalent to a repeated interaction scheme (or collision model) which has received attention recently in the field of open quantum systems \cite{campbell_collision_2021} including in the context of Landauer's principle \cite{Lorenzo2015}, and we leverage these existing results to quantify the dissipation which occurs as (possibly nonclassical) light propagates through the ALP. Notably, our analysis reveals a discontinuity in the slope of the energy-entropy curve as temperature approaches zero, which agrees with previous investigations of the random scattering of light \cite{aiello_physical_2005}.
The collision model approach also allows us to describe the dependence of the ALP's behavior on its temperature, both in terms of optical extinction and decoherence.  This permits us to make predictions concerning the relationship between temperature and the signature of entanglement observed in the quantum eraser \cite{scully_quantum_1982,kim_delayed_2000,walborn_double-slit_2002,kang_electronic_2007}. We predict that at higher temperatures the observed restoration of interference caused by measurement is suppressed.

\section{Landauer's principle for linear polarizers}

In the following, we will be deriving various statements of Landauer's principle. The principle expresses that a decrease in Shannon entropy of an information-bearing degree of freedom is accompanied by the dissipation of heat \cite{landauer_irreversibility_1961,bennett_notes_2003, piechocinska_information_2000}
\begin{equation}
\label{classical-landauer-principle}
    \dbar Q  \geq k_B T\, (-ds),
\end{equation}
where $s$ is the differential Shannon entropy \cite{bucholtz_handbook_2014}
\begin{equation}
\label{diff-shannon-def}
    s = -\sumint f_X(x) \lo{f_X(x)}.
\end{equation}
Here, $f_X(x)$ denotes the probability density function (PDF) of a discrete or continuous random variable $X$.

In our case, $X$ is the electric field, in which information is encoded. Then, an ALP has the effect of erasing this information stored in one component of the electric field. Therefore, by Landauer's principle the ALP should be required to dissipate heat. This is in fact the case, and we now provide the corresponding statement of Landauer's principle.

\subsection{Classical absorbing linear polarizer}

The ALP transmits horizontally polarized light and absorbs vertically polarized light. Consider a single mode of the electric field of frequency $\omega$. The vertical component of the electric field is \cite{grynberg_introduction_2010}
\begin{equation}
\label{single-mode-vert-field}
   E_\text{v}(z,t) = i E_{0 \omega} \left[\alpha_\text{v}(\omega)e^{i (\omega t - k z)}-\alpha_\text{v}(\omega)^*e^{i(kz -  \omega t )}\right],
\end{equation}
where the complex amplitude $\alpha_\text{v}(\omega)$ will be considered a random variable with PDF $f_\alpha$ and $E_{0\omega}$ is an arbitrary constant electric field strength. The distribution is assumed to have finite variance (and therefore finite energy) but is otherwise unspecified. 
The differential Shannon entropy of $\alpha_\text{v}(\omega)$ is 
\begin{equation}
\label{single-mode-diff-entropy}
    s[\alpha_\text{v}(\omega)] =-\int d^2 \alpha_\text{v}\, f_\alpha(\alpha_\text{v}) \lo{f_\alpha(\alpha_\text{v})} ,
\end{equation}
where the integral is over the entire complex plane. We assume the field decays rapidly outside of some region with volume $V$, and so the ensemble-averaged energy content of the field's vertical component is
\begin{equation}
\label{single-mode-energy}
    E = \frac{1}{2} V \epsilon_0 E_{0\omega}^2 \braket{ |\alpha_\text{v}|^2}.
\end{equation}
After the light has passed through the polarizer, we assume that the state of the field afterwards is a thermal state, given by the canonical ensemble, $f^\mrm{can}\propto \e{-E/k_BT_P}$, where $T_P$ is the temperature of the polarizer. Although it is well known that this assumption leads to an infinite total energy in the completely classical case when all frequencies are present \cite{ehrenfest_welche_1911}, this does not affect the present argument, which only concerns a single mode of the field. This thermal state has entropy $s^\mrm{can}$, and so the total erasure is simply the difference 
\begin{equation}
    \label{erasure-def}
   -\Delta s=s-s^\mrm{can}.
\end{equation}
We show in Appendix \ref{classical-landauer-appendix} that the following statement of Landauer's principle holds
\begin{equation}
\label{finite-landauer-alp}
    Q \geq k_B T_P \left(e^{-\Delta s}-1\right).
\end{equation}
Expanding about zero for small $\Delta s$, we obtain the Landauer bound
\begin{equation}
\label{diff-landauer-alp}
    \dbar Q \geq k_B T_P \left(- ds\right).
\end{equation}

Equations~\eqref{finite-landauer-alp} and \eqref{diff-landauer-alp} are our first main results, asserting Landauer's principle for the ALP. The assumption of finite variance of the field was necessary to avoid having a divergent amount of heat dissipated, and similarly it was necessary to assume that the ALP does not perfectly polarize the field (a thermal distribution at temperature $T_P$ remains in the vertical component) to avoid a divergent change in differential entropy. The assumption that all energy is converted to heat is justified on the basis that the ALP is a passive component with no mechanism for storing energy. If there are multiple frequencies $\omega_j$ in the source, we may treat the amplitude at each frequency as a random variable $\alpha_\text{v}(\omega_j)$, and we assume these are not correlated with each other. In this case due to the additivity of the Shannon entropy and the energy, the same result holds (in fact this argument holds even for a continuum of modes). We leave the case of correlated amplitudes for future work. It should also be noted that the quantity $-ds$ should not be interpreted as an amount of data which is erased, but simply as a decrease in Shannon entropy. There is a subtle distinction between these concepts given that it is possible that there is still some finite Shannon entropy even if all of the original data has been erased, if the original content of the signal has been replaced by random thermal fluctuations.

\subsection{Quantum absorbing linear polarizer}

The somewhat natural question is how things change if the light is treated quantum mechanically. In the quantum case the amplitude $\alpha$ is no longer an observable, and thus it does not have a probability density function. Instead we consider the dimensionless quadratures \cite{Schleich2011quantum}
\begin{equation}
    \label{q-def}
    \hat{q}=\frac{1}{\sqrt{2}}\left(\hat{a}^\dag + \hat{a}\right),
\end{equation}
and
\begin{equation}
    \label{p-def}
    \hat{p}=\frac{i}{\sqrt{2}}\left(\hat{a}^\dag - \hat{a}\right),
\end{equation}
in terms of which the Hamiltonian is expressed as
\begin{equation}
    \hat{H}=\frac{1}{2}\hbar \omega \left(\hat{q}^2 + \hat{p}^2\right).
\end{equation}
Therefore the average energy is
\begin{equation}
\label{harmonic-energy-qp}
    E = \frac{1}{2}\hbar \omega \left(\braket{q^2}+\braket{p^2}\right)
\end{equation}
Since $\hat{q}$ and $\hat{p}$ do not commute, the interpretation of a joint probability distribution over these variables is ambiguous, so it is not immediately clear how to define the entropy. 

Especially in quantum optics \cite{Schleich2011quantum} it has proven particularly useful to express quantum states in their Wigner representation
\begin{equation}
\label{eq16}
W(q,p)=\frac{1}{\pi\hbar}\,\int dy\,\bra{q+y}\rho\ket{q-y}\,e^{-2 i p \,y/\hbar}\,.
\end{equation}
Correspondingly, a Wigner entropy can be defined as \cite{van_herstraeten_quantum_2021}
\begin{equation}
\label{wigner-ent-def}
    s_\text{W}[W] = -\int  dq dp\, W(q,p) \lo{W(q,p)},
\end{equation}
which is nothing but the Weierstrass transform of the Wehrl entropy. The latter has been shown to be thermodynamically significant \cite{Santos2017PRL}. However, the Wigner entropy is only defined for states with strictly positive Wigner functions. Nonetheless, we are able to define a ``maximal Wigner entropy" for arbitrary states by the following argument: the sub-additivity of classical Shannon entropy dictates that for a joint distribution $f_{X,Y}$ with marginal distributions $f_X$ and $f_Y$, \cite{witten_mini-introduction_2020}
\begin{equation}
    \label{sub-additivity}
    s[f_{X,Y}]\leq s[f_X]+s[f_Y].
\end{equation}
When the Wigner entropy is defined, it is the same as the Shannon entropy over a joint distribution. Also, recall that the marginals of the Wigner distribution are the distributions for the quadratures $q$ and $p$ \cite{Schleich2011quantum}. Therefore, the following sub-additivity rule is obeyed
\begin{equation}
    \label{sub-additivity-wigner}
    s_\text{W}[W]\leq s[f_q]+s[f_p].
\end{equation}
Even when $s_\text{W}[W]$ is not defined, we can still evaluate $s[f_q]$ and $s[f_p]$, and we define the maximal Wigner entropy as
\begin{equation}
    \label{maximal-wigner-ent}
    s_\text{WM}[W]= s[f_q]+s[f_p].
\end{equation}
Then erasure is bounded above by
\begin{equation}
    -\Delta s_\text{W} \leq s_\text{WM}[W] - s_\text{W}[W'].
\end{equation}
It is shown in Appendix~\ref{quantum-landauer-appendix} that the following version of Landauer's principle holds
\begin{equation}
\label{finite-landauer-alp-quantum}
    Q \geq \frac{\hbar \omega}{2} \coth \left(\frac{\hbar\omega}{2 k_B T} \right) \left(e^{-\Delta s_\text{W}}-1\right).
\end{equation}
Again, expanding for small $\Delta s$ we have
\begin{equation}
\label{diff-landauer-alp-quantum}
    \dbar Q \geq\coth \left(\frac{\hbar\omega}{2 k_B T} \right) \left(- ds_\text{W}\right),
\end{equation}

\begin{figure}
\centering
\includegraphics[width=3.4in]{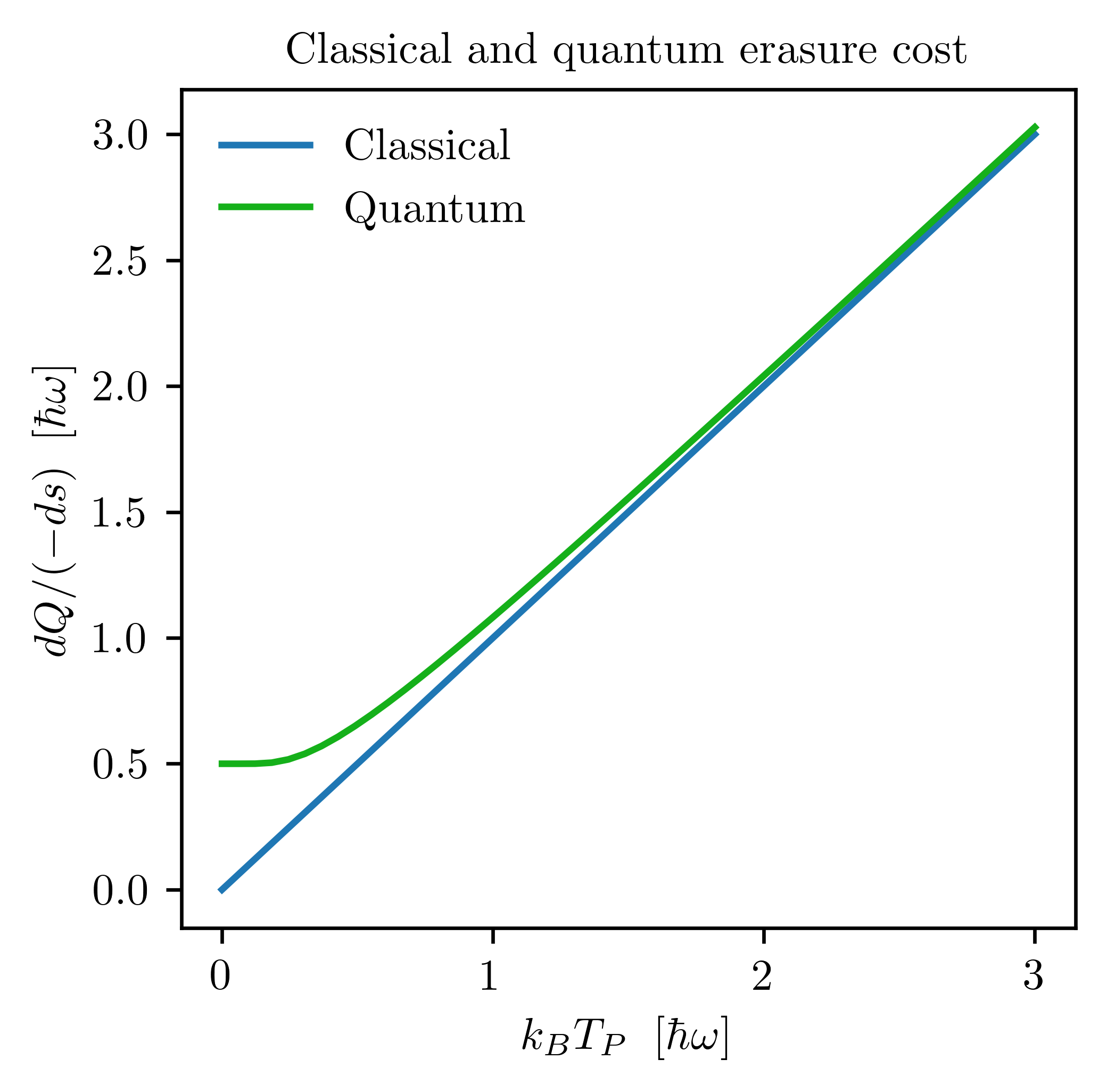}
\caption{Comparison of the classical, Eq. \eqref{diff-landauer-alp}, and quantum, Eq. \eqref{diff-landauer-alp-quantum}, Landauer bounds for the ALP.}
\label{landauer-comparison}
\end{figure}

In Fig.~\ref{landauer-comparison} we plot the two Landauer bounds \eqref{diff-landauer-alp} and \eqref{diff-landauer-alp-quantum} as a function of temperature. Observe that they agree at high temperatures, and at low temperature (when $\hbar \omega \approx k_B T)$, the quantum heat cost per bit asymptotically approaches a constant value $\ln(2)/2$ while the classical heat cost per bit goes to zero.

\section{Landauer's principle for polarizing beamsplitters}

\begin{figure}
\centering
\includegraphics[width=3.4in]{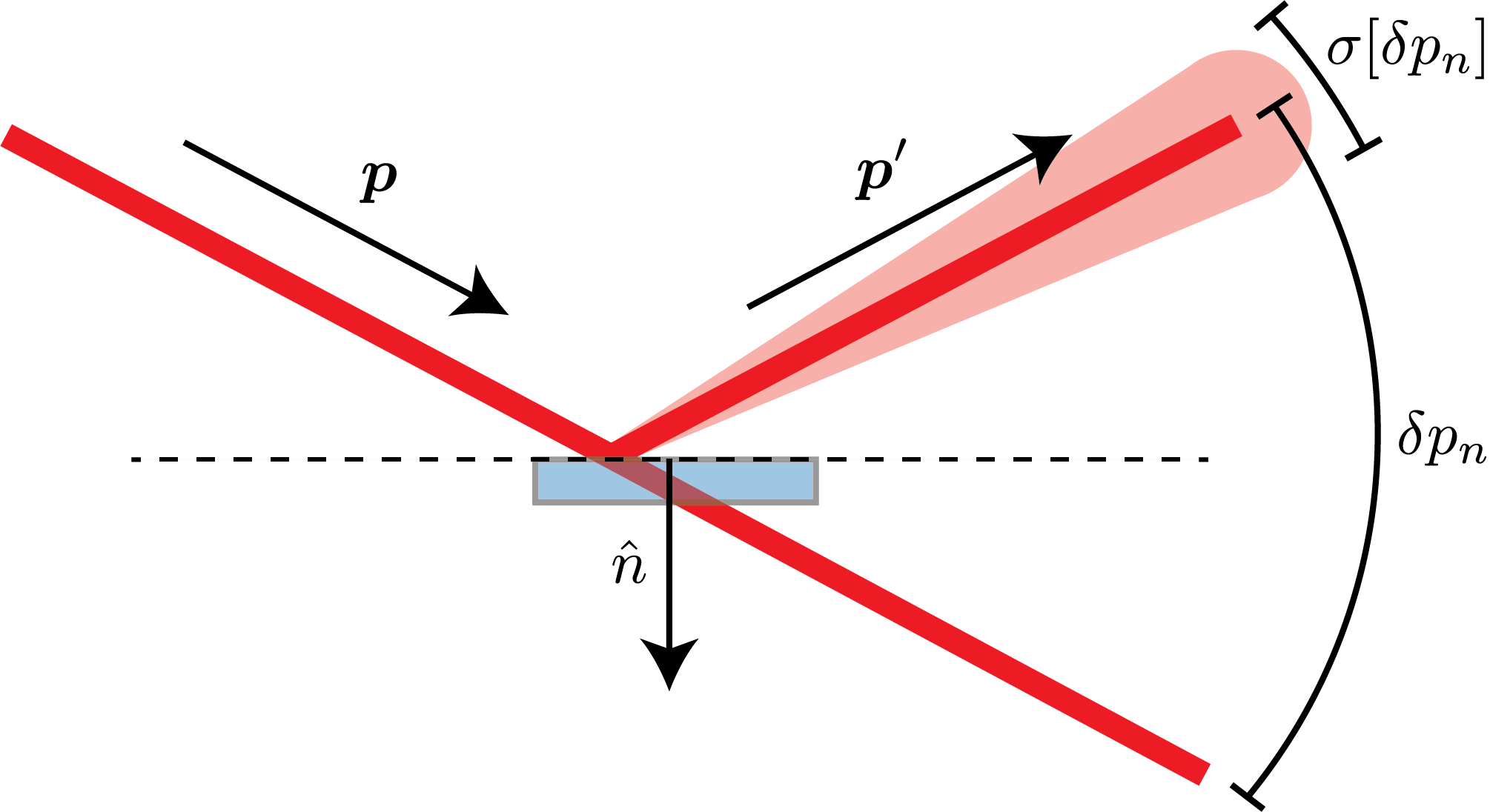}
\caption{Schematic representation of a polarizing beamsplitter. The incident beam is separated into into its horizontal and vertical polarization components.}
\label{reflection-fig}
\end{figure}

Another optical element that is subject to dissipation is a polarizing beamsplitters (PBS). Note that it is possible to produce linearly polarized light with a PBS which transmits horizontally polarized light and reflects vertically polarized light in a different direction. In this case, rather than being absorbed, the vertically polarized output is simply sent elsewhere, cf. Fig.~\ref{reflection-fig}. It is not clear that there is any nonzero amount of heat which must be generated in this process, which is often assumed to be non-dissipative \cite{Fearn1987OC}. In what follows we show that there is a minimal amount of dissipation which scales with temperature. 

\subsection{Semi-classical description}

We first treat the PBS semi-classically, meaning we do not involve the quantum state of light, although we consider it to be made up of pointlike particles; in other words we are using the Newtonian corpuscular model \cite{aspect_huygens_2017}. Each photon is assumed to have either horizontal or vertical polarization, with each photon's polarization an independent random variable, have probability one half of being vertically polarized and probability one half of being horizontally polarized. The PBS itself is assumed to have mass $m$ and temperature $T$, and the photons are assigned a frequency $\omega=E/\hbar$. We show (see Appendix \ref{PBS-appendix-classical}), under the assumption that the reflected photons are locally unavailable, that the following lower bound on dissipation holds
\begin{equation}
\label{PBS-landauer-semiclassical}
\dbar Q\geq \frac{\hbar^2\omega^2}{2 m^2 c^4}\, k_B T (-ds).
\end{equation}
Equation \eqref{PBS-landauer-semiclassical} relates logical information processed to dissipation cost for the semiclassical PBS.

This result is similar to Landauer's principle in that it is a lower bound on heat generation which is proportional to the amount of logical information processed. However it differs in that it does not require any information to be ``erased" per se. In our treatment we have assumed that light which is reflected by the beamsplitter is no longer accessible locally, which leads to dissipation. This is an instance of the more general phenomenon of modularity dissipation \cite{boyd_thermodynamics_2018}, whereby locally accessible information is transformed into global correlations across different parts of a system, which cannot be exploited due to physical constraints.

\subsection{Quantum non-conservation cost}

The result of the previous section was based on the global conservation of momentum and on the assumption of modularity, which prevents the exploitation of global correlations. This reasoning can be generalized to any information processing scheme which locally does not conserve quantities which are globally conserved, even if the desired logical operation is invertible. Here we give an example of this non-conservation cost in a quantum system which is equivalent to the PBS.

Consider the action of the polarizing beamsplitter (PBS) in the one photon subspace. The basis states of this subspace are
\begin{equation}
    \ket{\text{h1}},\ket{\text{h2}},\ket{\text{v1}},\ket{\text{v2}}.
\end{equation}
With this basis ordering the PBS unitary is then given by the following matrix
\begin{equation}
    \label{PBS-one-photon}
    U_{\text{PBS}}=\begin{pmatrix}
    1&0&0&0\\
    0&1&0&0\\
    0&0&0&1\\
    0&0&1&0
    \end{pmatrix}.
\end{equation}
The PBS can therefore be thought of as a CNOT acting on a two qubit system, whose control qubit is the polarization degree of freedom and whose target degree of freedom is the path, $U_\text{PBS}=U_\text{CN}$ for the one-photon subspace \cite{okamoto_demonstration_2005}. However, since the components of angular momentum are globally conserved, it is not possible to implement this operation deterministically unless there is access to a bath which can absorb the change in angular momentum. It has been realized that global conservation laws constrain the possible fidelity of quantum gates \cite{karasawa_gate_2009, ozawa_conservative_2002}, although to our knowledge this has not been applied to finding the thermodynamic cost of quantum gates under an assumption of modularity.

We assume that the two-qubit register and the bath are initially separable. The target qubit is known to be in the zero state, the control qubit is in the completely mixed state, and the bath is in an arbitrary state
\begin{equation}
    \rho = \frac{1}{2}(\ket{0}\bra{0} + \ket{1}\bra{1})\otimes \ket{0}\bra{0} \otimes \rho_B,
\end{equation}
where the bath is a register of $N$ qubits. As it is not possible to actually implement the CNOT exactly, by the previous arguments, we will assume it is implemented with some error bound $\epsilon$. This is similar to the C-maybe interaction that has been studied in the context of quantum Darwinism \cite{Touil2022PRL,Girolami2022PRL,touil2022branching}. Explicitly, we have
\begin{equation}
   \sqrt{ \tr{\left(\rho_S' - U_\text{CN}\, \rho_S\, U_\text{CN}^\dag\right)^2}}\leq \epsilon.
\end{equation}
We show in Appendix \ref{quantum-pbs-appendix} that the loss of purity on the bath is bounded below by
\begin{equation}
\label{quantum-pbs-purity}
\tr{\rho_B'^2} \leq \text{tr}\{\rho_B^2\}-\frac{1}{N2^N}\,\left(1 - 2\sqrt{2}\epsilon \right)^2.
\end{equation}
This implies that for error $\epsilon<(2\sqrt{2})^{-1}$ there is a nonzero loss of purity on the bath as a result of the CNOT operation, and the minimal loss of purity decreases exponentially with the size of the bath.

\begin{figure}
\centering
\includegraphics[width=9cm]{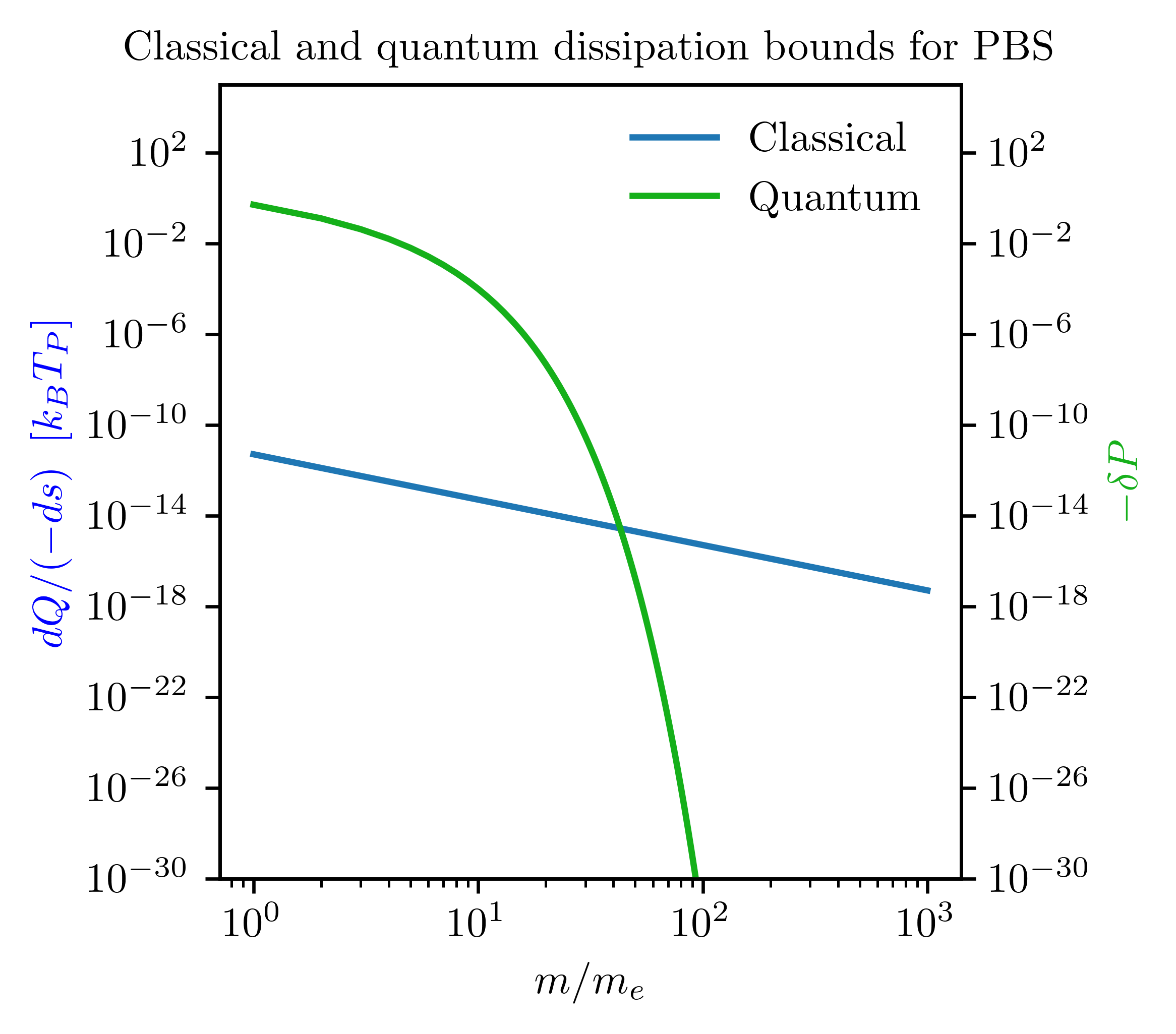}
\caption{Comparison of the classical, Eq. \eqref{PBS-landauer-semiclassical}, and quantum, Eq. \eqref{quantum-pbs-purity}, dissipation bounds for the PBS.}
\label{landauer-comparison-pbs}
\end{figure}

Figure \ref{landauer-comparison-pbs} gives a comparison of the bounds given by Eqs.~\eqref{PBS-landauer-semiclassical} and \eqref{quantum-pbs-purity}, plotted as a function of system size. It is assumed that in the quantum case the register is made up of electrons, so the number of spins is simply $m/m_e$. However, the two equations give lower bounds on the change in different quantities (heat and purity), so they cannot be compared directly. We still may get an idea of how the order of magnitude of the quantum and classical bounds compare at different system sizes. The plot shows that there for $m/m_e > 100$ the classical bound dominates and for $m/m_e < 100$ the quantum bound becomes more relevant.

\section{Quantum Master Equation for Linear Polarizers}

The above analysis makes it apparent that a more rigorous treatment of optical elements as genuinely quantum devices is required. To this end, we now consider the propagation of light through an ALP, and model this as a dynamical process using a quantum master equation. The ALP is conceptualized as a series of layers through which the light propagates, interacting with each layer in turn. Each layer is modeled as a PBS, which allows for noise photons to enter from the environment as well as for losses of photons to the environment due to attenuation. To set up this model, we first give the quantum description of the PBS.

\subsection{Quantum polarizing beamsplitter}

\begin{figure}
\centering
\includegraphics[width=8cm]{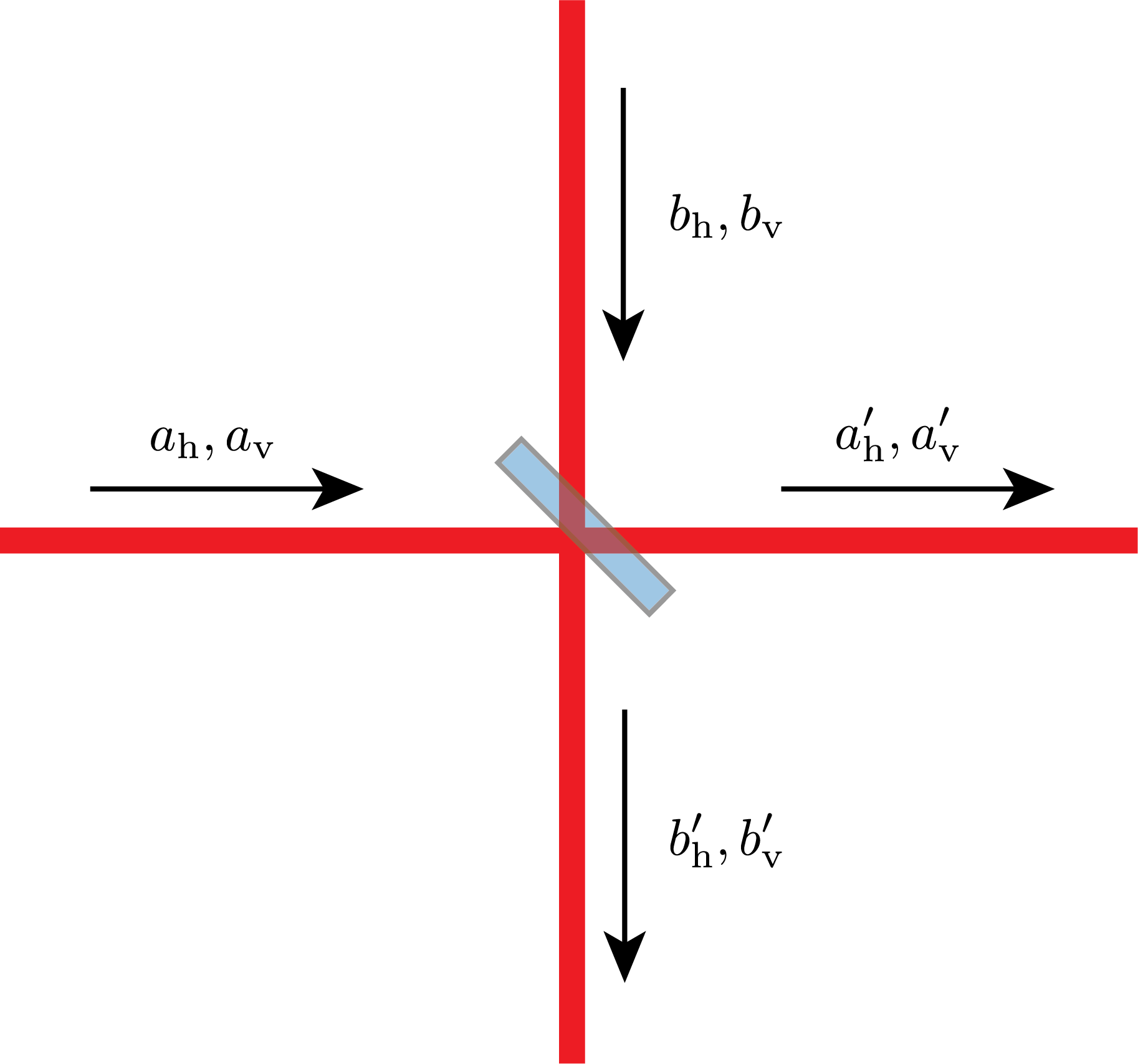}
\caption{Port labelling convention for the polarizing beamsplitter.}
\label{pbs-ports}
\end{figure}

The PBS has two input ports, each with two polarization modes. We label the two polarization modes of one input port $a_\text{h}$ an $a_\text{v}$, and similarly use $b_\text{h}$ and $b_\text{v}$ for the other input port (these will later be the annihilation operators for the input ports). The corresponding output ports are labeled $a_\text{h}'$, $a_\text{v}'$, $b_\text{h}'$, and $b_\text{v}'$ (See Fig. \ref{pbs-ports}).

The behavior of the PBS is easily described in the Heisenberg picture, where the operators are transformed by a scattering matrix $S$
\begin{equation}
\label{heisenberg-transformation}
\begin{pmatrix}
a_\text{h}' \\
a_\text{v}' \\
b_\text{h}' \\
b_\text{v}' 
\end{pmatrix}
= S\begin{pmatrix}
a_\text{h} \\
a_\text{v} \\
b_\text{h} \\
b_\text{v} 
\end{pmatrix}.
\end{equation}
The PBS scattering matrix for the PBS may be taken to be
\begin{equation}
\label{pbs-scattering-matrix}
S=\begin{pmatrix}
1 & 0 & 0 & 0 \\
0& t & 0 & r \\ 
0 & 0 & 1 & 0 \\
0 & -r & 0& t  \end{pmatrix},
\end{equation}
where $t$ and $r$ are the transmission and reflection coefficients, satisfying $t^2+r^2=1$. We will therefore use the parameterization $t=\cos(\phi)$, $r=\sin(\phi)$. We can also describe PBS whose transmission and reflection axes are rotated by an angle $\theta$. In this case the scattering matrix is
\begin{equation}
    S_\theta = R_\theta S R_\theta^\dag,
\end{equation}
where
\begin{equation}
R_\theta=\begin{pmatrix}
\cos(\theta) & -\sin(\theta) & 0 & 0 \\
\sin(\theta)& \cos(\theta) & 0 & 0 \\ 
0 & 0 & \cos(\theta) & -\sin(\theta) \\
0 & 0 & \sin(\theta)& \cos(\theta) \end{pmatrix}.
\end{equation}
As is shown in Appendix~\ref{evo-appendix}, the Schr\"odinger picture evolution operator on the Hilbert space of quantum states can be evaluated as
\begin{equation}
\label{evo-formula}
U= \exp \left( \sum_{ij} \ln (S)_{ij} a_i^\dag a_j\right).
\end{equation} 
In this case we find
\begin{equation} 
\label{U-def}
U(\theta,\phi)= \exp \left(\phi [\sin^2(\theta) S_\text{h} +\cos^2(\theta) S_\text{v}+\sin(2\theta)S_c]\right),
\end{equation}
where
\begin{equation}
\label{s-h-def}
S_\text{h}=a_\text{h}^\dag b_\text{h}-b_\text{h}^\dag a_\text{h} ,
\end{equation}

\begin{equation}
\label{s-v-def}
S_\text{v}= a_\text{v}^\dag b_\text{v}- b_\text{v}^\dag a_\text{v} ,
\end{equation}
and
\begin{equation}
\label{s-c-def}
S_c= \frac{1}{2}(b_\text{v} ^\dag a_\text{h} +  b_\text{h}^\dag a_\text{v}  - a_\text{v} ^\dag b_\text{h} - a_\text{h}^\dag b_\text{v}).
\end{equation}
The density matrix is then transformed by a unitary map
\begin{equation}
    \rho' = U(\theta,\phi)\, \rho\, U^\dag(\theta,\phi).
\end{equation}
Finally, the quantum state exiting the PBS is given formally by
 \begin{equation}
 \label{cptp-def}
\rho'=F_\theta(\rho)=\ptr{b}{U_\theta \rho \otimes \eta_T U_\theta ^\dag},
\end{equation}
where $\eta_T$ is the Gibbs state at the temperature $T$ of the environment. 

\subsection{Multilayer model of the linear polarizer}

\begin{figure}
\centering
\includegraphics[width=85mm]{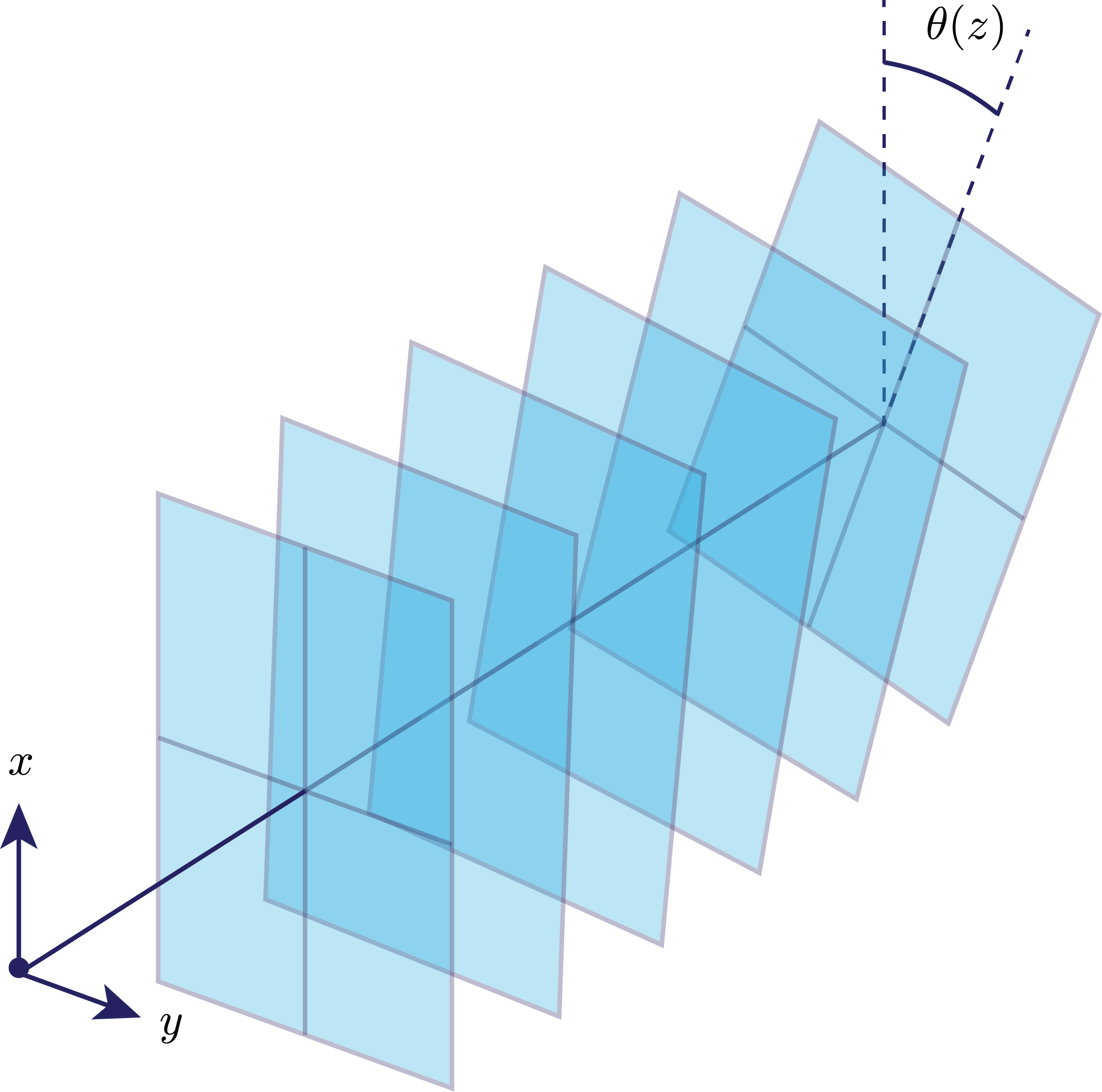}
\caption{A polarizer conceptualized as a sequence of planes with varying polarization axes.}
\label{planes-fig}
\end{figure}

\begin{figure*}
\label{pbs-chain-fig}
\centering
\includegraphics[width=16cm]{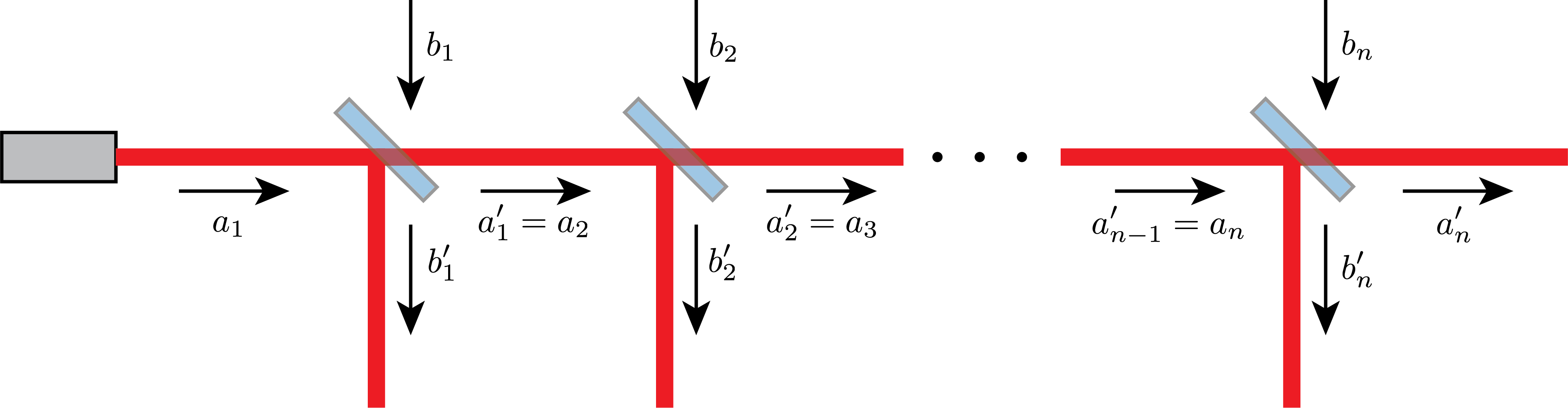}
\caption{A linear polarizer modeled as a chain of polarizing beamsplitters.}
\end{figure*}

We model the linear polarizer as a sequence of PBS elements at randomized angles which are drawn from a thermal distribution. These PBS elements represent physical objects (for example, the nanoparticles in a nanoparticle-based polarizer), which are in uncertain configurations due to thermal energy. Initially the system is in some mixed state $\rho_0$ of the form of Eq.~\eqref{zero-one-subspace}. At each layer the state is transformed using Eq.~\eqref{cptp-def}. We also allow the angle of each polarizer to be different, so the $n$th step of the evolution is given by
\begin{equation}
\label{collision-model-def}
\rho_{n+1}=\ptr{b_n}{U_n\rho_{n}\otimes \eta_T U_n^\dag} = F_{n}(\rho_n)
\end{equation}
Equation~\eqref{collision-model-def} is of the form of a so-called collision model or repeated interaction scheme \cite{campbell_collision_2021, cusumano_quantum_2022, rau_relaxation_1963}.

We see from Eq. \eqref{U-def} that there is a continuum limit as long as $\phi$ is taken to zero as $dt$ goes to zero. We set $\phi = c_\phi dt$. Then to first order in $dt$, we have
\begin{equation}
U_\theta = 1 +  c_\phi dt (\sin^2 (\theta) S_\text{h} + \cos^2 (\theta) S_\text{v} +\sin (2 \theta) S_\text{c} )
\end{equation}
This leads to the following master equation
\begin{equation}
\label{qme-general}
\frac{d\rho}{dt} = c_\phi\, \ptr{b}{[ \mathcal{H}_\theta , \rho \otimes \eta_T]}  ,
\end{equation}
where
\begin{equation}
\mathcal{H}_\theta = \sin^2 (\theta) S_\text{h} + \cos^2 (\theta) S_\text{v} +\sin (2 \theta) S_\text{c}.
\end{equation}
It has been shown by Lorenzo \etal \cite{lorenzo_landauers_2015} that for collision models of this form, the following form of Landauer's principle is observed 
\begin{equation}
\label{landauer-quantum-alp}
   \dbar Q \geq k_B T (- ds),
\end{equation}
which is again the desired result.

\subsection{Single-photon states at low temperature}

To make contact with experimentally realistic scenarios we are concerned with photons of optical frequency ($\sim 10^{14}$ Hz), so at room temperature ($k_B T \approx 25$ meV) the probability of finding $n$ photons in a given mode is proportional to $\exp(-17 n)$. We therefore assume that the number of photons in the vacuum ports is always $0$ initially. We are effectively treating the environment as being at zero temperature for the purposes of thermal radiation. However, we will not assume zero temperature for the mechanical degrees of freedom of the polarizer itself which contribute to uncertainty in the transmission axis.

Suppose that incident on the input port of a PBS there is a general mixed state belonging to the subspace with at most one photon,
\begin{equation}
\label{zero-one-subspace}
\rho = \sum_{\alpha, \alpha' \in \{0,\text{h},\text{v} \}} c_{\alpha \alpha'} \ket{\alpha} \bra{\alpha'}.
\end{equation} 
The state of the two input ports $a$ and $b$ combined is then
\begin{equation}
\sigma = \rho \otimes \ket{0}\bra{0}
\end{equation}
 We are concerned with the reduced density matrix for the input mode, $\rho$. The evolution of $\rho$ is given by
 \begin{equation}
\rho'=\ptr{b}{U_\theta \rho \otimes \ket{0}\bra{0}U_\theta ^\dag}\,.
\end{equation}
The evolution given by Eq.~\eqref{cptp-def} is a CPTP map \cite{Nielsen2010}, and therefore can be given in terms of Krauss operators. For this particular case, it can be expressed in terms of two Krauss operators which depend on $\theta$, $K_{1\theta}$ and $K_{2\theta}$
 \begin{equation}
 \label{cptp-krauss}
\rho'=F_\theta(\rho)=K_{1\theta}^\dag \rho K_{1\theta} + K_{2\theta}^\dag \rho K_{2\theta}.
\end{equation}
Explicitly, the Krauss operators are,
 \begin{equation}
 \label{krauss1}
K_{1\theta}=\begin{pmatrix}1&0&0\\0&t \cos^2(\theta)+\sin^2(\theta)&-(1-t)\sin(\theta)\cos(\theta)\\0&-(1-t)\sin(\theta)\cos(\theta)&\cos^2(\theta)+t\sin^2(\theta)\end{pmatrix}
\end{equation}
and,
 \begin{equation}
 \label{krauss2}
K_{2\theta}=\begin{pmatrix}0&r \cos(\theta)&r\sin(\theta)\\0&0&0\\0&0&0\end{pmatrix}
\end{equation}
Equations~\eqref{cptp-krauss}, \eqref{krauss1}, and \eqref{krauss2} completely specify the evolution of the density matrix as the single photon propagates through a single layer of the polarizer at low temperature. To simulate the evolution in the multilayer model, the angle $\theta$ for each polarizer is drawn from a thermal distribution
\begin{equation}
\label{theta-dist}
f(\theta)=\frac{1}{\sqrt{2\pi k_B T/\kappa}}\exp \left(-\frac{1}{2}\frac{\theta^2}{k_B T/\kappa}\right),
\end{equation}
where $\kappa$ is a physical parameter of the polarizing material which characterizes the energy required to change the orientation of polarizing elements from their (mechanical) equilibrium positions. The resulting maps $F_\theta$ are applied one after another. For $\theta=0$ we get a particularly simple evolution
\begin{equation}
F_0(\rho)=\begin{pmatrix}\rho_{00}+r^2\rho_{11}& t\rho_{01}& \rho_{02}\\\ t\rho_{10} & t^2 \rho_{11} &t\rho_{12}\\\rho_{20}&t\rho_{21}&\rho_{22}\end{pmatrix}
\end{equation}
In Appendix \ref{commutation-appendix}, we consider the Heisenberg picture description of a single application of the map $F_\theta$ with a random angle drawn from the distribution \eqref{theta-dist}. It is found that an adjoint map can be defined which gives evolved operators satisfying a modified commutation relation
\begin{equation}
\label{primed-commutators}
[a_\text{h}',a_\text{h}'^\dag] =[a_\text{v}',a_\text{v}'^\dag]  =1-2(1-t)(\chi-\chi^2).
\end{equation}
where $\chi=\frac{1}{2}\left[1-\exp(-2 k_B T/\kappa)\right]$. 

In Eq.~\eqref{primed-commutators} we observe the direct impact of the noise in the PBS on the optical signal. The quantum mechanical communtation relations ``degrade'' as a function of temperature, which is nothing but another way of looking at effects of decoherence.

\subsection{Numerical study of the low-temperature limit}

\begin{figure}
\centering
\includegraphics[width=3.4in]{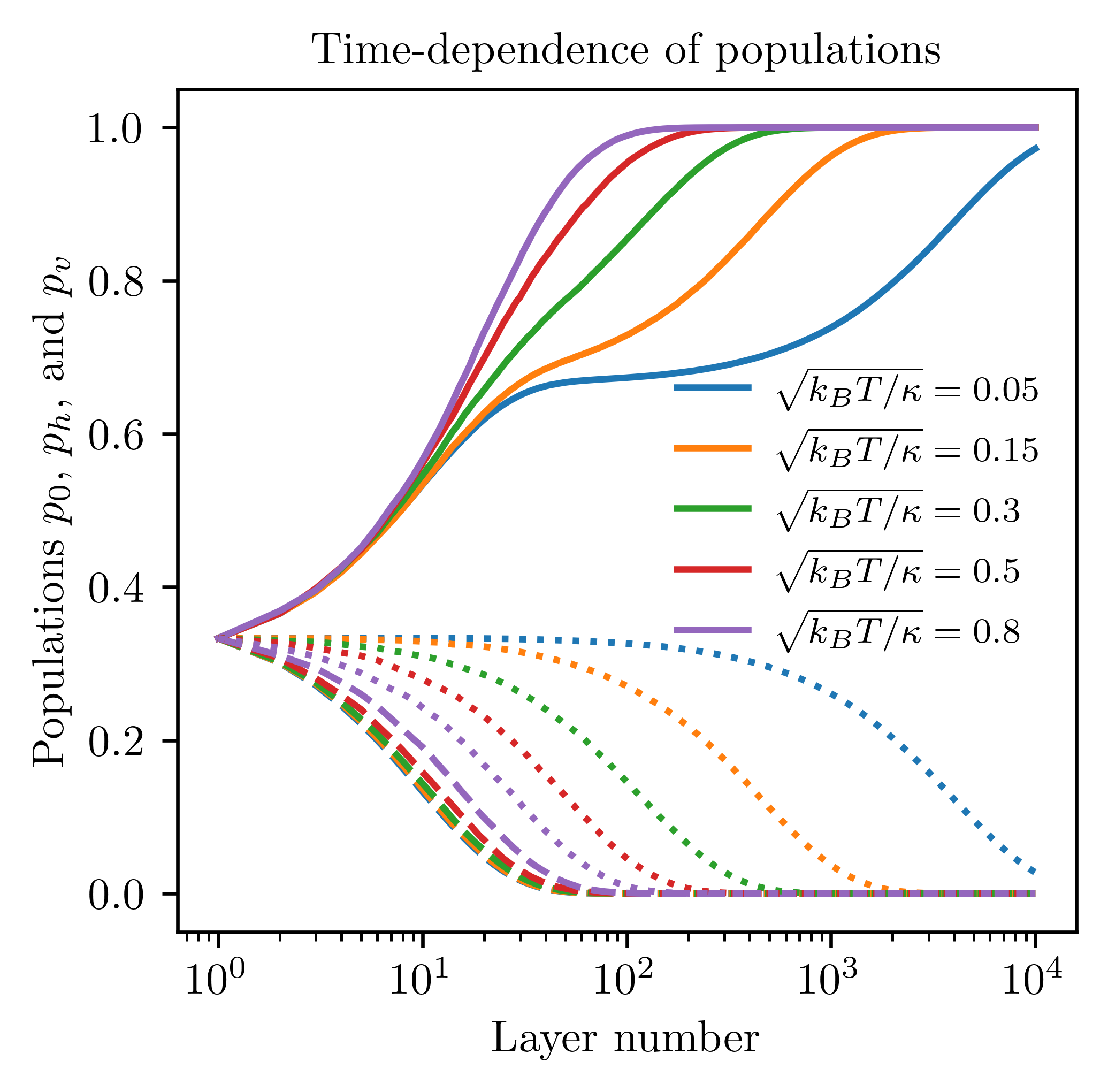}
\caption{Populations $p_0=\rho_{00}$ (solid lines), $p_\text{h}=\rho_{11}$ (dotted lines), and $p_\text{v}=\rho_{22}$ (dashed lines). The vertical population $p_\text{v}$ rapidly decays to zero but the horizontal component takes longer. As a result the vacuum population has a plateau that is flatter at lower temperatures.}
\label{time-pop-fig}
\end{figure}

Using the above formalism we simulate the evolution of the quantum state of light, limited to the $\leq 1$ photon subspace, and assuming low temperature, that is $k_B T \ll \hbar \omega$. This simulation allows us to discuss some qualitative aspects of the polarization process. We initialize the system in the state
\begin{equation}
\label{initial-condition-single-polarizer}
    \rho_0 = \begin{pmatrix}
    1/3 & 1/3 & 1/3\\
    1/3 & 1/3 & 1/3\\
    1/3 & 1/3 & 1/3
    \end{pmatrix}.
\end{equation}
Then the CPTP map $F_0$ is iterated for $N=10,000$ layers with $t=0.9$. A range of temperatures are chosen in the interval $\sqrt{k_B T/\kappa}\in[0.05,0.3]$. We find that the population $p_\text{v}=\rho_\text{vv}$ decays very quickly, as expected, falling to near zero after less than one hundred layers. The horizontal population $p_\text{h}$ takes a longer time to decay. As temperature increases, $p_\text{v}$ decays more slowly and $p_\text{h}$ decays more quickly, until the two polarizations behave near identically for $\sqrt{k_B T /\kappa} \approx 1$. At low temperature, due to the lag between absorption of vertically and horizontally polarized light, the vacuum population has a plateau at intermediate times (see Fig. \ref{time-pop-fig}).

\begin{figure}
\centering
\includegraphics[width=8.5cm]{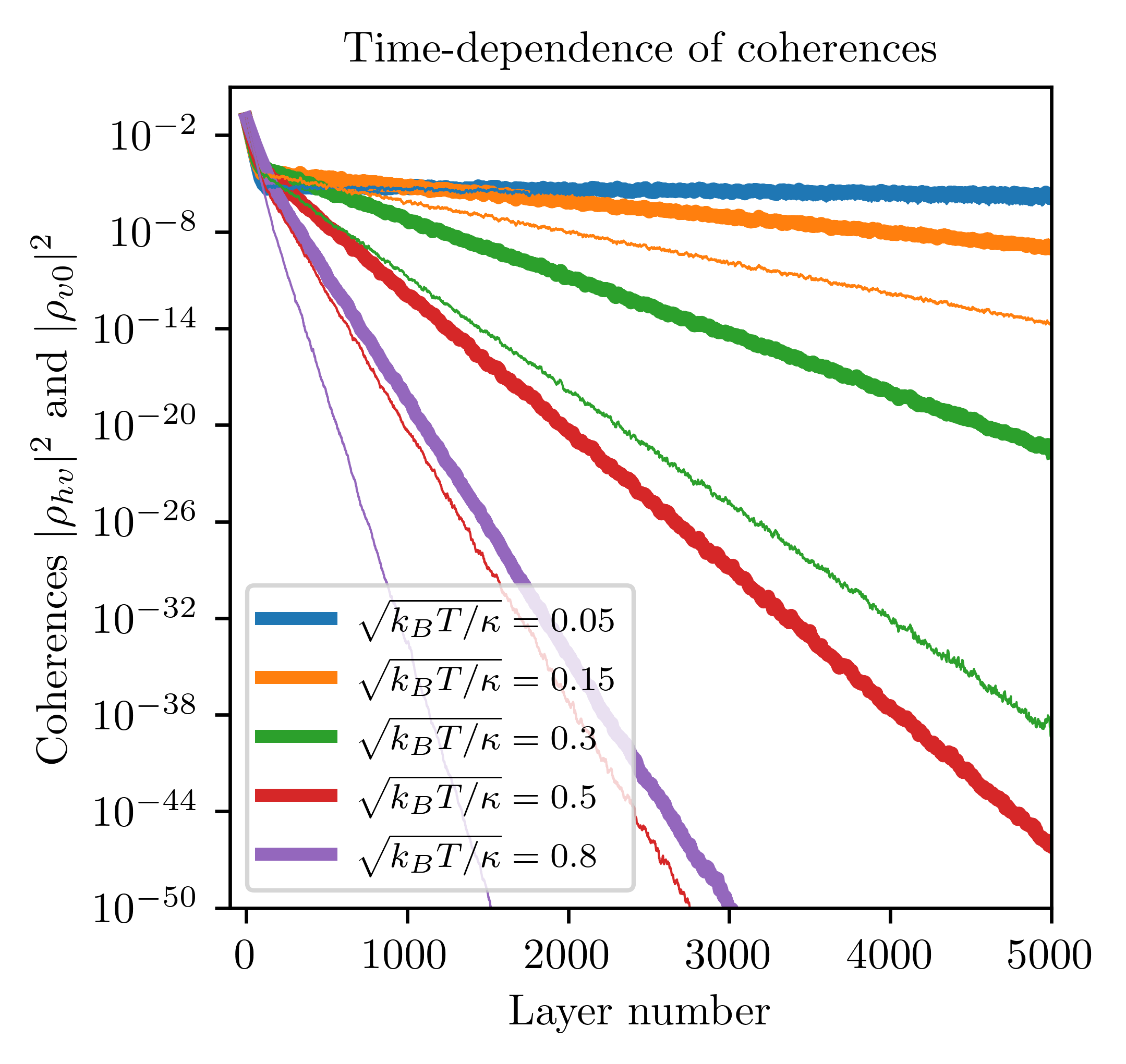}
\caption{Coherences $|\rho_\text{hv}|^2$ (thick lines) and $|\rho_\text{v0}|^2$ (thin lines). The coherences both decay exponentially, but $|\rho_\text{v0}|^2$ decays at a faster rate at all temperatures. Both coherences decay faster at higher temperatures.}
\label{time-coherence}
\end{figure}

We see from Fig.~\ref{time-coherence} that the coherences $|\rho_\text{v0}|^2$ and $|\rho_\text{hv}|^2$ decay exponentially, with the horizontal-vacuum coherence decaying faster. The decay of coherence occurs faster at higher temperatures. This result is in line with our expectations, given that decoherence has been found to be enhanced at higher temperatures in a variety of physical systems \cite{takahashi_coherent_2009,hansen_mesoscopic_2001,mohanty_decoherence_1997}.

\begin{figure}
\centering
\includegraphics[width=8.5cm]{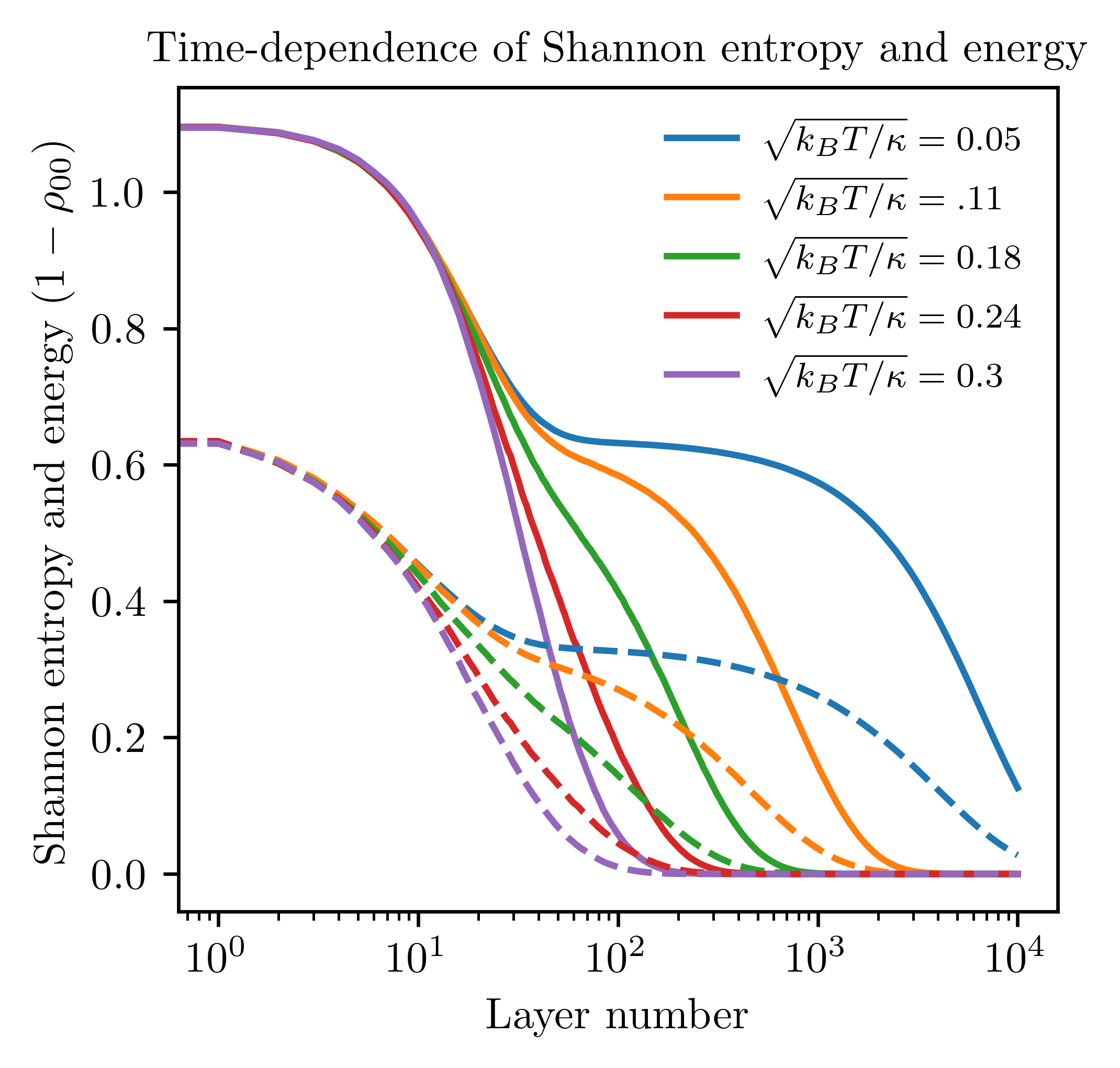}
\caption{The time dependence of Shannon entropy (solid lines) and energy $1-p_0$ (dashed lines). Both display plateaus at intermediate times which are made flatter at lower temperatures. This is due to the difference in timescales between absorption of horizontally and vertically polarized light. Both functions are decreasing, and their relative slopes are different before and after the plateau. }
\label{time-entropy-energy}
\end{figure}

The Shannon entropy is evaluated at all times, and it is found to decrease monotonically. The energy is proportional to $1-p_0$, and also decreases monotonically. Figure~\ref{time-entropy-energy} shows the time dependence of the energy and Shannon entropy, with both display the plateau feature at low temperatures discussed earlier.

Landauer's principle concerns the change in heat relative to the change in Shannon entropy, which can be expressed as the derivative $\dbar Q/ (-d s)$ along a system trajectory. Since we assume $\dbar Q=-dE$, we are therefore interested in the derivative $dE/ds$ along some system trajectory. Although this ratio is related to temperature, it is not given by any function of temperature given that the system is out of equilibrium. We plot the energy against the Shannon entropy for 100 realizations of the random process at various temperatures. For each realization, the angles of the polarizer layers have been drawn from a Boltzmann distribution at the given temperature, and the initial conditions are always those given by Eq.~\eqref{initial-condition-single-polarizer}.

\begin{figure}
\centering
\includegraphics[width=8.5cm]{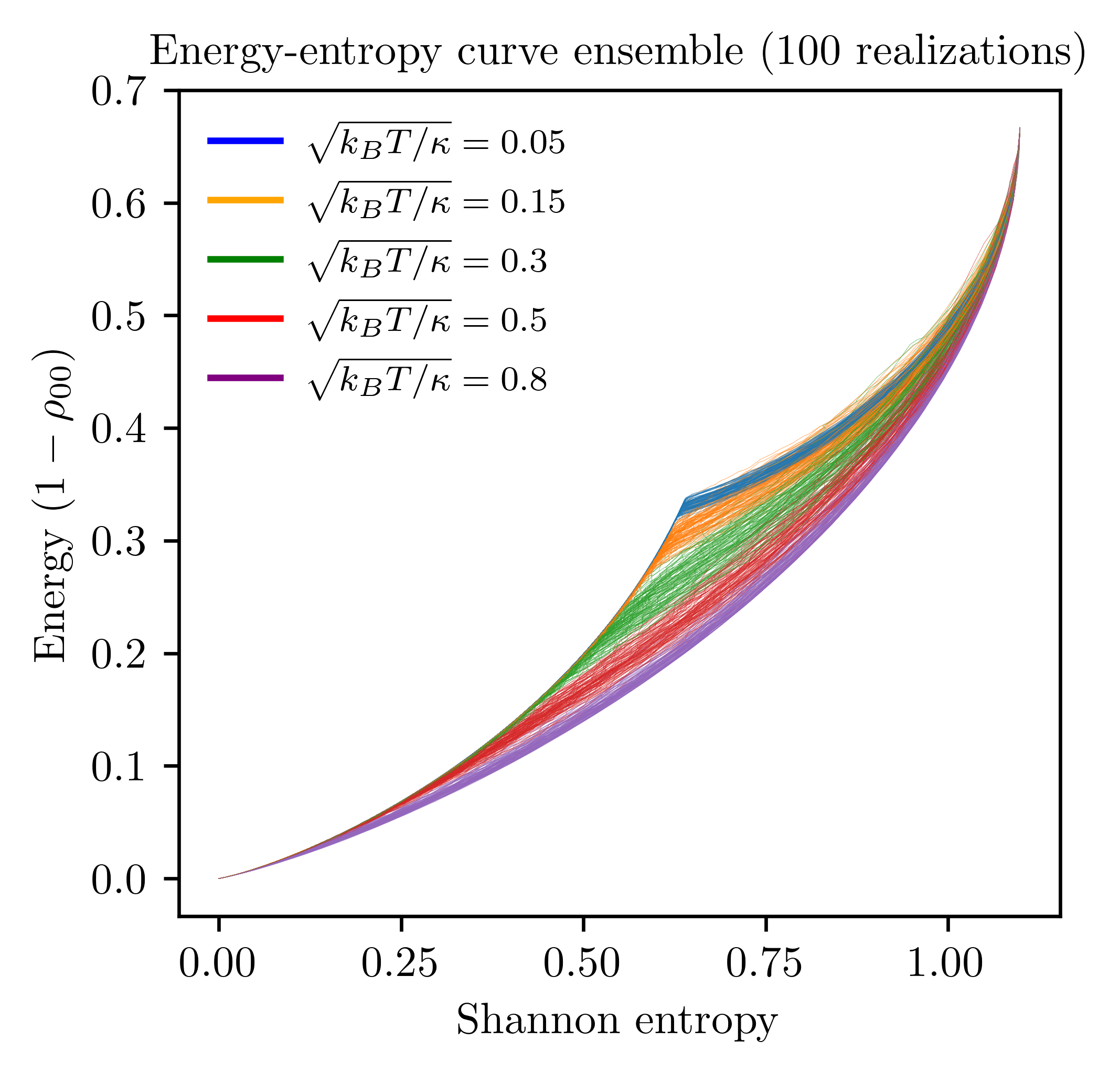}
\caption{One hundred realizations of the polarization process are plotted with the same initial conditions but randomized transmission angles for the individual layers drawn from the Boltzmann distribution. Energy is monotonically increasing with entropy, but displays a discontinuity in its slope at zero temperature.}
\label{energy-entropy-curve}
\end{figure}

Figure~\ref{energy-entropy-curve} shows that the energy always increases monotonically with Shannon entropy, as expected. Interestingly, as temperature approaches zero, a discontinuity of the slope appears near $s=0.6$. This can be explained by the plateaus appearing in Fig. \ref{time-entropy-energy} at low temperature. On the plateaus, both the entropy and energy are nearly constant, although many layers of the polarizer are traversed. After the plateau, the relative slope of the energy and entropy is different than what it was before, and so a discontinuity of the slope $dE/ds$ appears when the energy and Shannon entropy are plotted against each other. 

\section{Temperature-dependent quantum eraser}

\begin{figure*}
\centering
\includegraphics[width=16cm]{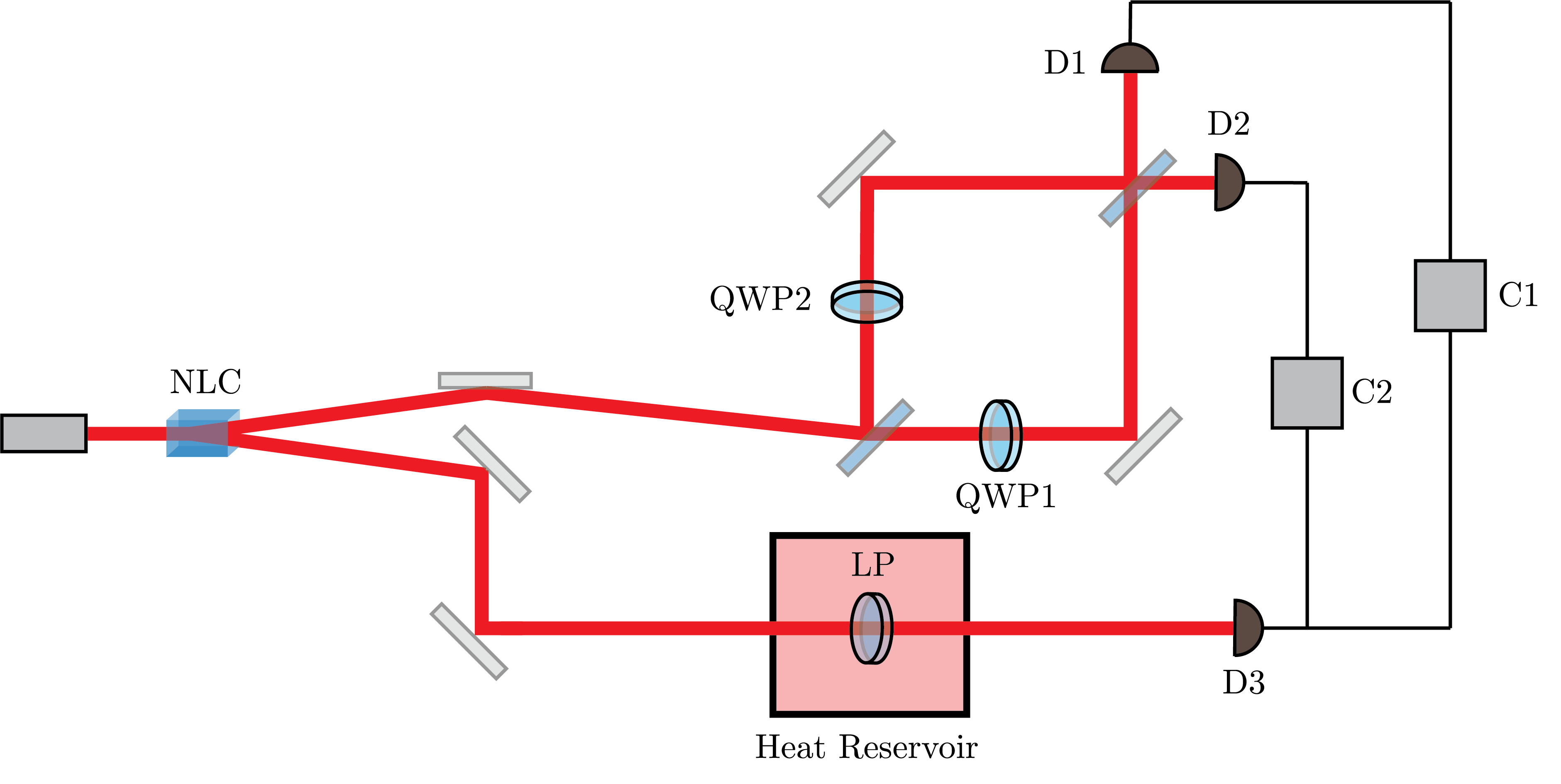}
\caption{Setup for the temperature-dependent quantum eraser experiment.}
\label{qe-setup-fig}
\end{figure*}

\begin{figure}
\centering
\includegraphics[width=3.4in]{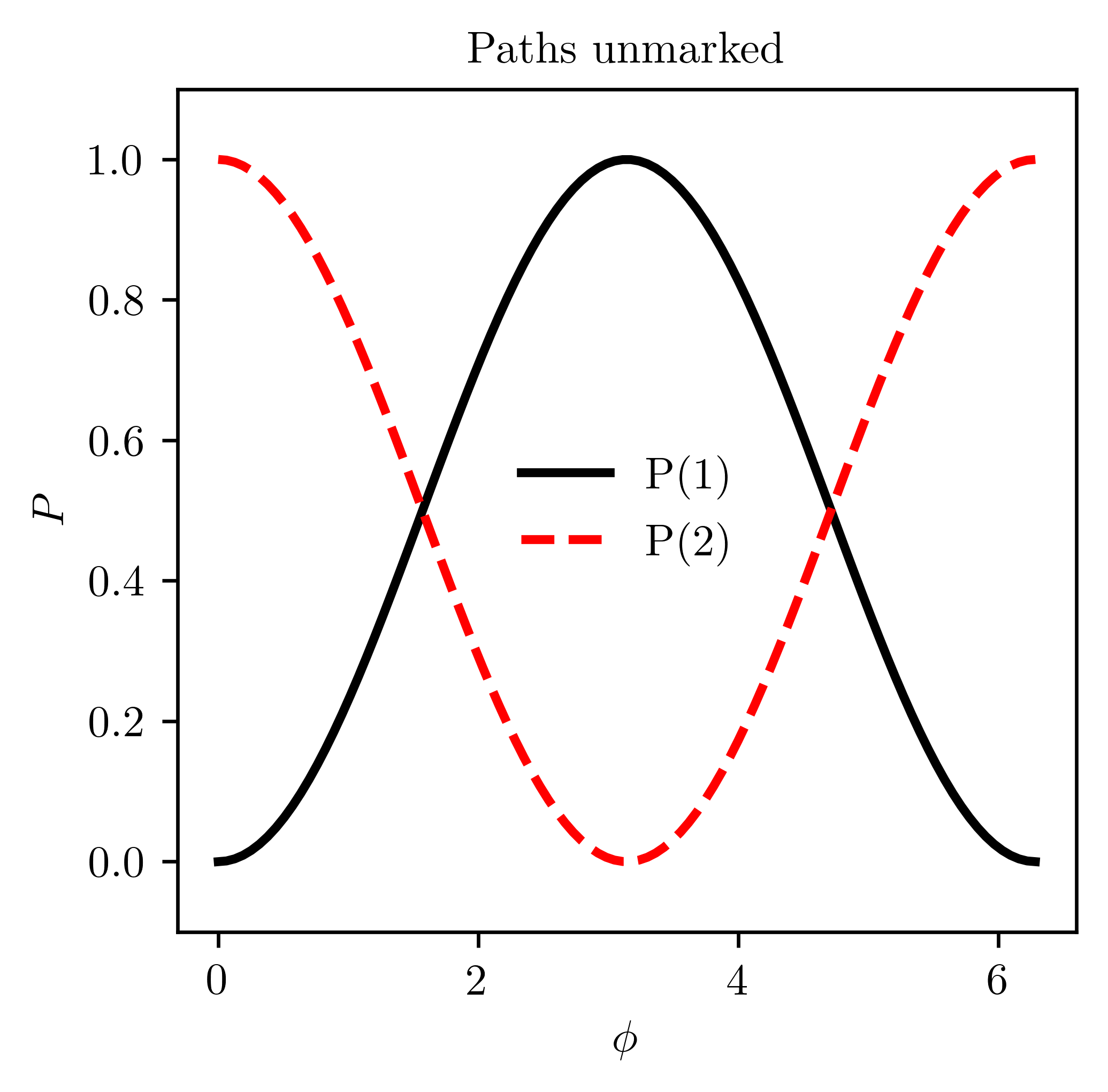}
\caption{Probability for the photon to exit each port as a function of relative phase difference for the two arms in the unmarked case.}
\label{qe-paths-unmarked}
\end{figure}

We conclude the analysis with an experimentally testable consequence of our findings. We have developed a theoretical model for the effect of temperature on the evolution of a single-photon state through a polarizing medium, and our model predicts that at higher temperature there is enhanced decoherence. We now propose an experimental test of this, which is a modified quantum eraser experiment where the linear polarizer used to measure one of the photon's polarization is in contact with a heat reservoir at temperature $T$.  In the quantum eraser experiment \cite{scully_quantum_1982, kang_electronic_2007, herzog_complementarity_1995, walborn_double-slit_2002, peng_delayed-choice_2014, kim_delayed_2000} there are two photons, called the signal and idler photons. The signal photon may be either horizontally or vertically polarized, or it may be absorbed by the polarizer so it lives in a three-dimensional Hilbert space spanned by the vectors $\ket{\text{0}},\ket{\text{h}},\ket{\text{v}}$. The idler photon may be either horizontally or vertically polarized, and it may go down path $1$ or path $2$ of the interferometer, so it lives in a four-dimensional Hilbert space spanned by the vectors $\ket{\text{h1}},\ket{\text{v1}},\ket{\text{h2}},\ket{\text{v2}}$. So the state of the two photons together belongs to a $12$-dimensional Hilbert space spanned by the following states
\begin{align}
\label{QE-basis-order}
\begin{split}
&\ket{\text{0h1}},\ket{\text{hh1}},\ket{\text{vh1}},\ket{\text{0v1}},\ket{\text{hv1}},\ket{\text{vv1}}, \\
&\ket{\text{0h2}},\ket{\text{hh2}},\ket{\text{vh2}},\ket{\text{0v2}},\ket{\text{hv2}},\ket{\text{vv2}}
\end{split}
\end{align}
These states can be expressed as $\ket{ijk}$, where $i\in \{ 0,\text{h}, \text{v}\}$, $j \in \{\text{h}, \text{v}\}$, and $k\in \{1,2 \}$. Then in general we have a density matrix of the form
\begin{equation}
\sigma = \sum_{ijk} c_{i j k i' j' k'} \ket{ijk}\bra{i'j'k'}.
\end{equation}
The full density matrix can be written as a $4\times 4$ array of $3 \times 3 $ submatrices, where the indices in the $4\times 4$ array correspond to the idler photon and the indices of each submatrix correspond to the signal photon.
\begin{equation}
\sigma=\begin{pmatrix}\rho_{\text{v1,v1}}&\rho_{\text{v1,h1}} &\rho_{\text{v1,v2}} &\rho_{\text{v1,h2}}\\ \rho_{\text{h1,v1}}&\rho_{\text{h1,h1}} &\rho_{\text{h1,v2}} &\rho_{\text{h1,h2}}\\\rho_{\text{v2,v1}}&\rho_{\text{v2,h1}} &\rho_{\text{v2,v2}} &\rho_{\text{v2,h2}} \\ \rho_{\text{h2,v1}}&\rho_{\text{h2,h1}} &\rho_{\text{h2,v2}} &\rho_{\text{h2,h2}}\end{pmatrix}
\end{equation}
To evolve the density matrix forward by one step, we use Eq.~\eqref{cptp-def}. This gives
\begin{equation}
\sigma' = G_\theta(\sigma)=\begin{pmatrix}F_\theta(\rho_{\text{v1,v1}})&\dots & F_\theta(\rho_{\text{v1,h2}})\\ \\ \\ F_\theta(\rho_{\text{h2,v1}})&\dots & F_\theta(\rho_{\text{h2,h2}})\end{pmatrix}
\end{equation}
The initial condition for the quantum eraser experiment is a Bell state in the polarization basis for the two photons:
\begin{equation}
\ket{\psi_0} = \frac{1}{\sqrt{2}} ( \ket{\text{hh1}} + \ket{\text{vv1}})
\end{equation}
First, a beamsplitter is applied to the idler photon. There are two distinct configurations for the interferometer, one in which the paths are ``marked" and the other the paths ``unmarked". For the marked configuration there is a quarter wave plate (QWP) in each path of the interferometer (see Fig.~\ref{qe-setup-fig}). One of the QWPs is at an angle of $\pi/4$ radians, the other at an angle of $-\pi/4$ radians. For the unmarked configuration there are no QWPs. First we simulate what happens in the unmarked configuration when there is no measurement made on the signal photon, and recover the standard result of a simple Mach-Zehnder interferometer experiment (see Fig.~\ref{qe-paths-unmarked}) \cite{zetie_how_2000}.

\begin{figure}
\centering
\includegraphics[width=3.4in]{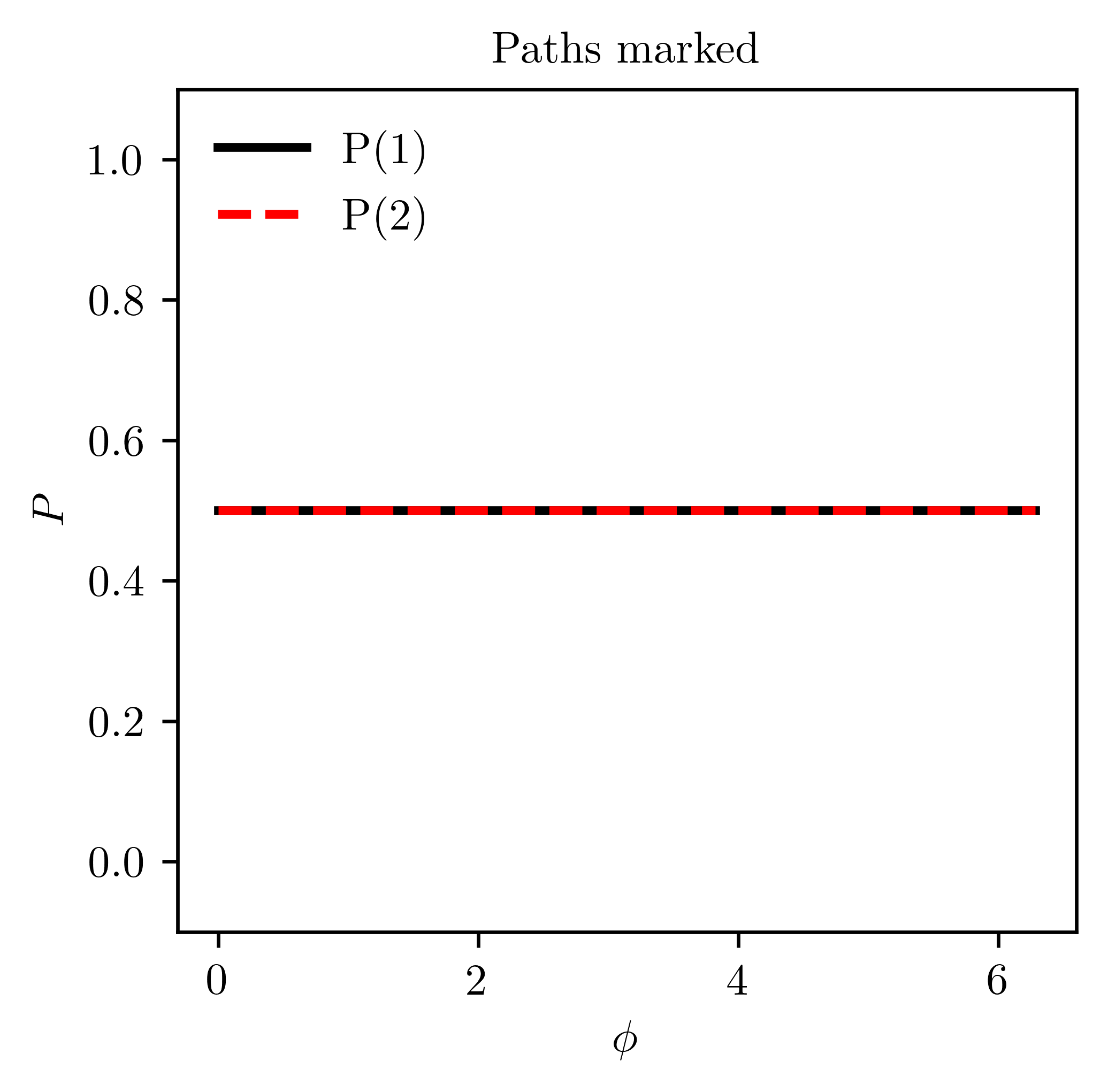}
\caption{Probability for the photon to exit each port as a function of relative phase difference for the two arms in the marked case.}
\label{qe-paths-marked}
\end{figure}

Next we look at the marked configuration, again with no measurement made on the signal photon, the result of which is shown in Fig.~\ref{qe-paths-marked}. We see that the interference pattern is now lost.

We then make a measurement on the signal photon with a linear polarizer at an angle $\theta$ from the vertical, resulting in a restoration of the interference pattern.

\begin{figure}
\centering
\includegraphics[width=3.4in]{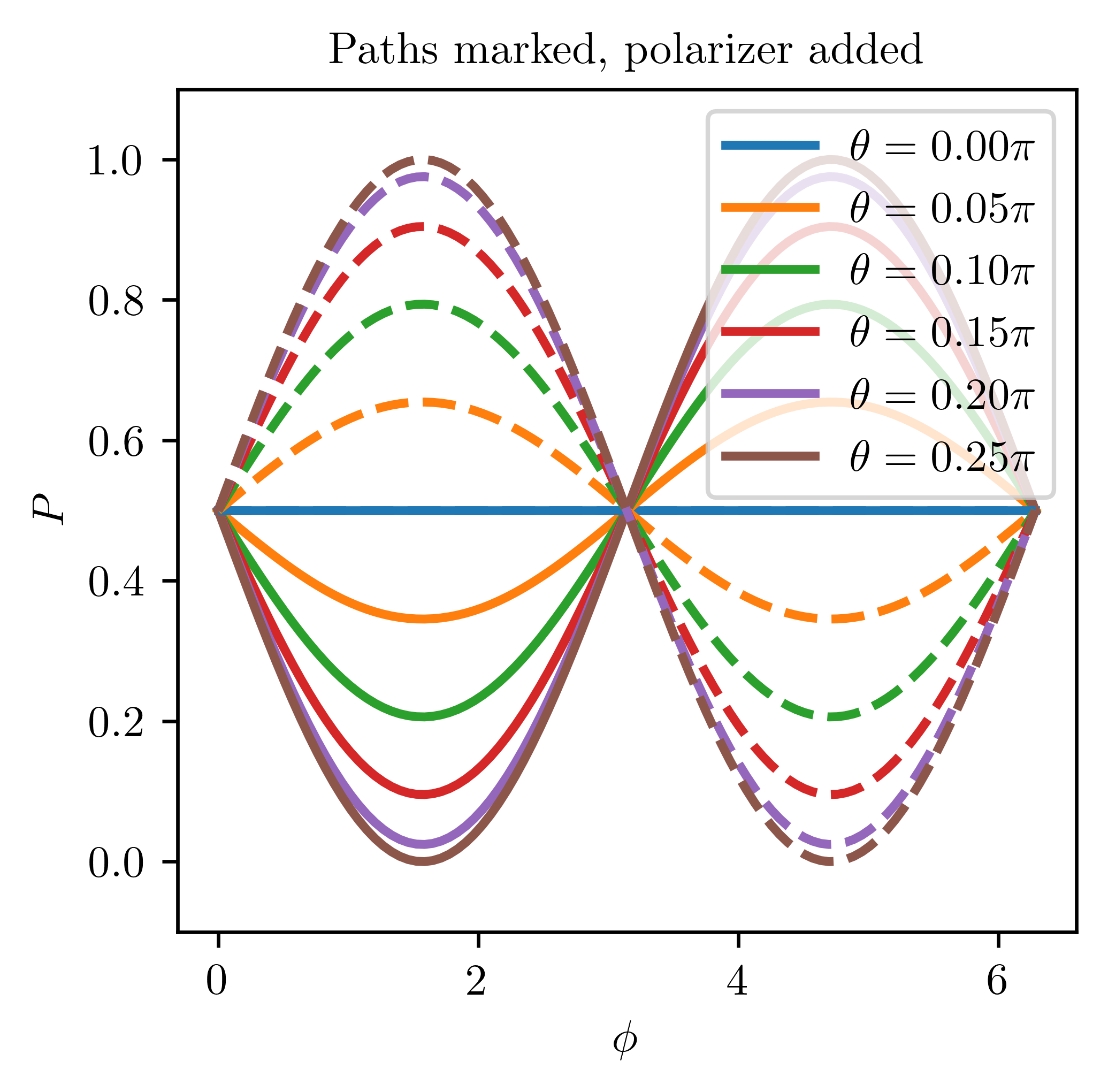}
\caption{Probability for the photon to exit each port as a function of relative phase difference for the two arms in the marked case, with a measurement made on the signal photon by a LP at angle $\theta$ from vertical. }
\label{qe-with-pol}
\end{figure}

We see that the interference pattern is gone when $\theta=0$, and is completely restored for $\theta=\pi/4$, which is the main result of the traditional quantum eraser experiment (see Fig. \ref{qe-with-pol}). The variation in the interference pattern is most pronounced at a path phase difference of $\phi=\pi/2$.

\begin{figure}
\centering
\includegraphics[width=3.4in]{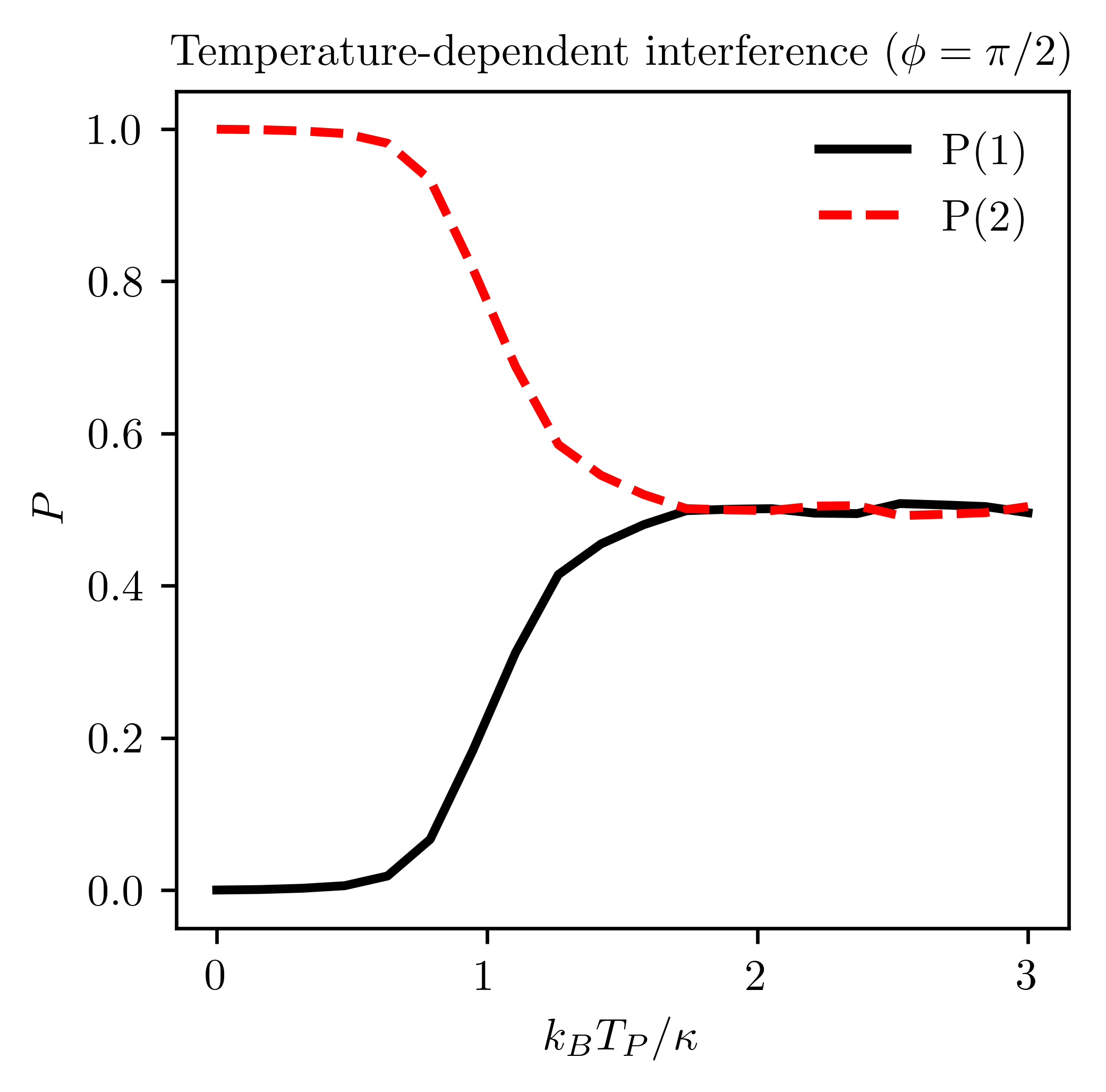}
\caption{Probability for the photon to each port as a function of temperature, with a relative phase of $\pi/2$ between the paths. Ensemble average of 1000 realizations. }
\label{temp-erasing}
\end{figure}

We see from Fig.~\ref{temp-erasing} that the effect of quantum erasing is temperature dependent. In particular, as temperature is increased, the ability of the polarizer to restore interference is suppressed, and eventually goes away entirely.

\section{Concluding remarks}

Our analysis has provided insight into the (quantum) thermodynamics of linear polarizers; we have given explicit forms of Landauer's principle for both absorbing linear polarizers and polarizing beamsplitters. We have also developed a formalism that incorporates the thermal energy contained in mechanical degrees of freedom of a polarizer, and used this model to investigate the time-dependent dynamics of the polarization process. We have provided a qualitative description of the dependence of this process on temperature, and proposed an experiment to test the temperature-dependence of decoherence via a quantum eraser apparatus.  

Some questions have been left unanswered which warrant further analysis. A full solution of the master equation derived for the quantum absorbing polarizer, even in the low temperature limit and the single photon subspace is lacking. We also have not yet written a master equation valid when $k_B T$ is on the order of at least $\hbar \omega$. Answering these and other questions will provide guidance for optimal control of optical polarization states. This is expected to be particularly challenging task when $k_B T$ is nearly of the order of $\hbar \omega$, since then environmental noise becomes significant. Therefore we hope to extend our results with special attention paid to deriving a master equation valid in the high temperature limit. It is then desirable to find bounds on the minimum dissipation required to perform various tasks in such an environment, including the version of Landauer's principle in Eq. \eqref{landauer-quantum-alp}.

\begin{acknowledgments}
We would like to thank Todd Pittman, Alejandro Rodriguez Perez, Steve Campbell, and Pat Eblen for enlighting discussions. M.A. gratefully acknowledges support from Harry Shaw of NASA Goddard Space Flight Center. N.M.M. acknowledges support from AFOSR (FA2386-21-1-4081, FA9550-19-1-0272, FA9550-23-1-0034) and ARO (W911NF2210247, W911NF2010013). S.D. acknowledges support from the John Templeton Foundation under Grant No. 62422.
\end{acknowledgments}

\appendix

\section{Derivation of classical Landauer's principle}
\label{classical-landauer-appendix}

Recall that for a random vector of dimension $D$, and with covariance matrix $\Sigma$, the entropy is upper bounded by that of the normal distribution \cite{dowson_maximum-entropy_1973, chung_bounds_2017}
\begin{equation}
\label{max-ent-bound}
    s[f]\leq \frac{1}{2}\lo{(2 \pi e)^D \left|\Sigma\right|},
\end{equation}
We regard $f_\alpha$ as a bivariate distribution over both the real and imaginary parts of $\alpha_\text{v}$. Note that in general $\sqrt{\left|\Sigma \right|} \leq \braket{|\alpha_\text{v}|^2}/2$, so
\begin{equation}
    s[f_\alpha]\leq \lo{ \pi e \braket{|\alpha_\text{v}|^2}} = \lo{\frac{2 e \pi E}{V \epsilon_0 E_{0\omega}^2}}.
\end{equation}
We assume that after the light has passed through the polarizer, it is in a thermal distribution with temperature $T_P$, which is associated with the polarizer itself. Therefore the energy is $E' =k_B T$ and the entropy is
\begin{equation}
   s[f_\alpha']=\lo{\frac{2 e\pi E'}{V \epsilon_0 E_{0\omega}^2}}.
\end{equation}
Consequently, the decrease in Shannon entropy is at most
\begin{equation}
    -\Delta s \leq  \lo{\frac{2 e\pi E}{V \epsilon_0 E_{0\omega}^2}}-\lo{\frac{2 e\pi E'}{V \epsilon_0 E_{0\omega}^2}}= \lo{\frac{E}{E'}}. 
\end{equation}
Since there is no work reservoir, all lost energy is dissipated as heat, and we have
\begin{equation}
    Q= E - E' = E' \left(\frac{E}{E'}- 1\right).
\end{equation}
Using $E'=k_B T_P$ and the upper bound on $\Delta s$ then gives
\begin{equation}
\label{finite-landauer-alp-app}
   Q \geq k_B T_P \left(e^{-\Delta s}-1\right).
\end{equation}

\section{Derivation of quantum Landauer's principle}

\label{quantum-landauer-appendix}
Again using Eq. \eqref{max-ent-bound}, we find that
\begin{equation}
    s_\text{WM}[W]=s[f_q]+ s[f_p]\leq \lo{2 \pi e \sqrt{\braket{q^2}\braket{p^2}}}.
\end{equation}
Due to Eq. \eqref{harmonic-energy-qp} this is bounded above by
\begin{equation}
    s_\text{WM}[W]\leq \lo{\frac{ 2\pi e E}{\hbar \omega}}.
\end{equation}
We again assume that the light is in a thermal state after leaving the polarizer. The thermal state has Wigner function \cite{bhusal_smart_2021, curtright_concise_2014}
\begin{equation}
\label{thermal-wigner}
    W'(q,p)=\frac{\hbar \omega}{\pi E'}\exp \left( -\frac{\hbar \omega(q^2 + p^2)}{2E'}\right),
\end{equation}
where $E'$ is given by
\begin{equation}
\label{thermal-state-energy}
    E'(T) = \hbar \omega \left( \frac{1}{\exp(\hbar \omega /k_B T)-1} +\frac{1}{2}\right),
\end{equation}
or alternatively
\begin{equation}
\label{thermal-state-energy-2}
    E'(T) = \frac{\hbar \omega}{2} \coth \left(\frac{\hbar \omega}{2 k_B T} \right).
\end{equation}
Since the thermal Wigner function \eqref{thermal-wigner} is positive, the final Wigner entropy $s_W[W']$ is defined, and is given by
\begin{equation}
    s_\text{W}[W']= \lo{\frac{2\pi e E'}{\hbar \omega}}.
\end{equation}
We can write
\begin{equation}
    -\Delta s_\text{W} \leq s_\text{WM}[W] - s_\text{W}[W'] \leq \ln \left( \frac{E}{E'}\right)
\end{equation}
We once again equate the heat to the lost energy
\begin{equation}
    Q=E' \left(\frac{E}{E'}- 1\right),
\end{equation}
and hence
\begin{equation}
\label{finite-landauer-alp-quantum-app}
   Q \geq \frac{\hbar \omega}{2} \coth \left(\frac{\hbar\omega}{2 k_B T} \right) \left(e^{-\Delta s_\text{W}}-1\right).
\end{equation}

\section{Minimal dissipation for the semi-classical PBS}
\label{PBS-appendix-classical}

Let $\hat{n}$ be the unit vector which is perpendicular to the PBS surface and is pointing \emph{into} the bulk. The PBS and the photon have momenta $\vec{p}_{b}$ and $\vec{p}$ respectively in the lab frame. We define $p_{bn} = \hat{n}\cdot\vec{p}_b$ and $p_n = \hat{n} \cdot \vec{p}$ as the inward normal components of $\vec{p}_{b}$ and $\vec{p}$. Using a Lorentz transformation, this component of the photon's momentum in the rest frame of the PBS is then \cite{french_special_2017}
\begin{equation}
p_n^{\text{(PBS)}}=\gamma(p_n-\hbar \omega p_{bn}/m c^2),
\end{equation}
where $\gamma$ is the Lorentz factor and $\omega$ is the frequency in the lab frame. Because the PBS momentum is nonrelativistic we set $\gamma=1$.
We assume that in the rest frame of the PBS, the photon's angle of reflection is the same as the angle of incidence, although this is only approximately true due to the transfer of energy from the photon to the PBS. That is, we set
\begin{equation}
p'^{\text{(PBS)}}_n = - p^{\text{(PBS)}}_n.
\end{equation}
Thus in the lab frame, the photon's momentum after the collision differs from its initial momentum by an amount
\begin{equation}
\label{delta-p-lab}
\delta p_n = p_n'-p_n =- 2 p_n  + 2 \hbar \omega p_{bn}/m c^2.
\end{equation}
To ensure that a reflected photon can be distinguished from a transmitted photon, the difference in momentum of the two paths should be greater than the uncertainty in momentum of the reflected photon, that is
\begin{equation}
 \left| \braket{\delta p_n}\right| \geq \sigma [\delta p_n].
\end{equation}
This condition is in effect a restriction on the statistical distance (as defined by Wootters \cite{wootters_statistical_1981}) between the momenta of the transmitted and reflected photons. And so from Eq. \eqref{delta-p-lab}, and the fact that $\sigma[p_n]=0$ we have
\begin{equation}
 \left| \delta p_n \right| \geq 2 \hbar \omega \sigma[ p_{bn}]/m c^2 ,
\end{equation}
Because the PBS is at finite temperature $T$, we assume its momentum is initially given by the canonical distribution
\begin{equation}
    p_{bn}\sim \mathcal{N}(0, m k_B T),
\end{equation}
and therefore
\begin{equation}
 \left|\delta p_{n}\right| \geq \frac{2\hbar \omega}{c^2}\sqrt{\frac{k_B T}{m}}.
\end{equation}
Next note that for momentum to be conserved, the momentum of the PBS must also change by an amount $\delta p_{bn}=-\delta p_n$. In our convention $\delta p_{bn}$ is positive by definition, so
\begin{equation}
\delta p_{bn} \geq  \frac{2 \hbar \omega}{c^2}\sqrt{\frac{k_B T}{m}}.
\end{equation}
Many photons are incident on the device, one after another. For each photon, the initial momentum $p_n$ is the same, and so $\delta p_{bn}$ is very nearly the same for each reflected photon, so we treat it as constant. We assume that each photon is vertically polarized with probability one half and horizontally polarized with probability one half. So with probability one half $p_{bn}$ is increased by $\delta p_{bn}$ as each photon passes. Therefore after $N$ photons have come, the total change in the momentum of the PBS is given by a binomial distribution
\begin{equation}
\frac{p_{bn}(t)-p_{bn}(0)}{\delta p_{bn}}\sim B(N, 1/2)
\end{equation}
By the de Moivre-Laplace theorem \cite{weisstein_moivre-laplace_nodate}, for large $N$ the distribution approaches a Gaussian with variance $\sigma^2=N/4$. Therefore, 
\begin{equation}
\sigma^2[p_{bn}(t)-p_{bn}(0)]=\delta p_{bn}^2 N/4.
\end{equation}
Recall that the sum of two Gaussian random variables is Gaussian, with variance equal to the sum of the two original distributions' variances.  Using the variance of $p_{bn}(0)$ then gives
\begin{equation}
\sigma^2[p_{bn}(t)]=\delta p_{bn}^2 N/4 + m k_B T.
\end{equation}
Then, using the formula for the Shannon entropy of a Gaussian, we have
\begin{equation}
s(t)-s(0)=\frac{1}{2} \lo{\frac{\delta p_{bn}^2 N/4 + m k_B T}{m k_B T}}
\end{equation}
\begin{equation}
s(t)-s(0)\geq \frac{1}{2} \lo{\frac{\hbar^2 \omega^2  }{m^2 c^4} N+ 1}.
\end{equation}
Our approximation is only valid for $\hbar \omega \ll m c^2$, and in this limit we have
\begin{equation}
s(t)-s(0) \geq \frac{1}{2}  \frac{\hbar^2 \omega^2  }{m^2 c^4} N.
\end{equation}
After many photons have passed, $p_{n}$ is in an equilibrium distribution, so we can equate the Shannon entropy with the thermal entropy. Then the heat dissipated as a single photon passes is
\begin{equation}
Q\geq \frac{\hbar^2\omega^2}{2 m^2 c^4} k_B T.
\end{equation}
The information carried by the photon is one bit, as per our assumption, so we may write this as
\begin{equation}
\label{PBS-landauer-semiclassical-app}
\dbar Q\geq \frac{\hbar^2\omega^2}{2 m^2 c^4}\, k_B T  (-ds).
\end{equation}

\section{Minimal loss of purity for quantum PBS}
\label{quantum-pbs-appendix}

The joint state of the system and the bath is acted on by a unitary operator $U$. Define the quantum state
\begin{equation}
    \rho_B^{i} = \ptr{S}{U[\ket{i0}\bra{i0} \otimes \rho_B ]U^\dag},
\end{equation}
and
\begin{equation}
    \rho_S^{i} = \ptr{B}{U[\ket{i0}\bra{i0} \otimes \rho_B ]U^\dag},
\end{equation}
where $i \in \{0,1\}$. Therefore, we write
\begin{equation}
    \rho_B ' = \frac{1}{2}\left( \rho_B^{0}+\rho_B^{1}\right).
\end{equation}
Now suppose that the bath is composed of $N$ qubits. For conservation of angular momentum to hold, we require that
\begin{equation}
\begin{split}
   &\tr{\rho_B^{i}\sum_{j=1}^N Z_B^j} + \tr{\rho_S^i (Z_S^1 + Z_S^2)} = \\ 
   &\tr{\rho_B\sum_{j=1}^N Z_B^j} + \tr{\ket{i0}\bra{i0} (Z_S^1 + Z_S^2)},
\end{split}
\end{equation}
where $Z$ is the Pauli Z-matrix. The noisy CNOT is implemented on the system qubits within accuracy $\epsilon$, meaning
\begin{equation}
   \sqrt{ \tr{(\rho_S' - U_\text{CN} \rho_S U^\dag_\text{CN})^2} }\leq \epsilon\,.
\end{equation}
Employing the Cauchy-Schwarz inequality we obtain
\begin{equation}
\begin{split}
    &|\tr{(\rho_S' - U_\text{CN} \rho_S U^\dag_\text{CN}) (Z_S^1 + Z_S^2)}| \leq\\
    &\quad \epsilon \sqrt{\tr{(Z_1 + Z_2)^2}} = 2\sqrt{2}\epsilon
\end{split}
\end{equation}
and hence
\begin{equation}
   | \tr{\rho_S^i (Z_S^1 + Z_S^2)} - 2(-1)^i| \leq  2\sqrt{2}\epsilon,
\end{equation}
and
\begin{equation}
   \tr{\ket{i0}\bra{i0} (Z_S^1 + Z_S^2)} = (-1)^i-1.
\end{equation}
Finally, we can then write
\begin{equation}
    \left|\tr{\left(\rho_B^{i}-\rho_B\right)\sum_j Z_B^j} - [-1-(-1)^i]\right| \leq  2\sqrt{2}\epsilon\,,
\end{equation}
which further leads to
\begin{equation}
    \tr{\left(\rho_B^{1}-\rho_B^{0}\right)\sum_j Z_B^j} \geq2 -4\sqrt{2}\epsilon.
\end{equation}
Thus, the purity of the final state $\rho_B'$ is
\begin{equation}
\tr{\rho_B'^2} = \frac{1}{4}\left(
\tr{(\rho_B^0)^2} +\tr{(\rho_B^1)^2} + 2 \tr{\rho_B^0 \rho_B^1} \right).
\end{equation}
Note that the last term can also be expressed as
 \begin{equation}
\begin{split}
&2\, \tr{\rho_B^0 \rho_B^1}    = \tr{(\rho_B^0)^2} +\tr{(\rho_B^1)^2} \\
&\qquad- \tr{(\rho_B^1 - \rho_B^0)^2}.
\end{split}
\end{equation}
Now, since the purity of the states $\rho_B^0$ and $\rho_B^1$ cannot exceed the purity of $\rho_B$, we have
\begin{equation}
    2\, \tr{\rho_B^0 \rho_B^1}    \leq 2 \tr{\rho_B^2}- \tr{(\rho_B^1 - \rho_B^0)^2}.
\end{equation}
Using the Cauchy-Schwarz inequality again we obtain
\begin{equation}
\begin{split}
    &\tr{\left(\rho_B^1 - \rho_B^0\right)\sum_{j=1}^N Z_B^j}^2 \leq \\
    &\qquad \tr{\left(\rho_B^1 - \rho_B^0\right)^2} \tr{\left(\sum_{j=1}^N Z_B^j \right)^2}
\end{split}
\end{equation}
and with $\text{tr}\{Z_j Z_i\}=0$ for $i\neq j$ we can have
\begin{equation}
\tr{\left(\sum_{j=1}^NZ_j\right)^2} = \tr{\sum_{j=1}^NZ_j^2}=N 2^N\,.
\end{equation}
Collecting expressions we can further write
\begin{equation}
\begin{split}
   &\tr{\left(\rho_B^1 - \rho_B^0\right)^2 } \geq \frac{1}{N2^N} \tr{\left(\rho_B^1 - \rho_B^0\right)\sum_j Z_B^j}^2\\
   &\qquad \geq \frac{(2-4\sqrt{2}\epsilon)^2}{N2^N}
\end{split}
\end{equation}
and finally
\begin{equation}
    2\, \tr{\rho_B^0 \rho_B^1}   \leq 2\,\tr{ \rho_B^2} - \frac{(2-4\sqrt{2}\epsilon)^2}{N2^N},
\end{equation}
which we rewrite (as in the main text) as
\begin{equation}
\tr{\rho_B'^2} \leq \tr{\rho_B^2}-\frac{1}{N2^N}\left(1 - 2\sqrt{2}\epsilon \right)^2. 
\end{equation}

\section{Derivation of modified commutation relation}
\label{commutation-appendix}

Consider the case where a PBS is at an unknown angle $\theta$, with some probability density $p(\theta)$. If the state of the two input modes is given some density matrix $\sigma_0=\rho_0\otimes \eta_0$, then the output density matrix is
\begin{equation}
\label{ensemble-average-schrodinger}
\sigma_1 = \int_{-\infty}^\infty d\theta\, p(\theta) U_\theta \sigma_0 U_\theta^\dag = \mathcal{E}(\sigma_0),
\end{equation}
where we define $\mathcal{E}$ as the quantum channel which propagates the density matrix in the Schr\"odinger picture. There is an adjoint map $\mathcal{E}^\dag$ which can be used to propagate operators in the Heisenberg picture, and satisfies \cite{wolf_quantum_2012}
\begin{equation}
\label{channel-adjoint-def}
\tr{\rho\, \mathcal{E}^\dag(X)} =\tr{\mathcal{E}(\rho)\,X} ,
\end{equation}
for all density matrices $\rho$ and operators $X$. In fact it is easily seen that $\mathcal{E}^\dag$ is given by
\begin{equation}
\label{adjoint-map}
\mathcal{E}^\dag(X)=\int_{-\infty}^\infty d\theta\, f(\theta)U_\theta^\dag X U_\theta.
\end{equation}

Then using Eq.~\eqref{heisenberg-transformation} we see that the adjoint map acting on the annihilation operators $x_i$ can be expressed as
\begin{equation}
\label{averaged-adjoint}
\mathcal{E}^\dag(x_i)=(\bar{S}\vec{x})_i,
\end{equation}
where we have defined the ensemble-averaged scattering matrix as
\begin{equation}
\label{averaged-scattering-matrix-def}
\bar{S}=\int_{-\infty}^\infty d\theta\, f(\theta)S_\theta.
\end{equation}
Note that the evolved operators $\mathcal{E}(x_i)$ do not necessarily obey the standard Dirac commutation relation anymore since $\bar{J}$ need not be unitary. While we can evaluate expectation values using Eq.~\eqref{channel-adjoint-def}, we cannot build new operators or states using the evolved operators. However, many useful operators are expressed in terms of products of annihilation operators. We will consider the class of operators which can be written as
\begin{equation}
\label{schwinger-operators}
X=\sum_{ij}X_{ij} x^\dag_i x_j.
\end{equation}
To see how such an operator evolves under these dynamics, we first note
\begin{equation}
U_\theta^\dag x_i^\dag x_j U_\theta = \sum_{k}S_{\theta ik}^* x_k^\dag \sum_{\ell}S_{\theta j\ell} x_\ell\ = \sum_{k \ell} \mathcal{S}^{ij}_{k\ell}(\theta) x^\dag_k x_\ell \,,
\end{equation}
where we defined
\begin{equation}
\mathcal{S}^{ij}_{k\ell}(\theta)=S_{\theta ik}^* S_{\theta j\ell}.
\end{equation}
Therefore by Eq.~\eqref{adjoint-map} we have
\begin{equation}
\mathcal{E}^\dag(X)=\sum_{i j k \ell}X_{ij} \overline{ \mathcal{S}}^{ij}_{k\ell} x^\dag_k x_\ell,
\end{equation}
where
\begin{equation}
\overline{ \mathcal{S}}^{ij}_{k\ell}=\int_{-\infty}^\infty d\theta\, p(\theta) \mathcal{S}^{ij}_{k\ell}(\theta).
\end{equation}

 We will now assume that $f(\theta)$ is Gaussian, with some temperature $T$
\begin{equation}
\label{theta-dist-app}
f(\theta)=\frac{1}{\sqrt{2\pi k_B T/\kappa}}\exp \left(-\frac{1}{2}\frac{\theta^2}{k_B T/\kappa}\right).
\end{equation}
Then evaluating Eq. \eqref{averaged-scattering-matrix-def}, we find that
\begin{widetext}
\begin{equation}
\bar{S} =
 \begin{pmatrix}
 1-(1-t)\chi & 0 & r\chi & 0 \\
 0 & t+(1-t)\chi  & 0 &  r- r\chi \\
 - r \chi & 0  & 1-(1-t)\chi & 0\\
 0 & r \chi - r & 0 & t + (1-t) \chi 
\end{pmatrix},
\end{equation}
\end{widetext}
where $\chi=\frac{1}{2}\left(1-\exp(-2 k_B T/\kappa)\right)$. We can then express the primed annihilation operators as
\begin{equation}
\label{ah-prime}
a_\text{h}'=\left( 1 - (1-t) \chi\right)a_\text{h} - r\chi\, b_\text{h}
\end{equation}
and
\begin{equation}
\label{av-prime}
a_\text{v}'=\left(t +(1-t) \chi \right) a_\text{v}   + r\left(\chi-1\right) b_\text{v}.
\end{equation}
Note that $\chi$ is a monotonically increasing function of temperature and $\chi=0$ at $T=0$, and hence Eqs.~\eqref{ah-prime} and \eqref{av-prime} clearly show the influence on increasing temperature on the propagation of the annihilation operators. For zero temperature we have
\begin{equation}
\label{ah-prime-inf}
\lim_{T\to 0}a_\text{h}'=a_\text{h}\quad \mrm{and}\quad \lim_{T\to 0} a_\text{v}'=t a_\text{v} - r b_\text{v}
\end{equation}
meaning the model behaves as a PBS at definite angle, as expected. In the limit of infinite temperature $\chi \to 1/2$, and we find that
\begin{equation}
\lim_{T\to \infty}a_\text{h}'=\frac{1+t}{2}a_\text{h} - \frac{r}{2} b_\text{h},
\end{equation}
and
\begin{equation}
\label{av-prime-inf}
\lim_{T\to \infty} a_\text{v}'=\frac{1+t}{2}a_\text{v} - \frac{r}{2} b_\text{v},
\end{equation}
meaning that horizontally polarized and vertically polarized light are not distinguished by the device. The primed annihilation operators do not obey the standard commutation relations. Instead we find that
\begin{equation}
\label{primed-commutators-app}
[a_\text{h}',a_\text{h}'^\dag] =[a_\text{v}',a_\text{v}'^\dag]  =1-2(1-t)(\chi-\chi^2).
\end{equation}

\begin{widetext}
\section{Construction of the Evolution Operator}
\label{evo-appendix}

In this last appendix, we show that the two treatments we gave in the Heisenberg and Schr\"odinger picture are in fact equivalent. This formulation is not a new result but we have included it for convenience (for similar calculations see e.g. \cite{leonhardt_essential_2010, leonhardt_quantum_2003}). Recall that the evolution operator was obtained as follows, for $t=1$
\begin{equation}
U(t)= \exp \left( t \sum_{ij} \braket{ i |\ln (S)|j}a_i^\dag a_j\right),
\end{equation}
where $S$ is the Unitary scattering matrix. Then 
\begin{equation}
\partial_t a_k' = (\partial_t  U^\dag) a_k U + U^\dag a_k ( \partial_t U),
\end{equation}
and
\begin{equation}
\partial_t a_k' = U^\dag \sum_{ij} \braket{ j |\ln (S)^\dag |i}a_j^\dag a_i a_k U + U^\dag \sum_{ij} \braket{ i |\ln (S)|j}a_k a_i^\dag a_j U. 
\end{equation}
Since $S$ is unitary, $\ln(S)$ is anti-Hermitian. Using this fact and interchanging $i$ and $j$ in the first summation, we get
\begin{equation}
\partial_t a_k' = U^\dag \left( -\sum_{ij} \braket{ i |\ln (S) |j}a_i^\dag a_j a_k + \sum_{ij} \braket{ i |\ln (S)|j}a_k a_i^\dag a_j \right)U, 
\end{equation}
and
\begin{equation}
\partial_t a_k' = U^\dag\sum_{ij} \braket{i | \ln(S)|j}[a_k, a_i^\dag a_j]U  = U^\dag\sum_{j} \braket{k| \ln(S)|j}a_jU \,.
\end{equation}
Thus, we have
\begin{equation}
\partial_t a_k' =\sum_{j} \braket{k| \ln(S)|j}a_j'.
\end{equation}
The unique solution to this equation is
\begin{equation}
 a_k' =\sum_{kj} \braket{k| \exp(t \ln S)|j}a_j' \,.
\end{equation}
\end{widetext}

\bibliography{ref}

\begin{thebibliography}{103}%
\makeatletter
\providecommand \@ifxundefined [1]{%
 \@ifx{#1\undefined}
}%
\providecommand \@ifnum [1]{%
 \ifnum #1\expandafter \@firstoftwo
 \else \expandafter \@secondoftwo
 \fi
}%
\providecommand \@ifx [1]{%
 \ifx #1\expandafter \@firstoftwo
 \else \expandafter \@secondoftwo
 \fi
}%
\providecommand \natexlab [1]{#1}%
\providecommand \enquote  [1]{``#1''}%
\providecommand \bibnamefont  [1]{#1}%
\providecommand \bibfnamefont [1]{#1}%
\providecommand \citenamefont [1]{#1}%
\providecommand \href@noop [0]{\@secondoftwo}%
\providecommand \href [0]{\begingroup \@sanitize@url \@href}%
\providecommand \@href[1]{\@@startlink{#1}\@@href}%
\providecommand \@@href[1]{\endgroup#1\@@endlink}%
\providecommand \@sanitize@url [0]{\catcode `\\12\catcode `\$12\catcode
  `\&12\catcode `\#12\catcode `\^12\catcode `\_12\catcode `\%12\relax}%
\providecommand \@@startlink[1]{}%
\providecommand \@@endlink[0]{}%
\providecommand \url  [0]{\begingroup\@sanitize@url \@url }%
\providecommand \@url [1]{\endgroup\@href {#1}{\urlprefix }}%
\providecommand \urlprefix  [0]{URL }%
\providecommand \Eprint [0]{\href }%
\providecommand \doibase [0]{https://doi.org/}%
\providecommand \selectlanguage [0]{\@gobble}%
\providecommand \bibinfo  [0]{\@secondoftwo}%
\providecommand \bibfield  [0]{\@secondoftwo}%
\providecommand \translation [1]{[#1]}%
\providecommand \BibitemOpen [0]{}%
\providecommand \bibitemStop [0]{}%
\providecommand \bibitemNoStop [0]{.\EOS\space}%
\providecommand \EOS [0]{\spacefactor3000\relax}%
\providecommand \BibitemShut  [1]{\csname bibitem#1\endcsname}%
\let\auto@bib@innerbib\@empty
\bibitem [{\citenamefont {Raymer}\ and\ \citenamefont
  {Monroe}(2019)}]{Raymer2019QST}%
  \BibitemOpen
  \bibfield  {author} {\bibinfo {author} {\bibfnamefont {M.~G.}\ \bibnamefont
  {Raymer}}\ and\ \bibinfo {author} {\bibfnamefont {C.}~\bibnamefont
  {Monroe}},\ }\bibfield  {title} {\bibinfo {title} {The {US} national quantum
  initiative},\ }\href {https://doi.org/10.1088/2058-9565/ab0441} {\bibfield
  {journal} {\bibinfo  {journal} {Quantum Sci. Technol.}\ }\textbf {\bibinfo
  {volume} {4}},\ \bibinfo {pages} {020504} (\bibinfo {year}
  {2019})}\BibitemShut {NoStop}%
\bibitem [{\citenamefont {Deffner}\ and\ \citenamefont
  {Campbell}(2019)}]{Deffner2019book}%
  \BibitemOpen
  \bibfield  {author} {\bibinfo {author} {\bibfnamefont {S.}~\bibnamefont
  {Deffner}}\ and\ \bibinfo {author} {\bibfnamefont {S.}~\bibnamefont
  {Campbell}},\ }\href@noop {} {\emph {\bibinfo {title} {Quantum
  Thermodynamics}}}\ (\bibinfo  {publisher} {Morgan \& Claypool Publishers},\
  \bibinfo {year} {2019})\BibitemShut {NoStop}%
\bibitem [{\citenamefont {Parrondo}\ \emph {et~al.}(2015)\citenamefont
  {Parrondo}, \citenamefont {Horowitz},\ and\ \citenamefont
  {Sagawa}}]{Parrondo2015NP}%
  \BibitemOpen
  \bibfield  {author} {\bibinfo {author} {\bibfnamefont {J.~M.~R.}\
  \bibnamefont {Parrondo}}, \bibinfo {author} {\bibfnamefont {J.~M.}\
  \bibnamefont {Horowitz}},\ and\ \bibinfo {author} {\bibfnamefont
  {T.}~\bibnamefont {Sagawa}},\ }\bibfield  {title} {\bibinfo {title}
  {Thermodynamics of information},\ }\href {https://doi.org/10.1038/nphys3230}
  {\bibfield  {journal} {\bibinfo  {journal} {Nature Physics}\ }\textbf
  {\bibinfo {volume} {11}},\ \bibinfo {pages} {131} (\bibinfo {year}
  {2015})}\BibitemShut {NoStop}%
\bibitem [{\citenamefont {Landauer}(1961)}]{landauer_irreversibility_1961}%
  \BibitemOpen
  \bibfield  {author} {\bibinfo {author} {\bibfnamefont {R.}~\bibnamefont
  {Landauer}},\ }\bibfield  {title} {\bibinfo {title} {Irreversibility and
  {Heat} {Generation} in the {Computing} {Process}},\ }\href
  {https://doi.org/10.1147/rd.53.0183} {\bibfield  {journal} {\bibinfo
  {journal} {IBM J. Res. \& Dev.}\ }\textbf {\bibinfo {volume} {5}},\ \bibinfo
  {pages} {183} (\bibinfo {year} {1961})}\BibitemShut {NoStop}%
\bibitem [{\citenamefont {Piechocinska}(2000)}]{piechocinska_information_2000}%
  \BibitemOpen
  \bibfield  {author} {\bibinfo {author} {\bibfnamefont {B.}~\bibnamefont
  {Piechocinska}},\ }\bibfield  {title} {\bibinfo {title} {Information
  erasure},\ }\href {https://doi.org/10.1103/PhysRevA.61.062314} {\bibfield
  {journal} {\bibinfo  {journal} {Phys. Rev. A}\ }\textbf {\bibinfo {volume}
  {61}},\ \bibinfo {pages} {062314} (\bibinfo {year} {2000})}\BibitemShut
  {NoStop}%
\bibitem [{\citenamefont {Deffner}\ and\ \citenamefont
  {Jarzynski}(2013)}]{deffner_information_2013}%
  \BibitemOpen
  \bibfield  {author} {\bibinfo {author} {\bibfnamefont {S.}~\bibnamefont
  {Deffner}}\ and\ \bibinfo {author} {\bibfnamefont {C.}~\bibnamefont
  {Jarzynski}},\ }\bibfield  {title} {\bibinfo {title} {Information
  {Processing} and the {Second} {Law} of {Thermodynamics}: {An} {Inclusive},
  {Hamiltonian} {Approach}},\ }\href
  {https://doi.org/10.1103/PhysRevX.3.041003} {\bibfield  {journal} {\bibinfo
  {journal} {Phys. Rev. X}\ }\textbf {\bibinfo {volume} {3}},\ \bibinfo {pages}
  {041003} (\bibinfo {year} {2013})}\BibitemShut {NoStop}%
\bibitem [{\citenamefont {Boyd}\ and\ \citenamefont
  {Crutchfield}(2016)}]{Boyd2016PRL}%
  \BibitemOpen
  \bibfield  {author} {\bibinfo {author} {\bibfnamefont {A.~B.}\ \bibnamefont
  {Boyd}}\ and\ \bibinfo {author} {\bibfnamefont {J.~P.}\ \bibnamefont
  {Crutchfield}},\ }\bibfield  {title} {\bibinfo {title} {Maxwell demon
  dynamics: Deterministic chaos, the {S}zilard map, and the intelligence of
  thermodynamic systems},\ }\href
  {https://doi.org/10.1103/PhysRevLett.116.190601} {\bibfield  {journal}
  {\bibinfo  {journal} {Phys. Rev. Lett.}\ }\textbf {\bibinfo {volume} {116}},\
  \bibinfo {pages} {190601} (\bibinfo {year} {2016})}\BibitemShut {NoStop}%
\bibitem [{\citenamefont {Boyd}\ \emph {et~al.}(2016)\citenamefont {Boyd},
  \citenamefont {Mandal},\ and\ \citenamefont {Crutchfield}}]{Boyd2016NJP}%
  \BibitemOpen
  \bibfield  {author} {\bibinfo {author} {\bibfnamefont {A.~B.}\ \bibnamefont
  {Boyd}}, \bibinfo {author} {\bibfnamefont {D.}~\bibnamefont {Mandal}},\ and\
  \bibinfo {author} {\bibfnamefont {J.~P.}\ \bibnamefont {Crutchfield}},\
  }\bibfield  {title} {\bibinfo {title} {Identifying functional thermodynamics
  in autonomous {M}axwellian ratchets},\ }\href
  {https://doi.org/10.1088/1367-2630/18/2/023049} {\bibfield  {journal}
  {\bibinfo  {journal} {New J. Phys.}\ }\textbf {\bibinfo {volume} {18}},\
  \bibinfo {pages} {023049} (\bibinfo {year} {2016})}\BibitemShut {NoStop}%
\bibitem [{\citenamefont {Boyd}\ \emph
  {et~al.}(2018{\natexlab{a}})\citenamefont {Boyd}, \citenamefont {Mandal},\
  and\ \citenamefont {Crutchfield}}]{Boyd2018PRX}%
  \BibitemOpen
  \bibfield  {author} {\bibinfo {author} {\bibfnamefont {A.~B.}\ \bibnamefont
  {Boyd}}, \bibinfo {author} {\bibfnamefont {D.}~\bibnamefont {Mandal}},\ and\
  \bibinfo {author} {\bibfnamefont {J.~P.}\ \bibnamefont {Crutchfield}},\
  }\bibfield  {title} {\bibinfo {title} {Thermodynamics of modularity:
  Structural costs beyond the {Landauer} bound},\ }\href
  {https://doi.org/10.1103/PhysRevX.8.031036} {\bibfield  {journal} {\bibinfo
  {journal} {Phys. Rev. X}\ }\textbf {\bibinfo {volume} {8}},\ \bibinfo {pages}
  {031036} (\bibinfo {year} {2018}{\natexlab{a}})}\BibitemShut {NoStop}%
\bibitem [{\citenamefont {Hilt}\ \emph {et~al.}(2011)\citenamefont {Hilt},
  \citenamefont {Shabbir}, \citenamefont {Anders},\ and\ \citenamefont
  {Lutz}}]{hilt_landauers_2011}%
  \BibitemOpen
  \bibfield  {author} {\bibinfo {author} {\bibfnamefont {S.}~\bibnamefont
  {Hilt}}, \bibinfo {author} {\bibfnamefont {S.}~\bibnamefont {Shabbir}},
  \bibinfo {author} {\bibfnamefont {J.}~\bibnamefont {Anders}},\ and\ \bibinfo
  {author} {\bibfnamefont {E.}~\bibnamefont {Lutz}},\ }\bibfield  {title}
  {\bibinfo {title} {Landauer’s principle in the quantum regime},\ }\href
  {https://doi.org/10.1103/PhysRevE.83.030102} {\bibfield  {journal} {\bibinfo
  {journal} {Phys. Rev. E}\ }\textbf {\bibinfo {volume} {83}},\ \bibinfo
  {pages} {030102} (\bibinfo {year} {2011})}\BibitemShut {NoStop}%
\bibitem [{\citenamefont {Lorenzo}\ \emph
  {et~al.}(2015{\natexlab{a}})\citenamefont {Lorenzo}, \citenamefont
  {McCloskey}, \citenamefont {Ciccarello}, \citenamefont {Paternostro},\ and\
  \citenamefont {Palma}}]{lorenzo_landauers_2015}%
  \BibitemOpen
  \bibfield  {author} {\bibinfo {author} {\bibfnamefont {S.}~\bibnamefont
  {Lorenzo}}, \bibinfo {author} {\bibfnamefont {R.}~\bibnamefont {McCloskey}},
  \bibinfo {author} {\bibfnamefont {F.}~\bibnamefont {Ciccarello}}, \bibinfo
  {author} {\bibfnamefont {M.}~\bibnamefont {Paternostro}},\ and\ \bibinfo
  {author} {\bibfnamefont {G.}~\bibnamefont {Palma}},\ }\bibfield  {title}
  {\bibinfo {title} {Landauer’s {Principle} in {Multipartite} {Open}
  {Quantum} {System} {Dynamics}},\ }\href
  {https://doi.org/10.1103/PhysRevLett.115.120403} {\bibfield  {journal}
  {\bibinfo  {journal} {Phys. Rev. Lett.}\ }\textbf {\bibinfo {volume} {115}},\
  \bibinfo {pages} {120403} (\bibinfo {year} {2015}{\natexlab{a}})}\BibitemShut
  {NoStop}%
\bibitem [{\citenamefont {Van~Vu}\ and\ \citenamefont
  {Saito}(2022)}]{van_vu_finite-time_2022}%
  \BibitemOpen
  \bibfield  {author} {\bibinfo {author} {\bibfnamefont {T.}~\bibnamefont
  {Van~Vu}}\ and\ \bibinfo {author} {\bibfnamefont {K.}~\bibnamefont {Saito}},\
  }\bibfield  {title} {\bibinfo {title} {Finite-{Time} {Quantum} {Landauer}
  {Principle} and {Quantum} {Coherence}},\ }\href
  {https://doi.org/10.1103/PhysRevLett.128.010602} {\bibfield  {journal}
  {\bibinfo  {journal} {Phys. Rev. Lett.}\ }\textbf {\bibinfo {volume} {128}},\
  \bibinfo {pages} {010602} (\bibinfo {year} {2022})}\BibitemShut {NoStop}%
\bibitem [{\citenamefont {Goold}\ \emph {et~al.}(2015)\citenamefont {Goold},
  \citenamefont {Paternostro},\ and\ \citenamefont
  {Modi}}]{goold_nonequilibrium_2015}%
  \BibitemOpen
  \bibfield  {author} {\bibinfo {author} {\bibfnamefont {J.}~\bibnamefont
  {Goold}}, \bibinfo {author} {\bibfnamefont {M.}~\bibnamefont {Paternostro}},\
  and\ \bibinfo {author} {\bibfnamefont {K.}~\bibnamefont {Modi}},\ }\bibfield
  {title} {\bibinfo {title} {Nonequilibrium {Quantum} {Landauer} {Principle}},\
  }\href {https://doi.org/10.1103/PhysRevLett.114.060602} {\bibfield  {journal}
  {\bibinfo  {journal} {Phys. Rev. Lett.}\ }\textbf {\bibinfo {volume} {114}},\
  \bibinfo {pages} {060602} (\bibinfo {year} {2015})}\BibitemShut {NoStop}%
\bibitem [{\citenamefont {B{\'e}rut}\ \emph {et~al.}(2012)\citenamefont
  {B{\'e}rut}, \citenamefont {Arakelyan}, \citenamefont {Petrosyan},
  \citenamefont {Ciliberto}, \citenamefont {Dillenschneider},\ and\
  \citenamefont {Lutz}}]{Berut2012}%
  \BibitemOpen
  \bibfield  {author} {\bibinfo {author} {\bibfnamefont {A.}~\bibnamefont
  {B{\'e}rut}}, \bibinfo {author} {\bibfnamefont {A.}~\bibnamefont
  {Arakelyan}}, \bibinfo {author} {\bibfnamefont {A.}~\bibnamefont
  {Petrosyan}}, \bibinfo {author} {\bibfnamefont {S.}~\bibnamefont
  {Ciliberto}}, \bibinfo {author} {\bibfnamefont {R.}~\bibnamefont
  {Dillenschneider}},\ and\ \bibinfo {author} {\bibfnamefont {E.}~\bibnamefont
  {Lutz}},\ }\bibfield  {title} {\bibinfo {title} {Experimental verification of
  {Landauer's} principle linking information and thermodynamics},\ }\href
  {https://doi.org/10.1038/nature10872} {\bibfield  {journal} {\bibinfo
  {journal} {Nature}\ }\textbf {\bibinfo {volume} {483}},\ \bibinfo {pages}
  {187} (\bibinfo {year} {2012})}\BibitemShut {NoStop}%
\bibitem [{\citenamefont {Maroney}(2009)}]{maroney_generalizing_2009}%
  \BibitemOpen
  \bibfield  {author} {\bibinfo {author} {\bibfnamefont {O.~J.~E.}\
  \bibnamefont {Maroney}},\ }\bibfield  {title} {\bibinfo {title} {Generalizing
  {Landauer}’s principle},\ }\href
  {https://doi.org/10.1103/PhysRevE.79.031105} {\bibfield  {journal} {\bibinfo
  {journal} {Phys. Rev. E}\ }\textbf {\bibinfo {volume} {79}},\ \bibinfo
  {pages} {031105} (\bibinfo {year} {2009})}\BibitemShut {NoStop}%
\bibitem [{\citenamefont {Norton}(2011)}]{norton_waiting_2011}%
  \BibitemOpen
  \bibfield  {author} {\bibinfo {author} {\bibfnamefont {J.~D.}\ \bibnamefont
  {Norton}},\ }\bibfield  {title} {\bibinfo {title} {Waiting for {Landauer}},\
  }\href {https://doi.org/10.1016/j.shpsb.2011.05.002} {\bibfield  {journal}
  {\bibinfo  {journal} {Stud. Hist. Philos. Sci. Part B: Stud. Hist. Philos.
  Mod. Phys.}\ }\textbf {\bibinfo {volume} {42}},\ \bibinfo {pages} {184}
  (\bibinfo {year} {2011})}\BibitemShut {NoStop}%
\bibitem [{\citenamefont {Maroney}(2005)}]{maroney_absence_2005}%
  \BibitemOpen
  \bibfield  {author} {\bibinfo {author} {\bibfnamefont {O.}~\bibnamefont
  {Maroney}},\ }\bibfield  {title} {\bibinfo {title} {The (absence of a)
  relationship between thermodynamic and logical reversibility},\ }\href
  {https://doi.org/10.1016/j.shpsb.2004.11.006} {\bibfield  {journal} {\bibinfo
   {journal} {Stud. Hist. Philos. Sci. Part B: Stud. Hist. Philos. Mod. Phys.}\
  }\textbf {\bibinfo {volume} {36}},\ \bibinfo {pages} {355} (\bibinfo {year}
  {2005})}\BibitemShut {NoStop}%
\bibitem [{\citenamefont {Bylicka}\ \emph {et~al.}(2016)\citenamefont
  {Bylicka}, \citenamefont {Tukiainen}, \citenamefont
  {Chru{\'{s}}ci{\'{n}}ski}, \citenamefont {Piilo},\ and\ \citenamefont
  {Maniscalco}}]{Bylicka2016}%
  \BibitemOpen
  \bibfield  {author} {\bibinfo {author} {\bibfnamefont {B.}~\bibnamefont
  {Bylicka}}, \bibinfo {author} {\bibfnamefont {M.}~\bibnamefont {Tukiainen}},
  \bibinfo {author} {\bibfnamefont {D.}~\bibnamefont
  {Chru{\'{s}}ci{\'{n}}ski}}, \bibinfo {author} {\bibfnamefont
  {J.}~\bibnamefont {Piilo}},\ and\ \bibinfo {author} {\bibfnamefont
  {S.}~\bibnamefont {Maniscalco}},\ }\bibfield  {title} {\bibinfo {title}
  {Thermodynamic power of {non-Markovianity}},\ }\href
  {https://doi.org/10.1038/srep27989} {\bibfield  {journal} {\bibinfo
  {journal} {Sci. Rep.}\ }\textbf {\bibinfo {volume} {6}},\ \bibinfo {pages}
  {27989} (\bibinfo {year} {2016})}\BibitemShut {NoStop}%
\bibitem [{\citenamefont {Pezzutto}\ \emph {et~al.}(2016)\citenamefont
  {Pezzutto}, \citenamefont {Paternostro},\ and\ \citenamefont
  {Omar}}]{Pezzutto2016}%
  \BibitemOpen
  \bibfield  {author} {\bibinfo {author} {\bibfnamefont {M.}~\bibnamefont
  {Pezzutto}}, \bibinfo {author} {\bibfnamefont {M.}~\bibnamefont
  {Paternostro}},\ and\ \bibinfo {author} {\bibfnamefont {Y.}~\bibnamefont
  {Omar}},\ }\bibfield  {title} {\bibinfo {title} {Implications of
  {non-Markovian} quantum dynamics for the {Landauer} bound},\ }\href
  {https://doi.org/10.1088/1367-2630/18/12/123018} {\bibfield  {journal}
  {\bibinfo  {journal} {New J. Phys.}\ }\textbf {\bibinfo {volume} {18}},\
  \bibinfo {pages} {123018} (\bibinfo {year} {2016})}\BibitemShut {NoStop}%
\bibitem [{\citenamefont {Zhang}\ \emph {et~al.}(2021)\citenamefont {Zhang},
  \citenamefont {Man},\ and\ \citenamefont {Xia}}]{Zhang2021}%
  \BibitemOpen
  \bibfield  {author} {\bibinfo {author} {\bibfnamefont {Q.}~\bibnamefont
  {Zhang}}, \bibinfo {author} {\bibfnamefont {Z.-X.}\ \bibnamefont {Man}},\
  and\ \bibinfo {author} {\bibfnamefont {Y.-J.}\ \bibnamefont {Xia}},\
  }\bibfield  {title} {\bibinfo {title} {{Non-Markovianity} and the {Landauer}
  principle in composite thermal environments},\ }\href
  {https://doi.org/10.1103/PhysRevA.103.032201} {\bibfield  {journal} {\bibinfo
   {journal} {Phys. Rev. A}\ }\textbf {\bibinfo {volume} {103}},\ \bibinfo
  {pages} {032201} (\bibinfo {year} {2021})}\BibitemShut {NoStop}%
\bibitem [{\citenamefont {Hu}\ \emph {et~al.}(2022)\citenamefont {Hu},
  \citenamefont {Li}, \citenamefont {Zou},\ and\ \citenamefont
  {Liu}}]{hu_relation_2022}%
  \BibitemOpen
  \bibfield  {author} {\bibinfo {author} {\bibfnamefont {H.-R.}\ \bibnamefont
  {Hu}}, \bibinfo {author} {\bibfnamefont {L.}~\bibnamefont {Li}}, \bibinfo
  {author} {\bibfnamefont {J.}~\bibnamefont {Zou}},\ and\ \bibinfo {author}
  {\bibfnamefont {W.-M.}\ \bibnamefont {Liu}},\ }\bibfield  {title} {\bibinfo
  {title} {Relation between non-{Markovianity} and {Landauer}'s principle},\
  }\href {https://doi.org/10.1103/PhysRevA.105.062429} {\bibfield  {journal}
  {\bibinfo  {journal} {Phys. Rev. A}\ }\textbf {\bibinfo {volume} {105}},\
  \bibinfo {pages} {062429} (\bibinfo {year} {2022})}\BibitemShut {NoStop}%
\bibitem [{\citenamefont {Man}\ \emph {et~al.}(2019)\citenamefont {Man},
  \citenamefont {Xia},\ and\ \citenamefont {Lo~Franco}}]{man_validity_2019}%
  \BibitemOpen
  \bibfield  {author} {\bibinfo {author} {\bibfnamefont {Z.-X.}\ \bibnamefont
  {Man}}, \bibinfo {author} {\bibfnamefont {Y.-J.}\ \bibnamefont {Xia}},\ and\
  \bibinfo {author} {\bibfnamefont {R.}~\bibnamefont {Lo~Franco}},\ }\bibfield
  {title} {\bibinfo {title} {Validity of the {Landauer} principle and quantum
  memory effects via collisional models},\ }\href
  {https://doi.org/10.1103/PhysRevA.99.042106} {\bibfield  {journal} {\bibinfo
  {journal} {Phys. Rev. A}\ }\textbf {\bibinfo {volume} {99}},\ \bibinfo
  {pages} {042106} (\bibinfo {year} {2019})}\BibitemShut {NoStop}%
\bibitem [{\citenamefont {Bremermann}(1967)}]{Bremermann1967}%
  \BibitemOpen
  \bibfield  {author} {\bibinfo {author} {\bibfnamefont {H.~J.}\ \bibnamefont
  {Bremermann}},\ }\bibfield  {title} {\bibinfo {title} {Quantum noise and
  information},\ }in\ \href {http://projecteuclid.org/euclid.bsmsp/1200513783}
  {\emph {\bibinfo {booktitle} {Proceedings of the Fifth Berkeley Symposium on
  Mathematical Statistics and Probability, Volume 4: Biology and Problems of
  Health}}}\ (\bibinfo  {publisher} {University of California Press},\ \bibinfo
  {address} {Berkeley, Calif.},\ \bibinfo {year} {1967})\ pp.\ \bibinfo {pages}
  {15--20}\BibitemShut {NoStop}%
\bibitem [{\citenamefont {Heisenberg}(1927)}]{Heisenberg1927}%
  \BibitemOpen
  \bibfield  {author} {\bibinfo {author} {\bibfnamefont {W.}~\bibnamefont
  {Heisenberg}},\ }\bibfield  {title} {\bibinfo {title} {{\"U}ber den
  anschaulichen {Inhalt} der quantentheoretischen {Kinematik und Mechanik}},\
  }\href {https://doi.org/10.1007/BF01397280} {\bibfield  {journal} {\bibinfo
  {journal} {Z. f{\"u}r Phys.}\ }\textbf {\bibinfo {volume} {43}},\ \bibinfo
  {pages} {172} (\bibinfo {year} {1927})}\BibitemShut {NoStop}%
\bibitem [{\citenamefont {Bekenstein}(1981{\natexlab{a}})}]{Bekenstein1981}%
  \BibitemOpen
  \bibfield  {author} {\bibinfo {author} {\bibfnamefont {J.~D.}\ \bibnamefont
  {Bekenstein}},\ }\bibfield  {title} {\bibinfo {title} {{Energy Cost of
  Information Transfer}},\ }\href {https://doi.org/10.1103/PhysRevLett.46.623}
  {\bibfield  {journal} {\bibinfo  {journal} {Phys. Rev. Lett.}\ }\textbf
  {\bibinfo {volume} {46}},\ \bibinfo {pages} {623} (\bibinfo {year}
  {1981}{\natexlab{a}})}\BibitemShut {NoStop}%
\bibitem [{\citenamefont {Bekenstein}(1973)}]{Bekenstein1973}%
  \BibitemOpen
  \bibfield  {author} {\bibinfo {author} {\bibfnamefont {J.~D.}\ \bibnamefont
  {Bekenstein}},\ }\bibfield  {title} {\bibinfo {title} {Black holes and
  entropy},\ }\href {https://doi.org/10.1103/PhysRevD.7.2333} {\bibfield
  {journal} {\bibinfo  {journal} {Phys. Rev. D}\ }\textbf {\bibinfo {volume}
  {7}},\ \bibinfo {pages} {2333} (\bibinfo {year} {1973})}\BibitemShut
  {NoStop}%
\bibitem [{\citenamefont {Bekenstein}(1974)}]{Bekenstein1974}%
  \BibitemOpen
  \bibfield  {author} {\bibinfo {author} {\bibfnamefont {J.~D.}\ \bibnamefont
  {Bekenstein}},\ }\bibfield  {title} {\bibinfo {title} {Generalized second law
  of thermodynamics in black-hole physics},\ }\href
  {https://journals.aps.org/prd/abstract/10.1103/PhysRevD.9.3292} {\bibfield
  {journal} {\bibinfo  {journal} {Phys. Rev. D}\ }\textbf {\bibinfo {volume}
  {9}},\ \bibinfo {pages} {3292} (\bibinfo {year} {1974})}\BibitemShut
  {NoStop}%
\bibitem [{\citenamefont {Hawking}(1975)}]{Hawking1975}%
  \BibitemOpen
  \bibfield  {author} {\bibinfo {author} {\bibfnamefont {S.~W.}\ \bibnamefont
  {Hawking}},\ }\bibfield  {title} {\bibinfo {title} {Particle creation by
  black holes},\ }\href {https://doi.org/10.1007/BF02345020} {\bibfield
  {journal} {\bibinfo  {journal} {Comm. Math. Phys.}\ }\textbf {\bibinfo
  {volume} {43}},\ \bibinfo {pages} {199} (\bibinfo {year} {1975})}\BibitemShut
  {NoStop}%
\bibitem [{\citenamefont {Bekenstein}(1981{\natexlab{b}})}]{Bekenstein1981PRD}%
  \BibitemOpen
  \bibfield  {author} {\bibinfo {author} {\bibfnamefont {J.~D.}\ \bibnamefont
  {Bekenstein}},\ }\bibfield  {title} {\bibinfo {title} {Universal upper bound
  on the entropy-to-energy ratio for bounded systems},\ }\href
  {https://doi.org/10.1103/PhysRevD.23.287} {\bibfield  {journal} {\bibinfo
  {journal} {Phys. Rev. D}\ }\textbf {\bibinfo {volume} {23}},\ \bibinfo
  {pages} {287} (\bibinfo {year} {1981}{\natexlab{b}})}\BibitemShut {NoStop}%
\bibitem [{\citenamefont {Bekenstein}\ and\ \citenamefont
  {Schiffer}(1990)}]{Bekenstein1990}%
  \BibitemOpen
  \bibfield  {author} {\bibinfo {author} {\bibfnamefont {J.~D.}\ \bibnamefont
  {Bekenstein}}\ and\ \bibinfo {author} {\bibfnamefont {M.}~\bibnamefont
  {Schiffer}},\ }\bibfield  {title} {\bibinfo {title} {Quantum limitations on
  the storage and transmission of information},\ }\href
  {https://www.worldscientific.com/doi/abs/10.1142/S0129183190000207}
  {\bibfield  {journal} {\bibinfo  {journal} {Int. J. Mod. Phys. C}\ }\textbf
  {\bibinfo {volume} {1}},\ \bibinfo {pages} {355} (\bibinfo {year}
  {1990})}\BibitemShut {NoStop}%
\bibitem [{\citenamefont {Pendry}(1983)}]{Pendry1983}%
  \BibitemOpen
  \bibfield  {author} {\bibinfo {author} {\bibfnamefont {J.~B.}\ \bibnamefont
  {Pendry}},\ }\bibfield  {title} {\bibinfo {title} {Quantum limits to the flow
  of information and entropy},\ }\href
  {https://doi.org/10.1088/0305-4470/16/10/012} {\bibfield  {journal} {\bibinfo
   {journal} {J. Phys. A: Math. Gen.}\ }\textbf {\bibinfo {volume} {16}},\
  \bibinfo {pages} {2161} (\bibinfo {year} {1983})}\BibitemShut {NoStop}%
\bibitem [{\citenamefont {Landauer}(1987)}]{Landauer1987}%
  \BibitemOpen
  \bibfield  {author} {\bibinfo {author} {\bibfnamefont {R.}~\bibnamefont
  {Landauer}},\ }\bibfield  {title} {\bibinfo {title} {Energy requirements in
  communication},\ }\href {https://doi.org/10.1063/1.98291} {\bibfield
  {journal} {\bibinfo  {journal} {Appl. Phys. Lett.}\ }\textbf {\bibinfo
  {volume} {51}},\ \bibinfo {pages} {2056} (\bibinfo {year}
  {1987})}\BibitemShut {NoStop}%
\bibitem [{\citenamefont {Bekenstein}(1988)}]{Bekenstein1988}%
  \BibitemOpen
  \bibfield  {author} {\bibinfo {author} {\bibfnamefont {J.~D.}\ \bibnamefont
  {Bekenstein}},\ }\bibfield  {title} {\bibinfo {title} {Communication and
  energy},\ }\href {https://doi.org/10.1103/PhysRevA.37.3437} {\bibfield
  {journal} {\bibinfo  {journal} {Phys. Rev. A}\ }\textbf {\bibinfo {volume}
  {37}},\ \bibinfo {pages} {3437} (\bibinfo {year} {1988})}\BibitemShut
  {NoStop}%
\bibitem [{\citenamefont {Caves}\ and\ \citenamefont
  {Drummond}(1994)}]{Caves1994}%
  \BibitemOpen
  \bibfield  {author} {\bibinfo {author} {\bibfnamefont {C.~M.}\ \bibnamefont
  {Caves}}\ and\ \bibinfo {author} {\bibfnamefont {P.~D.}\ \bibnamefont
  {Drummond}},\ }\bibfield  {title} {\bibinfo {title} {Quantum limits on
  bosonic communication rates},\ }\href
  {https://doi.org/10.1103/RevModPhys.66.481} {\bibfield  {journal} {\bibinfo
  {journal} {Rev. Mod. Phys.}\ }\textbf {\bibinfo {volume} {66}},\ \bibinfo
  {pages} {481} (\bibinfo {year} {1994})}\BibitemShut {NoStop}%
\bibitem [{\citenamefont {Blencowe}\ and\ \citenamefont
  {Vitelli}(2000)}]{Blencowe2000}%
  \BibitemOpen
  \bibfield  {author} {\bibinfo {author} {\bibfnamefont {M.~P.}\ \bibnamefont
  {Blencowe}}\ and\ \bibinfo {author} {\bibfnamefont {V.}~\bibnamefont
  {Vitelli}},\ }\bibfield  {title} {\bibinfo {title} {Universal quantum limits
  on single-channel information, entropy, and heat flow},\ }\href
  {https://doi.org/10.1103/PhysRevA.62.052104} {\bibfield  {journal} {\bibinfo
  {journal} {Phys. Rev. A}\ }\textbf {\bibinfo {volume} {62}},\ \bibinfo
  {pages} {052104} (\bibinfo {year} {2000})}\BibitemShut {NoStop}%
\bibitem [{\citenamefont {Lloyd}\ \emph {et~al.}(2004)\citenamefont {Lloyd},
  \citenamefont {Giovannetti},\ and\ \citenamefont {Maccone}}]{Lloyd2004}%
  \BibitemOpen
  \bibfield  {author} {\bibinfo {author} {\bibfnamefont {S.}~\bibnamefont
  {Lloyd}}, \bibinfo {author} {\bibfnamefont {V.}~\bibnamefont {Giovannetti}},\
  and\ \bibinfo {author} {\bibfnamefont {L.}~\bibnamefont {Maccone}},\
  }\bibfield  {title} {\bibinfo {title} {Physical limits to communication},\
  }\href {https://doi.org/10.1103/PhysRevLett.93.100501} {\bibfield  {journal}
  {\bibinfo  {journal} {Phys. Rev. Lett.}\ }\textbf {\bibinfo {volume} {93}},\
  \bibinfo {pages} {100501} (\bibinfo {year} {2004})}\BibitemShut {NoStop}%
\bibitem [{\citenamefont {Garbaczewski}(2007)}]{Garbaczewski2007}%
  \BibitemOpen
  \bibfield  {author} {\bibinfo {author} {\bibfnamefont {P.}~\bibnamefont
  {Garbaczewski}},\ }\bibfield  {title} {\bibinfo {title} {Information dynamics
  in quantum theory},\ }\href
  {http://fizyka.uni.opole.pl/pgar/documents/amis07.pdf} {\bibfield  {journal}
  {\bibinfo  {journal} {Appl. Math. \& Information Sciences}\ }\textbf
  {\bibinfo {volume} {1}},\ \bibinfo {pages} {1} (\bibinfo {year}
  {2007})}\BibitemShut {NoStop}%
\bibitem [{\citenamefont {Pei-Rong}\ and\ \citenamefont
  {Di}(2010)}]{Pei_Rong2010}%
  \BibitemOpen
  \bibfield  {author} {\bibinfo {author} {\bibfnamefont {G.}~\bibnamefont
  {Pei-Rong}}\ and\ \bibinfo {author} {\bibfnamefont {L.}~\bibnamefont {Di}},\
  }\bibfield  {title} {\bibinfo {title} {Upper bound for the time derivative of
  entropy for a stochastic dynamical system with double singularities driven by
  non-{Gaussian} noise},\ }\href
  {https://doi.org/10.1088/1674-1056/19/3/030520} {\bibfield  {journal}
  {\bibinfo  {journal} {Chin. Phys. B}\ }\textbf {\bibinfo {volume} {19}},\
  \bibinfo {pages} {030520} (\bibinfo {year} {2010})}\BibitemShut {NoStop}%
\bibitem [{\citenamefont {Deffner}\ and\ \citenamefont
  {Lutz}(2010)}]{Deffner2010PRL}%
  \BibitemOpen
  \bibfield  {author} {\bibinfo {author} {\bibfnamefont {S.}~\bibnamefont
  {Deffner}}\ and\ \bibinfo {author} {\bibfnamefont {E.}~\bibnamefont {Lutz}},\
  }\bibfield  {title} {\bibinfo {title} {Generalized {Clausius} inequality for
  nonequilibrium quantum processes},\ }\href
  {https://doi.org/10.1103/PhysRevLett.105.170402} {\bibfield  {journal}
  {\bibinfo  {journal} {Phys. Rev. Lett.}\ }\textbf {\bibinfo {volume} {105}},\
  \bibinfo {pages} {170402} (\bibinfo {year} {2010})}\BibitemShut {NoStop}%
\bibitem [{\citenamefont {Guo}\ \emph {et~al.}(2011)\citenamefont {Guo},
  \citenamefont {Xu}, \citenamefont {Liu}, \citenamefont {Li},\ and\
  \citenamefont {Wang}}]{Guo2011}%
  \BibitemOpen
  \bibfield  {author} {\bibinfo {author} {\bibfnamefont {Y.}~\bibnamefont
  {Guo}}, \bibinfo {author} {\bibfnamefont {W.}~\bibnamefont {Xu}}, \bibinfo
  {author} {\bibfnamefont {H.}~\bibnamefont {Liu}}, \bibinfo {author}
  {\bibfnamefont {D.}~\bibnamefont {Li}},\ and\ \bibinfo {author}
  {\bibfnamefont {L.}~\bibnamefont {Wang}},\ }\bibfield  {title} {\bibinfo
  {title} {Upper bound of time derivative of entropy for a dynamical system
  driven by quasimonochromatic noise},\ }\href
  {https://doi.org/https://doi.org/10.1016/j.cnsns.2010.04.013} {\bibfield
  {journal} {\bibinfo  {journal} {Comm. Nonlin. Sci. Num. Sim.}\ }\textbf
  {\bibinfo {volume} {16}},\ \bibinfo {pages} {522} (\bibinfo {year}
  {2011})}\BibitemShut {NoStop}%
\bibitem [{\citenamefont {Guo}\ and\ \citenamefont {Tan}(2012)}]{Guo2012}%
  \BibitemOpen
  \bibfield  {author} {\bibinfo {author} {\bibfnamefont {Y.-F.}\ \bibnamefont
  {Guo}}\ and\ \bibinfo {author} {\bibfnamefont {J.-G.}\ \bibnamefont {Tan}},\
  }\bibfield  {title} {\bibinfo {title} {Time evolution of information entropy
  for a stochastic system with double singularities driven by
  quasimonochromatic noise},\ }\href
  {https://doi.org/10.1088/1674-1056/21/12/120501} {\bibfield  {journal}
  {\bibinfo  {journal} {Chin. Phys. B}\ }\textbf {\bibinfo {volume} {21}},\
  \bibinfo {pages} {120501} (\bibinfo {year} {2012})}\BibitemShut {NoStop}%
\bibitem [{\citenamefont {Bousso}(2017)}]{Bousso2017}%
  \BibitemOpen
  \bibfield  {author} {\bibinfo {author} {\bibfnamefont {R.}~\bibnamefont
  {Bousso}},\ }\bibfield  {title} {\bibinfo {title} {Universal limit on
  communication},\ }\href {https://doi.org/10.1103/PhysRevLett.119.140501}
  {\bibfield  {journal} {\bibinfo  {journal} {Phys. Rev. Lett.}\ }\textbf
  {\bibinfo {volume} {119}},\ \bibinfo {pages} {140501} (\bibinfo {year}
  {2017})}\BibitemShut {NoStop}%
\bibitem [{\citenamefont {Lewis-Swan}\ \emph {et~al.}(2019)\citenamefont
  {Lewis-Swan}, \citenamefont {Safavi-Naini}, \citenamefont {Kaufman},\ and\
  \citenamefont {Rey}}]{Lewis2019}%
  \BibitemOpen
  \bibfield  {author} {\bibinfo {author} {\bibfnamefont {R.~J.}\ \bibnamefont
  {Lewis-Swan}}, \bibinfo {author} {\bibfnamefont {A.}~\bibnamefont
  {Safavi-Naini}}, \bibinfo {author} {\bibfnamefont {A.~M.}\ \bibnamefont
  {Kaufman}},\ and\ \bibinfo {author} {\bibfnamefont {A.~M.}\ \bibnamefont
  {Rey}},\ }\bibfield  {title} {\bibinfo {title} {Dynamics of quantum
  information},\ }\href {https://doi.org/10.1038/s42254-019-0090-y} {\bibfield
  {journal} {\bibinfo  {journal} {Nat. Rev. Phys.}\ }\textbf {\bibinfo {volume}
  {1}},\ \bibinfo {pages} {627} (\bibinfo {year} {2019})}\BibitemShut {NoStop}%
\bibitem [{\citenamefont {Deffner}(2020)}]{Deffner2020PRR}%
  \BibitemOpen
  \bibfield  {author} {\bibinfo {author} {\bibfnamefont {S.}~\bibnamefont
  {Deffner}},\ }\bibfield  {title} {\bibinfo {title} {Quantum speed limits and
  the maximal rate of information production},\ }\href
  {https://doi.org/10.1103/PhysRevResearch.2.013161} {\bibfield  {journal}
  {\bibinfo  {journal} {Phys. Rev. Research}\ }\textbf {\bibinfo {volume}
  {2}},\ \bibinfo {pages} {013161} (\bibinfo {year} {2020})}\BibitemShut
  {NoStop}%
\bibitem [{\citenamefont {Gisin}\ and\ \citenamefont
  {Thew}(2007)}]{Gisin2007NP}%
  \BibitemOpen
  \bibfield  {author} {\bibinfo {author} {\bibfnamefont {N.}~\bibnamefont
  {Gisin}}\ and\ \bibinfo {author} {\bibfnamefont {R.}~\bibnamefont {Thew}},\
  }\bibfield  {title} {\bibinfo {title} {Quantum communication},\ }\href
  {https://doi.org/10.1038/nphoton.2007.22} {\bibfield  {journal} {\bibinfo
  {journal} {Nat. Phot.}\ }\textbf {\bibinfo {volume} {1}},\ \bibinfo {pages}
  {165} (\bibinfo {year} {2007})}\BibitemShut {NoStop}%
\bibitem [{\citenamefont {Chen}(2021)}]{Chen2021JP}%
  \BibitemOpen
  \bibfield  {author} {\bibinfo {author} {\bibfnamefont {J.}~\bibnamefont
  {Chen}},\ }\bibfield  {title} {\bibinfo {title} {Review on quantum
  communication and quantum computation},\ }\href
  {https://doi.org/10.1088/1742-6596/1865/2/022008} {\bibfield  {journal}
  {\bibinfo  {journal} {J. Physics: Conf. Ser.}\ }\textbf {\bibinfo {volume}
  {1865}},\ \bibinfo {pages} {022008} (\bibinfo {year} {2021})}\BibitemShut
  {NoStop}%
\bibitem [{\citenamefont {Wehner}\ \emph {et~al.}(2018)\citenamefont {Wehner},
  \citenamefont {Elkouss},\ and\ \citenamefont {Hanson}}]{wehner_quantum_2018}%
  \BibitemOpen
  \bibfield  {author} {\bibinfo {author} {\bibfnamefont {S.}~\bibnamefont
  {Wehner}}, \bibinfo {author} {\bibfnamefont {D.}~\bibnamefont {Elkouss}},\
  and\ \bibinfo {author} {\bibfnamefont {R.}~\bibnamefont {Hanson}},\
  }\bibfield  {title} {\bibinfo {title} {Quantum internet: {A} vision for the
  road ahead},\ }\href {https://doi.org/10.1126/science.aam9288} {\bibfield
  {journal} {\bibinfo  {journal} {Science}\ }\textbf {\bibinfo {volume}
  {362}},\ \bibinfo {pages} {eaam9288} (\bibinfo {year} {2018})}\BibitemShut
  {NoStop}%
\bibitem [{\citenamefont {Gisin}\ \emph {et~al.}(2002)\citenamefont {Gisin},
  \citenamefont {Ribordy}, \citenamefont {Tittel},\ and\ \citenamefont
  {Zbinden}}]{gisin_quantum_2002}%
  \BibitemOpen
  \bibfield  {author} {\bibinfo {author} {\bibfnamefont {N.}~\bibnamefont
  {Gisin}}, \bibinfo {author} {\bibfnamefont {G.}~\bibnamefont {Ribordy}},
  \bibinfo {author} {\bibfnamefont {W.}~\bibnamefont {Tittel}},\ and\ \bibinfo
  {author} {\bibfnamefont {H.}~\bibnamefont {Zbinden}},\ }\bibfield  {title}
  {\bibinfo {title} {Quantum cryptography},\ }\href
  {https://doi.org/10.1103/RevModPhys.74.145} {\bibfield  {journal} {\bibinfo
  {journal} {Rev. Mod. Phys.}\ }\textbf {\bibinfo {volume} {74}},\ \bibinfo
  {pages} {145} (\bibinfo {year} {2002})}\BibitemShut {NoStop}%
\bibitem [{\citenamefont {Lu}\ \emph {et~al.}(2022)\citenamefont {Lu},
  \citenamefont {Cao}, \citenamefont {Peng},\ and\ \citenamefont
  {Pan}}]{Lu2022RMP}%
  \BibitemOpen
  \bibfield  {author} {\bibinfo {author} {\bibfnamefont {C.-Y.}\ \bibnamefont
  {Lu}}, \bibinfo {author} {\bibfnamefont {Y.}~\bibnamefont {Cao}}, \bibinfo
  {author} {\bibfnamefont {C.-Z.}\ \bibnamefont {Peng}},\ and\ \bibinfo
  {author} {\bibfnamefont {J.-W.}\ \bibnamefont {Pan}},\ }\bibfield  {title}
  {\bibinfo {title} {Micius quantum experiments in space},\ }\href
  {https://doi.org/10.1103/RevModPhys.94.035001} {\bibfield  {journal}
  {\bibinfo  {journal} {Rev. Mod. Phys.}\ }\textbf {\bibinfo {volume} {94}},\
  \bibinfo {pages} {035001} (\bibinfo {year} {2022})}\BibitemShut {NoStop}%
\bibitem [{\citenamefont {Briegel}\ \emph {et~al.}(1998)\citenamefont
  {Briegel}, \citenamefont {Dür}, \citenamefont {Cirac},\ and\ \citenamefont
  {Zoller}}]{briegel_quantum_1998}%
  \BibitemOpen
  \bibfield  {author} {\bibinfo {author} {\bibfnamefont {H.-J.}\ \bibnamefont
  {Briegel}}, \bibinfo {author} {\bibfnamefont {W.}~\bibnamefont {Dür}},
  \bibinfo {author} {\bibfnamefont {J.~I.}\ \bibnamefont {Cirac}},\ and\
  \bibinfo {author} {\bibfnamefont {P.}~\bibnamefont {Zoller}},\ }\bibfield
  {title} {\bibinfo {title} {Quantum {Repeaters}: {The} {Role} of {Imperfect}
  {Local} {Operations} in {Quantum} {Communication}},\ }\href
  {https://doi.org/10.1103/PhysRevLett.81.5932} {\bibfield  {journal} {\bibinfo
   {journal} {Phys. Rev. Lett.}\ }\textbf {\bibinfo {volume} {81}},\ \bibinfo
  {pages} {5932} (\bibinfo {year} {1998})}\BibitemShut {NoStop}%
\bibitem [{\citenamefont {Munro}\ \emph {et~al.}(2015)\citenamefont {Munro},
  \citenamefont {Azuma}, \citenamefont {Tamaki},\ and\ \citenamefont
  {Nemoto}}]{munro_inside_2015}%
  \BibitemOpen
  \bibfield  {author} {\bibinfo {author} {\bibfnamefont {W.~J.}\ \bibnamefont
  {Munro}}, \bibinfo {author} {\bibfnamefont {K.}~\bibnamefont {Azuma}},
  \bibinfo {author} {\bibfnamefont {K.}~\bibnamefont {Tamaki}},\ and\ \bibinfo
  {author} {\bibfnamefont {K.}~\bibnamefont {Nemoto}},\ }\bibfield  {title}
  {\bibinfo {title} {Inside {Quantum} {Repeaters}},\ }\href
  {https://doi.org/10.1109/JSTQE.2015.2392076} {\bibfield  {journal} {\bibinfo
  {journal} {IEEE J. Select. Topics Quantum Electron.}\ }\textbf {\bibinfo
  {volume} {21}},\ \bibinfo {pages} {78} (\bibinfo {year} {2015})}\BibitemShut
  {NoStop}%
\bibitem [{\citenamefont {Jiang}\ \emph {et~al.}(2007)\citenamefont {Jiang},
  \citenamefont {Taylor}, \citenamefont {Khaneja},\ and\ \citenamefont
  {Lukin}}]{jiang_optimal_2007}%
  \BibitemOpen
  \bibfield  {author} {\bibinfo {author} {\bibfnamefont {L.}~\bibnamefont
  {Jiang}}, \bibinfo {author} {\bibfnamefont {J.~M.}\ \bibnamefont {Taylor}},
  \bibinfo {author} {\bibfnamefont {N.}~\bibnamefont {Khaneja}},\ and\ \bibinfo
  {author} {\bibfnamefont {M.~D.}\ \bibnamefont {Lukin}},\ }\bibfield  {title}
  {\bibinfo {title} {Optimal approach to quantum communication using dynamic
  programming},\ }\href {https://doi.org/10.1073/pnas.0703284104} {\bibfield
  {journal} {\bibinfo  {journal} {Proc. Natl. Acad. Sci. U.S.A.}\ }\textbf
  {\bibinfo {volume} {104}},\ \bibinfo {pages} {17291} (\bibinfo {year}
  {2007})}\BibitemShut {NoStop}%
\bibitem [{\citenamefont {Eldada}(2004)}]{eldada_optical_2004}%
  \BibitemOpen
  \bibfield  {author} {\bibinfo {author} {\bibfnamefont {L.}~\bibnamefont
  {Eldada}},\ }\bibfield  {title} {\bibinfo {title} {Optical communication
  components},\ }\href {https://doi.org/10.1063/1.1647701} {\bibfield
  {journal} {\bibinfo  {journal} {Rev. Sci. Instrum.}\ }\textbf {\bibinfo
  {volume} {75}},\ \bibinfo {pages} {575} (\bibinfo {year} {2004})}\BibitemShut
  {NoStop}%
\bibitem [{\citenamefont {Jones}(1962)}]{jones_ultimate_1962}%
  \BibitemOpen
  \bibfield  {author} {\bibinfo {author} {\bibfnamefont {R.~C.}\ \bibnamefont
  {Jones}},\ }\bibfield  {title} {\bibinfo {title} {Ultimate {Performance} of
  {Polarizers} for {Visible} {Light}},\ }\href
  {https://doi.org/10.1364/JOSA.52.000747} {\bibfield  {journal} {\bibinfo
  {journal} {J. Opt. Soc. Am.}\ }\textbf {\bibinfo {volume} {52}},\ \bibinfo
  {pages} {747} (\bibinfo {year} {1962})}\BibitemShut {NoStop}%
\bibitem [{\citenamefont {Im}\ \emph {et~al.}(2018)\citenamefont {Im},
  \citenamefont {Sim},\ and\ \citenamefont {Kim}}]{im_microscale_2018}%
  \BibitemOpen
  \bibfield  {author} {\bibinfo {author} {\bibfnamefont {S.}~\bibnamefont
  {Im}}, \bibinfo {author} {\bibfnamefont {E.}~\bibnamefont {Sim}},\ and\
  \bibinfo {author} {\bibfnamefont {D.}~\bibnamefont {Kim}},\ }\bibfield
  {title} {\bibinfo {title} {Microscale heat transfer and thermal extinction of
  a wire-grid polarizer},\ }\href {https://doi.org/10.1038/s41598-018-33347-5}
  {\bibfield  {journal} {\bibinfo  {journal} {Sci. Rep.}\ }\textbf {\bibinfo
  {volume} {8}},\ \bibinfo {pages} {14973} (\bibinfo {year}
  {2018})}\BibitemShut {NoStop}%
\bibitem [{\citenamefont {Lounis}\ and\ \citenamefont
  {Orrit}(2005)}]{lounis_single-photon_2005}%
  \BibitemOpen
  \bibfield  {author} {\bibinfo {author} {\bibfnamefont {B.}~\bibnamefont
  {Lounis}}\ and\ \bibinfo {author} {\bibfnamefont {M.}~\bibnamefont {Orrit}},\
  }\bibfield  {title} {\bibinfo {title} {Single-photon sources},\ }\href
  {https://doi.org/10.1088/0034-4885/68/5/R04} {\bibfield  {journal} {\bibinfo
  {journal} {Rep. Prog. Phys.}\ }\textbf {\bibinfo {volume} {68}},\ \bibinfo
  {pages} {1129} (\bibinfo {year} {2005})}\BibitemShut {NoStop}%
\bibitem [{\citenamefont {Edamatsu}(2007)}]{edamatsu_entangled_2007}%
  \BibitemOpen
  \bibfield  {author} {\bibinfo {author} {\bibfnamefont {K.}~\bibnamefont
  {Edamatsu}},\ }\bibfield  {title} {\bibinfo {title} {Entangled {Photons}:
  {Generation}, {Observation}, and {Characterization}},\ }\href
  {https://doi.org/10.1143/JJAP.46.7175} {\bibfield  {journal} {\bibinfo
  {journal} {Jpn. J. Appl. Phys.}\ }\textbf {\bibinfo {volume} {46}},\ \bibinfo
  {pages} {7175} (\bibinfo {year} {2007})}\BibitemShut {NoStop}%
\bibitem [{\citenamefont {Takeuchi}(2014)}]{takeuchi_recent_2014}%
  \BibitemOpen
  \bibfield  {author} {\bibinfo {author} {\bibfnamefont {S.}~\bibnamefont
  {Takeuchi}},\ }\bibfield  {title} {\bibinfo {title} {Recent progress in
  single-photon and entangled-photon generation and applications},\ }\href
  {https://doi.org/10.7567/JJAP.53.030101} {\bibfield  {journal} {\bibinfo
  {journal} {Jpn. J. Appl. Phys.}\ }\textbf {\bibinfo {volume} {53}},\ \bibinfo
  {pages} {030101} (\bibinfo {year} {2014})}\BibitemShut {NoStop}%
\bibitem [{\citenamefont {Konrad}\ and\ \citenamefont
  {Forbes}(2019)}]{konrad_quantum_2019}%
  \BibitemOpen
  \bibfield  {author} {\bibinfo {author} {\bibfnamefont {T.}~\bibnamefont
  {Konrad}}\ and\ \bibinfo {author} {\bibfnamefont {A.}~\bibnamefont
  {Forbes}},\ }\bibfield  {title} {\bibinfo {title} {Quantum mechanics and
  classical light},\ }\href {https://doi.org/10.1080/00107514.2019.1580433}
  {\bibfield  {journal} {\bibinfo  {journal} {Contemp. Phys.}\ }\textbf
  {\bibinfo {volume} {60}},\ \bibinfo {pages} {1} (\bibinfo {year}
  {2019})}\BibitemShut {NoStop}%
\bibitem [{\citenamefont {Kenfack}\ and\ \citenamefont
  {\.{Z}yczkowski}(2004)}]{kenfack_negativity_2004}%
  \BibitemOpen
  \bibfield  {author} {\bibinfo {author} {\bibfnamefont {A.}~\bibnamefont
  {Kenfack}}\ and\ \bibinfo {author} {\bibfnamefont {K.}~\bibnamefont
  {\.{Z}yczkowski}},\ }\bibfield  {title} {\bibinfo {title} {Negativity of the
  {Wigner} function as an indicator of non-classicality},\ }\href
  {https://doi.org/10.1088/1464-4266/6/10/003} {\bibfield  {journal} {\bibinfo
  {journal} {J. Opt. B: Quantum Semiclass. Opt.}\ }\textbf {\bibinfo {volume}
  {6}},\ \bibinfo {pages} {396} (\bibinfo {year} {2004})}\BibitemShut {NoStop}%
\bibitem [{\citenamefont {Campbell}\ and\ \citenamefont
  {Vacchini}(2021)}]{campbell_collision_2021}%
  \BibitemOpen
  \bibfield  {author} {\bibinfo {author} {\bibfnamefont {S.}~\bibnamefont
  {Campbell}}\ and\ \bibinfo {author} {\bibfnamefont {B.}~\bibnamefont
  {Vacchini}},\ }\bibfield  {title} {\bibinfo {title} {Collision models in open
  system dynamics: {A} versatile tool for deeper insights?},\ }\href
  {https://doi.org/10.1209/0295-5075/133/60001} {\bibfield  {journal} {\bibinfo
   {journal} {Europhys. Lett.}\ }\textbf {\bibinfo {volume} {133}},\ \bibinfo
  {pages} {60001} (\bibinfo {year} {2021})}\BibitemShut {NoStop}%
\bibitem [{\citenamefont {Lorenzo}\ \emph
  {et~al.}(2015{\natexlab{b}})\citenamefont {Lorenzo}, \citenamefont
  {McCloskey}, \citenamefont {Ciccarello}, \citenamefont {Paternostro},\ and\
  \citenamefont {Palma}}]{Lorenzo2015}%
  \BibitemOpen
  \bibfield  {author} {\bibinfo {author} {\bibfnamefont {S.}~\bibnamefont
  {Lorenzo}}, \bibinfo {author} {\bibfnamefont {R.}~\bibnamefont {McCloskey}},
  \bibinfo {author} {\bibfnamefont {F.}~\bibnamefont {Ciccarello}}, \bibinfo
  {author} {\bibfnamefont {M.}~\bibnamefont {Paternostro}},\ and\ \bibinfo
  {author} {\bibfnamefont {G.~M.}\ \bibnamefont {Palma}},\ }\bibfield  {title}
  {\bibinfo {title} {{Landauer's} principle in multipartite open quantum system
  dynamics},\ }\href {https://doi.org/10.1103/PhysRevLett.115.120403}
  {\bibfield  {journal} {\bibinfo  {journal} {Phys. Rev. Lett.}\ }\textbf
  {\bibinfo {volume} {115}},\ \bibinfo {pages} {120403} (\bibinfo {year}
  {2015}{\natexlab{b}})}\BibitemShut {NoStop}%
\bibitem [{\citenamefont {Aiello}\ and\ \citenamefont
  {Woerdman}(2005)}]{aiello_physical_2005}%
  \BibitemOpen
  \bibfield  {author} {\bibinfo {author} {\bibfnamefont {A.}~\bibnamefont
  {Aiello}}\ and\ \bibinfo {author} {\bibfnamefont {J.~P.}\ \bibnamefont
  {Woerdman}},\ }\bibfield  {title} {\bibinfo {title} {Physical {Bounds} to the
  {Entropy}-{Depolarization} {Relation} in {Random} {Light} {Scattering}},\
  }\href {https://doi.org/10.1103/PhysRevLett.94.090406} {\bibfield  {journal}
  {\bibinfo  {journal} {Phys. Rev. Lett.}\ }\textbf {\bibinfo {volume} {94}},\
  \bibinfo {pages} {090406} (\bibinfo {year} {2005})}\BibitemShut {NoStop}%
\bibitem [{\citenamefont {Scully}\ and\ \citenamefont
  {Drühl}(1982)}]{scully_quantum_1982}%
  \BibitemOpen
  \bibfield  {author} {\bibinfo {author} {\bibfnamefont {M.~O.}\ \bibnamefont
  {Scully}}\ and\ \bibinfo {author} {\bibfnamefont {K.}~\bibnamefont
  {Drühl}},\ }\bibfield  {title} {\bibinfo {title} {Quantum eraser: {A}
  proposed photon correlation experiment concerning observation and ``delayed
  choice" in quantum mechanics},\ }\href
  {https://doi.org/10.1103/PhysRevA.25.2208} {\bibfield  {journal} {\bibinfo
  {journal} {Phys. Rev. A}\ }\textbf {\bibinfo {volume} {25}},\ \bibinfo
  {pages} {2208} (\bibinfo {year} {1982})}\BibitemShut {NoStop}%
\bibitem [{\citenamefont {Kim}\ \emph {et~al.}(2000)\citenamefont {Kim},
  \citenamefont {Yu}, \citenamefont {Kulik}, \citenamefont {Shih},\ and\
  \citenamefont {Scully}}]{kim_delayed_2000}%
  \BibitemOpen
  \bibfield  {author} {\bibinfo {author} {\bibfnamefont {Y.-H.}\ \bibnamefont
  {Kim}}, \bibinfo {author} {\bibfnamefont {R.}~\bibnamefont {Yu}}, \bibinfo
  {author} {\bibfnamefont {S.~P.}\ \bibnamefont {Kulik}}, \bibinfo {author}
  {\bibfnamefont {Y.}~\bibnamefont {Shih}},\ and\ \bibinfo {author}
  {\bibfnamefont {M.~O.}\ \bibnamefont {Scully}},\ }\bibfield  {title}
  {\bibinfo {title} {Delayed “{Choice}” {Quantum} {Eraser}},\ }\href
  {https://doi.org/10.1103/PhysRevLett.84.1} {\bibfield  {journal} {\bibinfo
  {journal} {Phys. Rev. Lett.}\ }\textbf {\bibinfo {volume} {84}},\ \bibinfo
  {pages} {1} (\bibinfo {year} {2000})}\BibitemShut {NoStop}%
\bibitem [{\citenamefont {Walborn}\ \emph {et~al.}(2002)\citenamefont
  {Walborn}, \citenamefont {Terra~Cunha}, \citenamefont {Pádua},\ and\
  \citenamefont {Monken}}]{walborn_double-slit_2002}%
  \BibitemOpen
  \bibfield  {author} {\bibinfo {author} {\bibfnamefont {S.~P.}\ \bibnamefont
  {Walborn}}, \bibinfo {author} {\bibfnamefont {M.~O.}\ \bibnamefont
  {Terra~Cunha}}, \bibinfo {author} {\bibfnamefont {S.}~\bibnamefont
  {Pádua}},\ and\ \bibinfo {author} {\bibfnamefont {C.~H.}\ \bibnamefont
  {Monken}},\ }\bibfield  {title} {\bibinfo {title} {Double-slit quantum
  eraser},\ }\href {https://doi.org/10.1103/PhysRevA.65.033818} {\bibfield
  {journal} {\bibinfo  {journal} {Phys. Rev. A}\ }\textbf {\bibinfo {volume}
  {65}},\ \bibinfo {pages} {033818} (\bibinfo {year} {2002})}\BibitemShut
  {NoStop}%
\bibitem [{\citenamefont {Kang}(2007)}]{kang_electronic_2007}%
  \BibitemOpen
  \bibfield  {author} {\bibinfo {author} {\bibfnamefont {K.}~\bibnamefont
  {Kang}},\ }\bibfield  {title} {\bibinfo {title} {Electronic {Mach}-{Zehnder}
  quantum eraser},\ }\href {https://doi.org/10.1103/PhysRevB.75.125326}
  {\bibfield  {journal} {\bibinfo  {journal} {Phys. Rev. B}\ }\textbf {\bibinfo
  {volume} {75}},\ \bibinfo {pages} {125326} (\bibinfo {year}
  {2007})}\BibitemShut {NoStop}%
\bibitem [{\citenamefont {Bennett}(2003)}]{bennett_notes_2003}%
  \BibitemOpen
  \bibfield  {author} {\bibinfo {author} {\bibfnamefont {C.~H.}\ \bibnamefont
  {Bennett}},\ }\bibfield  {title} {\bibinfo {title} {Notes on {Landauer}'s
  principle, reversible computation, and {Maxwell}'s {Demon}},\ }\href
  {https://doi.org/10.1016/S1355-2198(03)00039-X} {\bibfield  {journal}
  {\bibinfo  {journal} {Stud. Hist. Philos. Sci. Part B: Stud. Hist. Philos.
  Mod. Phys.}\ }\textbf {\bibinfo {volume} {34}},\ \bibinfo {pages} {501}
  (\bibinfo {year} {2003})}\BibitemShut {NoStop}%
\bibitem [{\citenamefont {Bucholtz}\ \emph {et~al.}(2014)\citenamefont
  {Bucholtz}, \citenamefont {Michalowicz}, \citenamefont {Nichols},\ and\
  \citenamefont {Bucholtz}}]{bucholtz_handbook_2014}%
  \BibitemOpen
  \bibfield  {author} {\bibinfo {author} {\bibfnamefont {F.}~\bibnamefont
  {Bucholtz}}, \bibinfo {author} {\bibfnamefont {J.~V.}\ \bibnamefont
  {Michalowicz}}, \bibinfo {author} {\bibfnamefont {J.~M.}\ \bibnamefont
  {Nichols}},\ and\ \bibinfo {author} {\bibfnamefont {F.}~\bibnamefont
  {Bucholtz}},\ }\href@noop {} {\emph {\bibinfo {title} {Handbook of
  differential entropy}}}\ (\bibinfo  {publisher} {CRC Press, Taylor \& Francis
  Group},\ \bibinfo {address} {Boca Raton},\ \bibinfo {year}
  {2014})\BibitemShut {NoStop}%
\bibitem [{\citenamefont {Grynberg}\ \emph {et~al.}(2010)\citenamefont
  {Grynberg}, \citenamefont {Aspect}, \citenamefont {Fabre},\ and\
  \citenamefont {Cohen-Tannoudji}}]{grynberg_introduction_2010}%
  \BibitemOpen
  \bibfield  {author} {\bibinfo {author} {\bibfnamefont {G.}~\bibnamefont
  {Grynberg}}, \bibinfo {author} {\bibfnamefont {A.}~\bibnamefont {Aspect}},
  \bibinfo {author} {\bibfnamefont {C.}~\bibnamefont {Fabre}},\ and\ \bibinfo
  {author} {\bibfnamefont {C.}~\bibnamefont {Cohen-Tannoudji}},\ }\href
  {https://doi.org/10.1017/CBO9780511778261} {\emph {\bibinfo {title}
  {Introduction to {Quantum} {Optics}: {From} the {Semi}-classical {Approach}
  to {Quantized} {Light}}}}\ (\bibinfo  {publisher} {Cambridge University
  Press},\ \bibinfo {address} {Cambridge},\ \bibinfo {year} {2010})\BibitemShut
  {NoStop}%
\bibitem [{\citenamefont {Ehrenfest}(1911)}]{ehrenfest_welche_1911}%
  \BibitemOpen
  \bibfield  {author} {\bibinfo {author} {\bibfnamefont {P.}~\bibnamefont
  {Ehrenfest}},\ }\bibfield  {title} {\bibinfo {title} {Welche {Z\"uge} der
  {Lichtquantenhypothese} spielen in der {Theorie} der {W\"armestrahlung} eine
  wesentliche {Rolle}?},\ }\href {https://doi.org/10.1002/andp.19113411106}
  {\bibfield  {journal} {\bibinfo  {journal} {Ann. Phys.}\ }\textbf {\bibinfo
  {volume} {341}},\ \bibinfo {pages} {91} (\bibinfo {year} {1911})}\BibitemShut
  {NoStop}%
\bibitem [{\citenamefont {Schleich}(2011)}]{Schleich2011quantum}%
  \BibitemOpen
  \bibfield  {author} {\bibinfo {author} {\bibfnamefont {W.~P.}\ \bibnamefont
  {Schleich}},\ }\href@noop {} {\emph {\bibinfo {title} {Quantum optics in
  phase space}}}\ (\bibinfo  {publisher} {John Wiley \& Sons},\ \bibinfo {year}
  {2011})\BibitemShut {NoStop}%
\bibitem [{\citenamefont {Van~Herstraeten}\ and\ \citenamefont
  {Cerf}(2021)}]{van_herstraeten_quantum_2021}%
  \BibitemOpen
  \bibfield  {author} {\bibinfo {author} {\bibfnamefont {Z.}~\bibnamefont
  {Van~Herstraeten}}\ and\ \bibinfo {author} {\bibfnamefont {N.~J.}\
  \bibnamefont {Cerf}},\ }\bibfield  {title} {\bibinfo {title} {Quantum
  {Wigner} entropy},\ }\href {https://doi.org/10.1103/PhysRevA.104.042211}
  {\bibfield  {journal} {\bibinfo  {journal} {Phys. Rev. A}\ }\textbf {\bibinfo
  {volume} {104}},\ \bibinfo {pages} {042211} (\bibinfo {year}
  {2021})}\BibitemShut {NoStop}%
\bibitem [{\citenamefont {Santos}\ \emph {et~al.}(2017)\citenamefont {Santos},
  \citenamefont {Landi},\ and\ \citenamefont {Paternostro}}]{Santos2017PRL}%
  \BibitemOpen
  \bibfield  {author} {\bibinfo {author} {\bibfnamefont {J.~P.}\ \bibnamefont
  {Santos}}, \bibinfo {author} {\bibfnamefont {G.~T.}\ \bibnamefont {Landi}},\
  and\ \bibinfo {author} {\bibfnamefont {M.}~\bibnamefont {Paternostro}},\
  }\bibfield  {title} {\bibinfo {title} {Wigner entropy production rate},\
  }\href {https://doi.org/10.1103/PhysRevLett.118.220601} {\bibfield  {journal}
  {\bibinfo  {journal} {Phys. Rev. Lett.}\ }\textbf {\bibinfo {volume} {118}},\
  \bibinfo {pages} {220601} (\bibinfo {year} {2017})}\BibitemShut {NoStop}%
\bibitem [{\citenamefont {Witten}(2020)}]{witten_mini-introduction_2020}%
  \BibitemOpen
  \bibfield  {author} {\bibinfo {author} {\bibfnamefont {E.}~\bibnamefont
  {Witten}},\ }\bibfield  {title} {\bibinfo {title} {A mini-introduction to
  information theory},\ }\href {https://doi.org/10.1007/s40766-020-00004-5}
  {\bibfield  {journal} {\bibinfo  {journal} {Riv. Nuovo Cim.}\ }\textbf
  {\bibinfo {volume} {43}},\ \bibinfo {pages} {187} (\bibinfo {year}
  {2020})}\BibitemShut {NoStop}%
\bibitem [{\citenamefont {Fearn}\ and\ \citenamefont
  {Loudon}(1987)}]{Fearn1987OC}%
  \BibitemOpen
  \bibfield  {author} {\bibinfo {author} {\bibfnamefont {H.}~\bibnamefont
  {Fearn}}\ and\ \bibinfo {author} {\bibfnamefont {R.}~\bibnamefont {Loudon}},\
  }\bibfield  {title} {\bibinfo {title} {Quantum theory of the lossless beam
  splitter},\ }\href
  {https://doi.org/https://doi.org/10.1016/0030-4018(87)90275-6} {\bibfield
  {journal} {\bibinfo  {journal} {Opt. Commun.}\ }\textbf {\bibinfo {volume}
  {64}},\ \bibinfo {pages} {485} (\bibinfo {year} {1987})}\BibitemShut
  {NoStop}%
\bibitem [{\citenamefont {Aspect}(2017)}]{aspect_huygens_2017}%
  \BibitemOpen
  \bibfield  {author} {\bibinfo {author} {\bibfnamefont {A.}~\bibnamefont
  {Aspect}},\ }\bibfield  {title} {\bibinfo {title} {From {Huygens}' waves to
  {Einstein}'s photons: {Weird} light},\ }\href
  {https://doi.org/10.1016/j.crhy.2017.11.005} {\bibfield  {journal} {\bibinfo
  {journal} {C.R. Phys.}\ }\textbf {\bibinfo {volume} {18}},\ \bibinfo {pages}
  {498} (\bibinfo {year} {2017})}\BibitemShut {NoStop}%
\bibitem [{\citenamefont {Boyd}\ \emph
  {et~al.}(2018{\natexlab{b}})\citenamefont {Boyd}, \citenamefont {Mandal},\
  and\ \citenamefont {Crutchfield}}]{boyd_thermodynamics_2018}%
  \BibitemOpen
  \bibfield  {author} {\bibinfo {author} {\bibfnamefont {A.~B.}\ \bibnamefont
  {Boyd}}, \bibinfo {author} {\bibfnamefont {D.}~\bibnamefont {Mandal}},\ and\
  \bibinfo {author} {\bibfnamefont {J.~P.}\ \bibnamefont {Crutchfield}},\
  }\bibfield  {title} {\bibinfo {title} {Thermodynamics of {Modularity}:
  {Structural} {Costs} {Beyond} the {Landauer} {Bound}},\ }\href
  {https://doi.org/10.1103/PhysRevX.8.031036} {\bibfield  {journal} {\bibinfo
  {journal} {Phys. Rev. X}\ }\textbf {\bibinfo {volume} {8}},\ \bibinfo {pages}
  {031036} (\bibinfo {year} {2018}{\natexlab{b}})}\BibitemShut {NoStop}%
\bibitem [{\citenamefont {Okamoto}\ \emph {et~al.}(2005)\citenamefont
  {Okamoto}, \citenamefont {Hofmann}, \citenamefont {Takeuchi},\ and\
  \citenamefont {Sasaki}}]{okamoto_demonstration_2005}%
  \BibitemOpen
  \bibfield  {author} {\bibinfo {author} {\bibfnamefont {R.}~\bibnamefont
  {Okamoto}}, \bibinfo {author} {\bibfnamefont {H.~F.}\ \bibnamefont
  {Hofmann}}, \bibinfo {author} {\bibfnamefont {S.}~\bibnamefont {Takeuchi}},\
  and\ \bibinfo {author} {\bibfnamefont {K.}~\bibnamefont {Sasaki}},\
  }\bibfield  {title} {\bibinfo {title} {Demonstration of an {Optical}
  {Quantum} {Controlled}-{NOT} {Gate} without {Path} {Interference}},\ }\href
  {https://doi.org/10.1103/PhysRevLett.95.210506} {\bibfield  {journal}
  {\bibinfo  {journal} {Phys. Rev. Lett.}\ }\textbf {\bibinfo {volume} {95}},\
  \bibinfo {pages} {210506} (\bibinfo {year} {2005})}\BibitemShut {NoStop}%
\bibitem [{\citenamefont {Karasawa}\ \emph {et~al.}(2009)\citenamefont
  {Karasawa}, \citenamefont {Gea-Banacloche},\ and\ \citenamefont
  {Ozawa}}]{karasawa_gate_2009}%
  \BibitemOpen
  \bibfield  {author} {\bibinfo {author} {\bibfnamefont {T.}~\bibnamefont
  {Karasawa}}, \bibinfo {author} {\bibfnamefont {J.}~\bibnamefont
  {Gea-Banacloche}},\ and\ \bibinfo {author} {\bibfnamefont {M.}~\bibnamefont
  {Ozawa}},\ }\bibfield  {title} {\bibinfo {title} {Gate fidelity of arbitrary
  single-qubit gates constrained by conservation laws},\ }\href
  {https://doi.org/10.1088/1751-8113/42/22/225303} {\bibfield  {journal}
  {\bibinfo  {journal} {J. Phys. A: Math. Theor.}\ }\textbf {\bibinfo {volume}
  {42}},\ \bibinfo {pages} {225303} (\bibinfo {year} {2009})}\BibitemShut
  {NoStop}%
\bibitem [{\citenamefont {Ozawa}(2002)}]{ozawa_conservative_2002}%
  \BibitemOpen
  \bibfield  {author} {\bibinfo {author} {\bibfnamefont {M.}~\bibnamefont
  {Ozawa}},\ }\bibfield  {title} {\bibinfo {title} {Conservative {Quantum}
  {Computing}},\ }\href {https://doi.org/10.1103/PhysRevLett.89.057902}
  {\bibfield  {journal} {\bibinfo  {journal} {Phys. Rev. Lett.}\ }\textbf
  {\bibinfo {volume} {89}},\ \bibinfo {pages} {057902} (\bibinfo {year}
  {2002})}\BibitemShut {NoStop}%
\bibitem [{\citenamefont {Touil}\ \emph
  {et~al.}(2022{\natexlab{a}})\citenamefont {Touil}, \citenamefont {Yan},
  \citenamefont {Girolami}, \citenamefont {Deffner},\ and\ \citenamefont
  {Zurek}}]{Touil2022PRL}%
  \BibitemOpen
  \bibfield  {author} {\bibinfo {author} {\bibfnamefont {A.}~\bibnamefont
  {Touil}}, \bibinfo {author} {\bibfnamefont {B.}~\bibnamefont {Yan}}, \bibinfo
  {author} {\bibfnamefont {D.}~\bibnamefont {Girolami}}, \bibinfo {author}
  {\bibfnamefont {S.}~\bibnamefont {Deffner}},\ and\ \bibinfo {author}
  {\bibfnamefont {W.~H.}\ \bibnamefont {Zurek}},\ }\bibfield  {title} {\bibinfo
  {title} {Eavesdropping on the decohering environment: Quantum darwinism,
  amplification, and the origin of objective classical reality},\ }\href
  {https://doi.org/10.1103/PhysRevLett.128.010401} {\bibfield  {journal}
  {\bibinfo  {journal} {Phys. Rev. Lett.}\ }\textbf {\bibinfo {volume} {128}},\
  \bibinfo {pages} {010401} (\bibinfo {year} {2022}{\natexlab{a}})}\BibitemShut
  {NoStop}%
\bibitem [{\citenamefont {Girolami}\ \emph {et~al.}(2022)\citenamefont
  {Girolami}, \citenamefont {Touil}, \citenamefont {Yan}, \citenamefont
  {Deffner},\ and\ \citenamefont {Zurek}}]{Girolami2022PRL}%
  \BibitemOpen
  \bibfield  {author} {\bibinfo {author} {\bibfnamefont {D.}~\bibnamefont
  {Girolami}}, \bibinfo {author} {\bibfnamefont {A.}~\bibnamefont {Touil}},
  \bibinfo {author} {\bibfnamefont {B.}~\bibnamefont {Yan}}, \bibinfo {author}
  {\bibfnamefont {S.}~\bibnamefont {Deffner}},\ and\ \bibinfo {author}
  {\bibfnamefont {W.~H.}\ \bibnamefont {Zurek}},\ }\bibfield  {title} {\bibinfo
  {title} {Redundantly amplified information suppresses quantum correlations in
  many-body systems},\ }\href {https://doi.org/10.1103/PhysRevLett.129.010401}
  {\bibfield  {journal} {\bibinfo  {journal} {Phys. Rev. Lett.}\ }\textbf
  {\bibinfo {volume} {129}},\ \bibinfo {pages} {010401} (\bibinfo {year}
  {2022})}\BibitemShut {NoStop}%
\bibitem [{\citenamefont {Touil}\ \emph
  {et~al.}(2022{\natexlab{b}})\citenamefont {Touil}, \citenamefont {Anza},
  \citenamefont {Deffner},\ and\ \citenamefont
  {Crutchfield}}]{touil2022branching}%
  \BibitemOpen
  \bibfield  {author} {\bibinfo {author} {\bibfnamefont {A.}~\bibnamefont
  {Touil}}, \bibinfo {author} {\bibfnamefont {F.}~\bibnamefont {Anza}},
  \bibinfo {author} {\bibfnamefont {S.}~\bibnamefont {Deffner}},\ and\ \bibinfo
  {author} {\bibfnamefont {J.~P.}\ \bibnamefont {Crutchfield}},\ }\bibfield
  {title} {\bibinfo {title} {Branching states as the emergent structure of a
  quantum universe},\ }\href@noop {} {\bibfield  {journal} {\bibinfo  {journal}
  {arXiv preprint arXiv:2208.05497}\ } (\bibinfo {year}
  {2022}{\natexlab{b}})}\BibitemShut {NoStop}%
\bibitem [{\citenamefont {Cusumano}(2022)}]{cusumano_quantum_2022}%
  \BibitemOpen
  \bibfield  {author} {\bibinfo {author} {\bibfnamefont {S.}~\bibnamefont
  {Cusumano}},\ }\bibfield  {title} {\bibinfo {title} {Quantum {Collision}
  {Models}: {A} {Beginner} {Guide}},\ }\href
  {https://doi.org/10.3390/e24091258} {\bibfield  {journal} {\bibinfo
  {journal} {Entropy}\ }\textbf {\bibinfo {volume} {24}},\ \bibinfo {pages}
  {1258} (\bibinfo {year} {2022})}\BibitemShut {NoStop}%
\bibitem [{\citenamefont {Rau}(1963)}]{rau_relaxation_1963}%
  \BibitemOpen
  \bibfield  {author} {\bibinfo {author} {\bibfnamefont {J.}~\bibnamefont
  {Rau}},\ }\bibfield  {title} {\bibinfo {title} {Relaxation {Phenomena} in
  {Spin} and {Harmonic} {Oscillator} {Systems}},\ }\href
  {https://doi.org/10.1103/PhysRev.129.1880} {\bibfield  {journal} {\bibinfo
  {journal} {Phys. Rev.}\ }\textbf {\bibinfo {volume} {129}},\ \bibinfo {pages}
  {1880} (\bibinfo {year} {1963})}\BibitemShut {NoStop}%
\bibitem [{\citenamefont {Nielsen}\ and\ \citenamefont
  {Chuang}(2010)}]{Nielsen2010}%
  \BibitemOpen
  \bibfield  {author} {\bibinfo {author} {\bibfnamefont {M.~A.}\ \bibnamefont
  {Nielsen}}\ and\ \bibinfo {author} {\bibfnamefont {I.~L.}\ \bibnamefont
  {Chuang}},\ }\href
  {http://scholar.google.com/scholar?hl=en\&btnG=Search\&q=intitle:Quantum+Computation+and+Quantum+Information\#0}
  {\emph {\bibinfo {title} {{Quantum Computation and Quantum Information}}}}\
  (\bibinfo  {publisher} {Cambridge University Press},\ \bibinfo {address}
  {Cambridge, UK},\ \bibinfo {year} {2010})\BibitemShut {NoStop}%
\bibitem [{\citenamefont {Takahashi}\ \emph {et~al.}(2009)\citenamefont
  {Takahashi}, \citenamefont {van Tol}, \citenamefont {Beedle}, \citenamefont
  {Hendrickson}, \citenamefont {Brunel},\ and\ \citenamefont
  {Sherwin}}]{takahashi_coherent_2009}%
  \BibitemOpen
  \bibfield  {author} {\bibinfo {author} {\bibfnamefont {S.}~\bibnamefont
  {Takahashi}}, \bibinfo {author} {\bibfnamefont {J.}~\bibnamefont {van Tol}},
  \bibinfo {author} {\bibfnamefont {C.~C.}\ \bibnamefont {Beedle}}, \bibinfo
  {author} {\bibfnamefont {D.~N.}\ \bibnamefont {Hendrickson}}, \bibinfo
  {author} {\bibfnamefont {L.-C.}\ \bibnamefont {Brunel}},\ and\ \bibinfo
  {author} {\bibfnamefont {M.~S.}\ \bibnamefont {Sherwin}},\ }\bibfield
  {title} {\bibinfo {title} {Coherent {Manipulation} and {Decoherence} of {S} =
  10 {Single}-{Molecule} {Magnets}},\ }\href
  {https://doi.org/10.1103/PhysRevLett.102.087603} {\bibfield  {journal}
  {\bibinfo  {journal} {Phys. Rev. Lett.}\ }\textbf {\bibinfo {volume} {102}},\
  \bibinfo {pages} {087603} (\bibinfo {year} {2009})}\BibitemShut {NoStop}%
\bibitem [{\citenamefont {Hansen}\ \emph {et~al.}(2001)\citenamefont {Hansen},
  \citenamefont {Kristensen}, \citenamefont {Pedersen}, \citenamefont
  {Sørensen},\ and\ \citenamefont {Lindelof}}]{hansen_mesoscopic_2001}%
  \BibitemOpen
  \bibfield  {author} {\bibinfo {author} {\bibfnamefont {A.~E.}\ \bibnamefont
  {Hansen}}, \bibinfo {author} {\bibfnamefont {A.}~\bibnamefont {Kristensen}},
  \bibinfo {author} {\bibfnamefont {S.}~\bibnamefont {Pedersen}}, \bibinfo
  {author} {\bibfnamefont {C.~B.}\ \bibnamefont {Sørensen}},\ and\ \bibinfo
  {author} {\bibfnamefont {P.~E.}\ \bibnamefont {Lindelof}},\ }\bibfield
  {title} {\bibinfo {title} {Mesoscopic decoherence in {Aharonov}-{Bohm}
  rings},\ }\href {https://doi.org/10.1103/PhysRevB.64.045327} {\bibfield
  {journal} {\bibinfo  {journal} {Phys. Rev. B}\ }\textbf {\bibinfo {volume}
  {64}},\ \bibinfo {pages} {045327} (\bibinfo {year} {2001})}\BibitemShut
  {NoStop}%
\bibitem [{\citenamefont {Mohanty}\ and\ \citenamefont
  {Webb}(1997)}]{mohanty_decoherence_1997}%
  \BibitemOpen
  \bibfield  {author} {\bibinfo {author} {\bibfnamefont {P.}~\bibnamefont
  {Mohanty}}\ and\ \bibinfo {author} {\bibfnamefont {R.~A.}\ \bibnamefont
  {Webb}},\ }\bibfield  {title} {\bibinfo {title} {Decoherence and quantum
  fluctuations},\ }\href {https://doi.org/10.1103/PhysRevB.55.R13452}
  {\bibfield  {journal} {\bibinfo  {journal} {Phys. Rev. B}\ }\textbf {\bibinfo
  {volume} {55}},\ \bibinfo {pages} {R13452} (\bibinfo {year}
  {1997})}\BibitemShut {NoStop}%
\bibitem [{\citenamefont {Herzog}\ \emph {et~al.}(1995)\citenamefont {Herzog},
  \citenamefont {Kwiat}, \citenamefont {Weinfurter},\ and\ \citenamefont
  {Zeilinger}}]{herzog_complementarity_1995}%
  \BibitemOpen
  \bibfield  {author} {\bibinfo {author} {\bibfnamefont {T.~J.}\ \bibnamefont
  {Herzog}}, \bibinfo {author} {\bibfnamefont {P.~G.}\ \bibnamefont {Kwiat}},
  \bibinfo {author} {\bibfnamefont {H.}~\bibnamefont {Weinfurter}},\ and\
  \bibinfo {author} {\bibfnamefont {A.}~\bibnamefont {Zeilinger}},\ }\bibfield
  {title} {\bibinfo {title} {Complementarity and the {Quantum} {Eraser}},\
  }\href {https://doi.org/10.1103/PhysRevLett.75.3034} {\bibfield  {journal}
  {\bibinfo  {journal} {Phys. Rev. Lett.}\ }\textbf {\bibinfo {volume} {75}},\
  \bibinfo {pages} {3034} (\bibinfo {year} {1995})}\BibitemShut {NoStop}%
\bibitem [{\citenamefont {Peng}\ \emph {et~al.}(2014)\citenamefont {Peng},
  \citenamefont {Chen}, \citenamefont {Shih},\ and\ \citenamefont
  {Scully}}]{peng_delayed-choice_2014}%
  \BibitemOpen
  \bibfield  {author} {\bibinfo {author} {\bibfnamefont {T.}~\bibnamefont
  {Peng}}, \bibinfo {author} {\bibfnamefont {H.}~\bibnamefont {Chen}}, \bibinfo
  {author} {\bibfnamefont {Y.}~\bibnamefont {Shih}},\ and\ \bibinfo {author}
  {\bibfnamefont {M.~O.}\ \bibnamefont {Scully}},\ }\bibfield  {title}
  {\bibinfo {title} {Delayed-{Choice} {Quantum} {Eraser} with {Thermal}
  {Light}},\ }\href {https://doi.org/10.1103/PhysRevLett.112.180401} {\bibfield
   {journal} {\bibinfo  {journal} {Phys. Rev. Lett.}\ }\textbf {\bibinfo
  {volume} {112}},\ \bibinfo {pages} {180401} (\bibinfo {year}
  {2014})}\BibitemShut {NoStop}%
\bibitem [{\citenamefont {Zetie}\ \emph {et~al.}(2000)\citenamefont {Zetie},
  \citenamefont {Adams},\ and\ \citenamefont {Tocknell}}]{zetie_how_2000}%
  \BibitemOpen
  \bibfield  {author} {\bibinfo {author} {\bibfnamefont {K.~P.}\ \bibnamefont
  {Zetie}}, \bibinfo {author} {\bibfnamefont {S.~F.}\ \bibnamefont {Adams}},\
  and\ \bibinfo {author} {\bibfnamefont {R.~M.}\ \bibnamefont {Tocknell}},\
  }\bibfield  {title} {\bibinfo {title} {How does a {Mach}-{Zehnder}
  interferometer work?},\ }\href {https://doi.org/10.1088/0031-9120/35/1/308}
  {\bibfield  {journal} {\bibinfo  {journal} {Phys. Educ.}\ }\textbf {\bibinfo
  {volume} {35}},\ \bibinfo {pages} {46} (\bibinfo {year} {2000})}\BibitemShut
  {NoStop}%
\bibitem [{\citenamefont {Dowson}\ and\ \citenamefont
  {Wragg}(1973)}]{dowson_maximum-entropy_1973}%
  \BibitemOpen
  \bibfield  {author} {\bibinfo {author} {\bibfnamefont {D.}~\bibnamefont
  {Dowson}}\ and\ \bibinfo {author} {\bibfnamefont {A.}~\bibnamefont {Wragg}},\
  }\bibfield  {title} {\bibinfo {title} {Maximum-entropy distributions having
  prescribed first and second moments ({Corresp}.)},\ }\href
  {https://doi.org/10.1109/TIT.1973.1055060} {\bibfield  {journal} {\bibinfo
  {journal} {IEEE Trans. Inform. Theory}\ }\textbf {\bibinfo {volume} {19}},\
  \bibinfo {pages} {689} (\bibinfo {year} {1973})}\BibitemShut {NoStop}%
\bibitem [{\citenamefont {Chung}\ \emph {et~al.}(2017)\citenamefont {Chung},
  \citenamefont {Sadler},\ and\ \citenamefont {Hero}}]{chung_bounds_2017}%
  \BibitemOpen
  \bibfield  {author} {\bibinfo {author} {\bibfnamefont {H.~W.}\ \bibnamefont
  {Chung}}, \bibinfo {author} {\bibfnamefont {B.~M.}\ \bibnamefont {Sadler}},\
  and\ \bibinfo {author} {\bibfnamefont {A.~O.}\ \bibnamefont {Hero}},\
  }\bibfield  {title} {\bibinfo {title} {Bounds on {Variance} for {Unimodal}
  {Distributions}},\ }\href {https://doi.org/10.1109/TIT.2017.2749310}
  {\bibfield  {journal} {\bibinfo  {journal} {IEEE Trans. Inform. Theory}\
  }\textbf {\bibinfo {volume} {63}},\ \bibinfo {pages} {6936} (\bibinfo {year}
  {2017})}\BibitemShut {NoStop}%
\bibitem [{\citenamefont {Bhusal}(2021)}]{bhusal_smart_2021}%
  \BibitemOpen
  \bibfield  {author} {\bibinfo {author} {\bibfnamefont {N.}~\bibnamefont
  {Bhusal}},\ }\href {http://arxiv.org/abs/2103.07081} {\bibinfo {title} {Smart
  {Quantum} {Technologies} using {Photons}}} (\bibinfo {year} {2021}),\
  \bibinfo {note} {arXiv:2103.07081 [physics, physics:quant-ph]}\BibitemShut
  {NoStop}%
\bibitem [{\citenamefont {Curtright}\ \emph {et~al.}(2014)\citenamefont
  {Curtright}, \citenamefont {Fairlie},\ and\ \citenamefont
  {Zachos}}]{curtright_concise_2014}%
  \BibitemOpen
  \bibfield  {author} {\bibinfo {author} {\bibfnamefont {T.}~\bibnamefont
  {Curtright}}, \bibinfo {author} {\bibfnamefont {D.}~\bibnamefont {Fairlie}},\
  and\ \bibinfo {author} {\bibfnamefont {C.}~\bibnamefont {Zachos}},\
  }\href@noop {} {\emph {\bibinfo {title} {A concise treatise on quantum
  mechanics in phase space}}}\ (\bibinfo  {publisher} {World Scientific},\
  \bibinfo {address} {New Jersey},\ \bibinfo {year} {2014})\BibitemShut
  {NoStop}%
\bibitem [{\citenamefont {French}(2017)}]{french_special_2017}%
  \BibitemOpen
  \bibfield  {author} {\bibinfo {author} {\bibfnamefont {A.~P.}\ \bibnamefont
  {French}},\ }\href {https://doi.org/10.1201/9781315272597} {\emph {\bibinfo
  {title} {Special relativity}}},\ \bibinfo {edition} {1st}\ ed.\ (\bibinfo
  {publisher} {CRC Press},\ \bibinfo {year} {2017})\BibitemShut {NoStop}%
\bibitem [{\citenamefont {Wootters}(1981)}]{wootters_statistical_1981}%
  \BibitemOpen
  \bibfield  {author} {\bibinfo {author} {\bibfnamefont {W.~K.}\ \bibnamefont
  {Wootters}},\ }\bibfield  {title} {\bibinfo {title} {Statistical distance and
  {Hilbert} space},\ }\href {https://doi.org/10.1103/PhysRevD.23.357}
  {\bibfield  {journal} {\bibinfo  {journal} {Phys. Rev. D}\ }\textbf {\bibinfo
  {volume} {23}},\ \bibinfo {pages} {357} (\bibinfo {year} {1981})}\BibitemShut
  {NoStop}%
\bibitem [{\citenamefont {Weisstein}()}]{weisstein_moivre-laplace_nodate}%
  \BibitemOpen
  \bibfield  {author} {\bibinfo {author} {\bibfnamefont {E.~W.}\ \bibnamefont
  {Weisstein}},\ }\href
  {https://mathworld.wolfram.com/deMoivre-LaplaceTheorem.html} {\bibinfo
  {title} {De {Moivre}-{Laplace} {Theorem}. {From} {MathWorld}—{A} {Wolfram}
  {Web} {Resource}}}\BibitemShut {NoStop}%
\bibitem [{\citenamefont {Wolf}(2012)}]{wolf_quantum_2012}%
  \BibitemOpen
  \bibfield  {author} {\bibinfo {author} {\bibfnamefont {M.~M.}\ \bibnamefont
  {Wolf}},\ }\href@noop {} {\emph {\bibinfo {title} {Quantum {Channels} \&
  {Operations} {Guided} {Tour}}}}\ (\bibinfo  {publisher} {Citeseer},\ \bibinfo
  {year} {2012})\BibitemShut {NoStop}%
\bibitem [{\citenamefont {Leonhardt}(2010)}]{leonhardt_essential_2010}%
  \BibitemOpen
  \bibfield  {author} {\bibinfo {author} {\bibfnamefont {U.}~\bibnamefont
  {Leonhardt}},\ }\href {https://doi.org/10.1017/CBO9780511806117} {\emph
  {\bibinfo {title} {Essential {Quantum} {Optics}: {From} {Quantum}
  {Measurements} to {Black} {Holes}}}},\ \bibinfo {edition} {1st}\ ed.\
  (\bibinfo  {publisher} {Cambridge University Press},\ \bibinfo {year}
  {2010})\BibitemShut {NoStop}%
\bibitem [{\citenamefont {Leonhardt}(2003)}]{leonhardt_quantum_2003}%
  \BibitemOpen
  \bibfield  {author} {\bibinfo {author} {\bibfnamefont {U.}~\bibnamefont
  {Leonhardt}},\ }\bibfield  {title} {\bibinfo {title} {Quantum {Physics} of
  {Simple} {Optical} {Instruments}},\ }\href
  {https://doi.org/10.1088/0034-4885/66/7/203} {\bibfield  {journal} {\bibinfo
  {journal} {Rep. Prog. Phys.}\ }\textbf {\bibinfo {volume} {66}},\ \bibinfo
  {pages} {1207} (\bibinfo {year} {2003})},\ \bibinfo {note}
  {arXiv:quant-ph/0305007}\BibitemShut {NoStop}%
\end{thebibliography}%


%
\end{document}